%% file: main.tex
\documentclass[a4paper,11pt]{article}
\pdfoutput=1 

\usepackage{jcappub} 

\usepackage[utf8]{inputenc}

\usepackage{siunitx}
\usepackage{framed}
\usepackage{lipsum}
\usepackage{xcolor}
\usepackage{hyperref}
\usepackage{graphicx}
\usepackage{float}
\usepackage{geometry}
\usepackage{dcolumn}
\usepackage{bm}
\usepackage{comment} 
\usepackage{subcaption}

\usepackage[section,above]{placeins}

\title{A First Look on 3D Effects in Open Axion Haloscopes}

\subheader{\hfill DESY 19-075\\
\hbox to\textwidth{\hfill MPP-2019-89}\\
\hbox to\textwidth{\hfill NORDITA-2019-049}}

\author[a,1]{Stefan~Knirck,\note{Corresponding author.}}
\author[b,1]{Jan~Sch{\"u}tte-Engel,}
\author[c,d]{Alexander~Millar,}
\author[a,e]{Javier~Redondo,}
\author[a]{Olaf~Reimann,}
\author[f]{Andreas~Ringwald,}
\author[a]{Frank~Steffen}

\affiliation[a]{Max-Planck-Institute for Physics, 80805 Munich, Germany}
\affiliation[b]{Hamburg University, 22761 Hamburg, Germany}
\affiliation[c]{The Oskar Klein Centre for Cosmoparticle Physics,
Department of Physics,
Stockholm University, AlbaNova, 10691 Stockholm, Sweden}
\affiliation[d]{Nordita, KTH Royal Institute of Technology and
Stockholm
  University, Roslagstullsbacken 23, 10691 Stockholm, Sweden}
\affiliation[e]{University of Zaragoza, 50009 Zaragoza, Spain}
\affiliation[f]{Deutsches Elektronen-Synchrotron, 22607 Hamburg, Germany}

\emailAdd{knirck@mpp.mpg.de}
\emailAdd{jan.schuette-engel@desy.de}

\abstract{\input{abstract.tex}}


\begin{document}

\maketitle
\flushbottom

\section{\label{sec:introduction}Introduction }
\input{introduction.tex}

\section{Methods for Axion-Electrodynamics}\label{sec:toolbox}
\input{maxwell_axion_equations}
\input{tools}

\section{Free Space}\label{sec:free_space}
\input{free_space}

\section{Dish Antenna}
\label{sec:pec}
\input{perfect_mirror}

\section{Dielectric Disk}
\label{sec:dieldisc}
\input{dielectric_disc}

\section{Tiled Dielectric Disk}
\label{sec:dieldisc_tiled}
\input{tilted_disc}

\section{Minimal Dielectric Haloscope}
\label{sec:dieldiscMirror}
\input{disk_mirror}

\section{Conclusions}
\label{sec:conclusion}
\input{conclusion}

\acknowledgments
We thank Erika Garutti, Béla Majorovits, Ken'ichi Saikawa, Alexander Schmidt and the MADMAX collaboration for inspiring discussions.
We are thankful to Jörn Schwandt for sharing with us his insights on finite element methods.
We thank the Max Planck Computing and Data Facility and the DESY Maxwell cluster for providing the infrastructure to run our simulations.
SK is supported by the Deutsche Forschungsgemeinschaft through Germany‘s collaborative research centre -- SFB~1258 ``Neutrinos and Dark Matter in Astro- and Particle Physics.'' JSE is supported through Germany‘s Excellence Strategy – EXC 2121 ``Quantum Universe'' – 390833306.
We thank for financial support SFB~676 ``Teilchen, Strings und das frühe Universum – Struktur von Materie und Raumzeit'' and PIER Helmholtz Graduate School. AM is supported by the
European Research Council under Grant No.~742104.

\bibliographystyle{JHEP}
\bibliography{sim_for_axion_experiments}

\end{document}

%% file: abstract.tex
We explore finite size 3D effects in open axion haloscopes
such as a dish antenna, a dielectric disk and a minimal dielectric haloscope consisting of a mirror and one dielectric disk.
Particularly dielectric haloscopes are a promising new method for detecting 
dark matter axions in the mass range above \SI{40}{\micro\electronvolt}.
By using two specialized independent approaches
--- based on finite element methods and Fourier optics ---
we compute the electromagnetic fields in these settings expected in the presence
of an axion dark matter field.
This allows us to study diffraction and near field effects for realistically sized experimental setups %
in contrast to earlier idealized 1D studies with infinitely extended mirrors and disks.
We also study axion velocity effects and disk tiling.
Diffraction effects are found to become less relevant 
towards larger axion masses and 
for the larger disk radii for example aimed at in full size dielectric haloscopes such as MADMAX.
The insights of our study not only provide a foundation for a realistic modelling 
of open axion dark matter search experiments in general, they are in particular also the first results taking into account 3D effects for dielectric haloscopes.

%% file: introduction.tex
Originally introduced to explain the absence of charge-parity violation in strong interactions (strong CP problem) \cite{PhysRevLett.38.1440,PhysRevLett.40.223,PhysRevLett.40.279}, the axion is one of the best motivated cold dark matter (CDM) candidates~\cite{DINE1983137,PRESKILL1983127,ABBOTT1983133,DAVIS1986225,LYTH1992189,Kawasaki:2014sqa,Fleury:2015aca,Ringwald:2015dsf,Fleury:2016xrz,Borsanyietal,Ballesteros:2016xej}. %
A huge amount of effort is being devoted to discovering the axion. Many of these efforts rely on the conversion of axions to photons under a strong magnetic field; for a review cf.~\cite{PhysRevD.98.030001,Graham:2015ouw,Irastorza:2018dyq}.
The first experimental efforts have been taken using resonant cavities to detect CDM axions with masses of ${m_a\sim1-\SI{40}{\micro\electronvolt}}$~\cite{PhysRevLett.51.1415,Du:2018uak,Zhong:2018rsr}. 
Such a mass range is well-motivated if the breaking of the Peccei-Quinn (PQ) symmetry happens before inflation. In this scenario the axion mass is poorly constrained with ${m_a \lesssim \SI{e-2}{\electronvolt}}$~\cite{PhysRevD.98.030001}, although high masses again require some fine-tuning in the theory~\cite{PhysRevD.82.123508}. 
However, a post-inflationary PQ symmetry breaking scenario is also possible but tends to require higher axion masses between ${\SI{20}{\micro\electronvolt} \lesssim m_a \lesssim \SI{200}{\micro\electronvolt}}$ to allow for the axion CDM matching the observed abundance~\cite{PhysRevD.85.105020,Kawasaki:2014sqa,Fleury:2015aca,Klaer:2017ond,Gorghetto:2018myk,Kawasaki:2018bzv}. 
Since CDM axions move non-relativistically, the energy of the converted photons is given by the axion mass up to small velocity corrections $\sim10^{-6} m_a$. Therefore, the photons have smaller wavelengths at higher $m_a$ which implies a smaller volume of resonant cavities. Since the output power is proportional to the cavity volume, this makes it difficult for cavity experiments to achieve good sensitivity for higher axion masses.
For this reason, various efforts are underway to develop large-scale and higher mode resonators for axion detection, see for example references~\cite{Horns:2012jf,TheMADMAXWorkingGroup:2016hpc,McAllister:2017lkb,Jeong:2017hqs,Melcon:2018dba,Baryakhtar:2018doz,Lawson:2019brd}.

In this paper we focus on effects relevant for axion searches employing volumes with typical length scales larger than a few photon wavelengths, such as dish antennas~\cite{Horns:2012jf,Suzuki:2015sza,FUNK:2017icw,Knirck:2018ojz,BRASS:website,Raaijmakers:2019hqj} or dielectric haloscopes like MADMAX~\cite{TheMADMAXWorkingGroup:2016hpc} or LAMPOST~\cite{Baryakhtar:2018doz}.
A dish antenna converts axion CDM to photons on a magnetized metallic surface.
Dielectric haloscopes generalize this idea by enhancing the axion-photon conversion by placing multiple dielectric layers in front of a metallic mirror inside a strong magnetic field. The ratio of the power emitted by such a haloscope compared to a single perfectly conducting mirror is called the power boost factor $\beta^2$. In MADMAX for example one aims to use around 80 dielectric lanthanum aluminate disks in order to achieve $\beta^2 \sim 10^4 - 10^5$~\cite{TheMADMAXWorkingGroup:2016hpc,Brun:2019lyf}.
A detailed one-dimensional discussion on the working principle of dielectric haloscopes
has been previously given~in~\cite{millar2017dielectric,Millar:2017eoc}~(``1D model'').
However, the 1D model can only consider interfaces with an infinite transverse extend. It cannot assess finite size effects such as diffraction and near fields and their effects on the search sensitivity. Therefore, in this paper we perform 3D calculations taking such effects explicitly into account.

We will first review the axion-Maxwell equations in 3D and introduce two methods to solve them for axion haloscopes in section~\ref{sec:toolbox}. First we consider specialized finite element methods (FEM) in section~\ref{subsec:fem-tools}. We present a second method in section~\ref{subsec:recFourierApproach} where we calculate the $E$-field solutions with a scalar diffraction theory based on Fourier optics.
In section~\ref{sec:free_space} we discuss the well understood solutions in free space. Afterwards we apply our methods to cases for open axion haloscopes: we consider a dish antenna in section~\ref{sec:pec}, already pointing out the primary 3D phenomena due to the finite size of the disks: diffraction and near field effects. In this context we also discuss the interplay with axion velocity effects.
We then study a dielectric disk in section~\ref{sec:dieldisc} and the effect of the tiling of multiple dielectrics into one large disk by simulating a disk glued together from two pieces in section~\ref{sec:dieldisc_tiled}.
Finally we discuss a minimal dielectric haloscope with a mirror and a dielectric disk in section~\ref{sec:dieldiscMirror}. We conclude in section~\ref{sec:conclusion}.

%% file: maxwell_axion_equations.tex
In the following we will discuss the axion-Maxwell equations and methods in order to solve them in the case of open axion haloscopes. To this end we will pursue two different methods -- specialized finite element methods and a scalar diffraction theory based on Fourier optics. 
This will later enable us to validate them with each other.
Moreover, we will cross check them against other well-understood analytical approaches specialized for a free space situation in section~\ref{sec:free_space} and for a dish antenna in section~\ref{sec:pec}.

\subsection{Axion-Maxwell Equations}
The macroscopic axion-Maxwell equations~\cite{PhysRevLett.51.1415} are
\begin{eqnarray}
\nabla \cdot \bm{D} &=& \rho_f -g_{a\gamma} \bm{B}\cdot \nabla a,\label{Maxwell_axion_inhomogeneous_matter_gauss_electric}\\ 
\nabla \times \bm{H}-\partial_t \bm{D} &=& \bm{J}_f+g_{a\gamma}(\bm{B}\partial_ta-\bm{E}\times\nabla a),\quad\label{Maxwell_axion_inhomogeneous_matter_Ampere}\\
\nabla \cdot \bm{B} &=& 0,\label{Maxwell_axion_homogeneous_matter_gauss_magnetic}\\
\nabla \times \bm{E} +\partial_t \bm{B} &=& 0,\label{Maxwell_axion_homogeneous_matter_Faraday}\\
(\Box+m_a^2)a &=& g_{a\gamma}\bm{E}\cdot \bm{B},\label{Maxwell_axion_Klein_Gordon_matter}
\end{eqnarray}
where $g_{a\gamma}$ is the axion-photon coupling, $a$ the pseudo-scalar axion field, $\bm{E}$ is the electric field, $\bm{B}$ the magnetic flux density, $\bm{D}$ the displacement field, $\bm{H}$ the magnetic field strength, $\rho_f$ the free charge density and $\bm{J}_f$ the free current density which fulfill the continuity equation ${\nabla \cdot \bm{J}_f+\dot{\rho}_f = 0}$ as in usual electrodynamics. The axion photon coupling is also often expressed in terms of the dimensionless constant $C_{a\gamma}$, the axion decay constant $f_a$ and the fine structure constant $\alpha$ as~\cite{millar2017dielectric}
\begin{equation}
g_{a\gamma}=-\frac{\alpha}{2\pi f_a}C_{a\gamma}=-1.16 \times 10^{-12}\text{GeV}^{-1}\bigg(\frac{10^9 \text{GeV}}{f_a}\bigg)C_{a\gamma},
\end{equation} 
where $C_{a\gamma}$~\cite{diCortona:2015ldu} is a model dependent quantity of order unity.

The axion-Maxwell equations \eqref{Maxwell_axion_inhomogeneous_matter_gauss_electric}--\eqref{Maxwell_axion_Klein_Gordon_matter} are a coupled system of partial differential equations (PDEs) which can be solved by existing numerical algorithms. However, it is computationally very expensive to solve this highly coupled system of PDEs. To obtain uncoupled equations, we extend the perturbation approach~\cite{Hoang:2017bwp,Kim:2018sci} to all fields, i.e., we expand all fields in $g_{a\gamma}$:
\begin{eqnarray}
X(\bm{x},t) &=& \sum_{i=0}^{\infty} g_{a\gamma}^i m_a^iX^{(i)}(\bm{x},t),
\label{perturbation_Ansatz}
\end{eqnarray}
with $X^{(i)}=\bm{E}^{(i)},\bm{D}^{(i)},\bm{B}^{(i)},\bm{H}^{(i)},\bm{J}_f^{(i)},\rho_f^{(i)}$, and $a^{(i)}$. In the expansion we have introduced a further factor of $m_a$ such that all expansion factors $g_{a\gamma}m_a$ are dimensionless.
 
The equations corresponding to zeroth order in $g_{a\gamma}$ are the well known unmodified Maxwell equations plus a free Klein-Gordon equation for the axion field.
The free current density and charge density appearing in the zeroth order case can be chosen such that one generates the desired zeroth order $\bm{E}^{(0)}$ and $\bm{B}^{(0)}$ fields.

As the coupling $g_{a\gamma}$ is very small for the viable values of $f_a \gtrsim 10^8 \text{GeV}$, it is sufficient to consider only the linear order corrections, which gives
\begin{eqnarray}
\nabla \cdot \bm{D}^{(1)}&=&\rho_f^{(1)}+\rho_a^{(1)},
\label{Maxwell_axion_inhomogeneous_matter_gauss_e_1storder}
\\
\nabla \times \bm{H}^{(1)}-\partial_t\bm{D}^{(1)}&=&\bm{J}_f^{(1)}+\bm
J_a^{(1)},
\label{Maxwell_axion_inhomogeneous_matter_ampere_1storder}
\\
\nabla \cdot \bm{B}^{(1)}&=&0,
\label{Maxwell_axion_homogeneous_matter_gauss_b_1storder}
\\
\nabla \times \bm{E}^{(1)} +\partial_t \bm{B}^{(1)}&=&0,
\label{Maxwell_axion_homogeneous_matter_faraday_1storder}
\\
 (\Box+m_a^2)a^{(1)}&=&\frac{1}{m_a}\bm{E}^{(0)}\cdot \bm{B}^{(0)},
\label{Maxwell_axion_Klein_Gordon_matter_1storder}
\end{eqnarray}
with 
\begin{eqnarray}
\rho_a^{(1)}&=& -\frac{1}{m_a} \bm{B}^{(0)}\cdot \nabla a^{(0)}, \label{eq:maxwell:axion-chargedensity} \\
\bm{J}_a^{(1)}&=&\frac{1}{m_a}(\bm{B}^{(0)}\partial_ta^{(0)}-\bm{E}^{(0)}\times\nabla a^{(0)}). \label{eq:maxwell:axion-currentdensity}
\end{eqnarray}
The perturbative approach leads to a decoupling of the first order Klein-Gordon equation from the other first order equations. It also guarantees that the free charge continuity equation $\nabla\bm{J}_f^{(1)}+\partial_t\rho_f^{(1)}=0$ and axion continuity equation $\nabla\bm{J}_a^{(1)}+\partial_t\rho_a^{(1)}=0$ hold. Of course this statement is correct for all orders. Physically this means that the zeroth order fields induce first order fields, e.g.,~by inducing first order charge and current densities. The first order charge and current density source the first order $E$ and $B$-fields, which then induce second order fields and so on. Furthermore it becomes clear that axions are sourced only at first order in $g_{a\gamma}$ if $\bm{E}^{(0)}\cdot\bm{B}^{(0)}\neq 0$. The back reaction of photons to sourced axions will always be at higher order, and will thus be negligible.

We consider perfect conductors as inner boundaries of the simulation domain. Objects with finite conductivity $\sigma$ can be included by setting ${\bm{J}_f^{(1)}=\sigma \bm{E}^{(1)}}$ inside.
In this work we further assume that linear constitutive relations are fulfilled ${\bm{D}=\epsilon \bm{E}}$, ${\bm{H}=\mu^{-1} \bm{B}}$, where $\epsilon$ is the relative electric permittivity and $\mu$ the relative magnetic permeability. If we insert them into the equations \eqref{Maxwell_axion_inhomogeneous_matter_gauss_e_1storder}--\eqref{Maxwell_axion_Klein_Gordon_matter_1storder} and combine\footnote{Equations~\eqref{Maxwell_axion_inhomogeneous_matter_gauss_e_1storder} and \eqref{Maxwell_axion_homogeneous_matter_gauss_b_1storder} can be treated as initial conditions to be valid at $t=t_0$ if the two continuity equations hold separately. In the case of harmonic time dependence  equations~\eqref{Maxwell_axion_inhomogeneous_matter_gauss_e_1storder} and \eqref{Maxwell_axion_homogeneous_matter_gauss_b_1storder} are trivially fulfilled.} equation \eqref{Maxwell_axion_inhomogeneous_matter_ampere_1storder} with equation \eqref{Maxwell_axion_homogeneous_matter_faraday_1storder} we obtain a wave equation for $\bm{E}^{(1)}$:
\begin{eqnarray}
\nabla\times (\mu^{-1}\nabla \times \bm{E}^{(1)} )+ \partial_t^2 \epsilon\bm{E}^{(1)}+\sigma\partial_t\bm{E}^{(1)}+\partial_t\bm{J}^{(1)}_a=0.
\label{vectorized_E_wave_equation_lin_media}
\end{eqnarray}
This wave equation is a PDE for $\bm{E}^{(1)}$ which can be solved with standard numerical methods.

As dark matter must be highly non-relativistic, the axion de Broglie wavelength is $\lambda_{\rm dB} \sim \SI{10}{\metre}\,(10^{-3} / v)\, (\SI{100}{\micro\electronvolt} / m_a)$ with $v$ the axion velocity.
Therefore, the axion field $a^{(0)}$ can be treated as spatially constant over the size of the experiment.
For small axion velocities one can model the background axion field as the real part of $a^{(0)}(\bm{x},t)=a^{(0)}(t)=a_0 e^{-i\omega t}$, with a constant $a_0$. Throughout this paper we will always assume that the axion velocity and therefore the gradient of the axion field is negligible. Only in section~\ref{sec:pec} we describe what happens when we also take velocity effects into account.
To simulate an axion haloscope we assume a strong and static external $B$-field $\bm{B}^{(0)}(\bm{x})$ and no external $E$-field\footnote{More generally $\bm{E}^{(0)}$ only appears in~\eqref{vectorized_E_wave_equation_lin_media_zerovel_EBharmonic} if one assumes that the axions have a velocity.} $\bm{E}^{(0)}=0$. Defining the axion induced field as
\begin{eqnarray}
\bm{E}_a(\bm{x})\equiv -g_{a\gamma}a_0\bm{B}^{(0)}(\bm{x}),\label{eq:axionmaxwell:axion_induced_field}
\end{eqnarray}
equation~\eqref{vectorized_E_wave_equation_lin_media} reduces in this case to
\begin{eqnarray}
\nabla\times (\mu^{-1}\nabla \times \bm{E} ) - \omega^2 \tilde{\epsilon}\bm{E} =- m_a^2 \bm{E}_a,
\label{vectorized_E_wave_equation_lin_media_zerovel_EBharmonic}
\end{eqnarray}
with $\omega=m_a$ (from the zeroth order Klein-Gordon equation), the complex permittivity $\tilde{\epsilon}\equiv \epsilon(1+\frac{i\sigma}{\omega\epsilon})$ and the physical first order $E$-field $\bm{E} = g_{a\gamma} m_a \bm{E}^{\rm (1)}$. Furthermore, $\bm{E}$ has harmonic time dependence, because of our time harmonic choice of $a^{(0)}$ and the assumption that time and spatial coordinates in $\bm{E}$ are separable.

When discussing the free space solution of equation~\eqref{vectorized_E_wave_equation_lin_media_zerovel_EBharmonic} in section~\ref{sec:free_space} we will give the axion induced field a physical meaning. In this work we define the external magnetic field as  ${\bm{B}^{(0)}=\hat{B}^{(0)}\hat{\bm{B}}^{(0)}(\bm{x})}$, where $\hat{B}^{(0)}$ is the constant magnitude and $\hat{\bm{B}}^{(0)}(\bm{x})$ contains the spatial dependency and is of order one. We use the symbol $E_a$ both for the constant field ${E_a=g_{a\gamma}a_0\hat{B}^{(0)}}$ as well as for the magnitude $|\bm{E}_a(\bm{x})|={E_a=g_{a\gamma}a_0B^{(0)}}(\bm{x})$, since from the context it will be clear what case we mean. 
In this paper we further assume only pure dielectric materials ($\mu=1$) without losses ($\sigma=0$, $\tilde{\epsilon}=\epsilon$) or perfect electrical conductors (PEC) as described above. 

We would like to stress, that the formalism applied here can also be applied to hidden photons~\cite{Horns:2012jf}, where we would have to expand the the dark $E$ and $B$-fields instead of the pseudo-scalar axion field. The source term in equation~\eqref{vectorized_E_wave_equation_lin_media_zerovel_EBharmonic} would then include the dark $E$ and $B$-fields. As for the axion field, they can be treated as spatially constant for non-relativistic CDM velocities and sufficiently small masses.

%% file: tools.tex
\subsection{Specialized Finite Element Methods (FEM)}\label{subsec:fem-tools}

We explicitly solve equation~\eqref{vectorized_E_wave_equation_lin_media_zerovel_EBharmonic} for complex geometries numerically with the finite element method (FEM)~\cite{zienkiewicz1977finite}. In order to resolve the wave structure sufficiently precise, the space discretization (``mesh'') needs 5 to 10 mesh points per photon wavelength. 
In this paper we consider the commercially available COMSOL Multiphysics~\textregistered~\cite{COMSOL} (wave-optics module) and the open-source package Elmer \cite{elmer,elmer-waveguide-paper} (VectorHelmholtz module).
Using two different FEM tools enables us to cross-check the results obtained with both tools and to possibly disentangle systematics from the different algorithms or meshes from physical effects. Moreover, in this paper we will test our simulations against analytical well understood cases and semi-analytical methods described in the next sections.

Previous axion-electrodynamics studies~\cite{Hoang:2017bwp,Krawczyk:2018wma,Jeong:2017hqs} for cavity setups aim to find modes of closed cavities. In the case of a dish antenna and dielectric haloscopes we are faced with open situations.
Therefore in Elmer we use Robin boundary conditions to describe an open system
\begin{align}
\mathbf{n} \times \nabla \times \mathbf{E} - \alpha \mathbf{n} \times (\mathbf{n} \times \mathbf{E}) = g \quad \text{on} ~ \Gamma, \label{eq:robin-bc}
\end{align}
with $\mathbf{n}$ the normal vector on the simulation boundary $\Gamma$ and $\alpha = ik, g = 0$ (impedance boundary conditions)~\cite{elmer-models-manual}.  In COMSOL we use more sophisticated perfectly matched layer (PML) boundary conditions \cite{BERENGER1994185,COMSOL}, which absorb a large amount of impinging radiation even under a large incidence angle. 

\begin{figure}
\centering \includegraphics[width=0.7\textwidth]{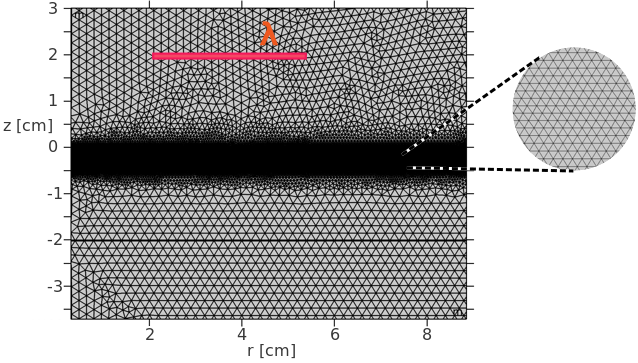}
  \caption{Exemplary mesh for a minimal dielectric haloscope, which consists of a circular PEC at $z=\SI{0.02}{\metre}$ and a dielectric disk of thickness $\SI{0.5}{\centi\metre}$, as considered in section~\ref{sec:dieldiscMirror}. The upper side of the disk is located around $z=0$. The shown segment is in the $r,z$ plane in which we calculate the fields in the 2D3D approach. $r=\sqrt{x^2+y^2}$ is the radial coordinate. }
  \label{fig:mesh} 
\end{figure}

FEM can handle 3D geometries which have typical length scales of a few wavelengths in all three spatial dimensions. In this case both direct and iterative solvers may be used.
In Elmer for example we obtain the 3D solution always by using a tuned ``Biconjugate Gradient Stabilized" iterative solver \cite{barrett1994templates,freund1993transpose,elmer-models-manual}, while in COMSOL we use direct solvers~\cite{COMSOL, Antmannetal}.
While iterative solvers may suffer from insufficient convergence, a direct solver directly calculates the solution to the matrix system. On the other hand a direct solver is much more memory intensive and therefore computationally demanding.

In order to reduce the size of the problem by one dimension we exploit the radial symmetry (``2D3D approach''). The reduction by one dimension reduces the computational costs when using a direct solver~\cite{Antmannetal}. This makes it possible to do parameter sweeps in a reasonable time (cf. section~\ref{sec:dieldisc} and \ref{sec:dieldiscMirror}) and also to use the FEM in the future for much larger geometries than considered in this paper. Furthermore we can now use a finer mesh. An example for a setup of a dielectric disk and a circular PEC is shown in figure~\ref{fig:mesh}. The usage of the radial symmetry is possible even though the external $B$-field -- and therefore also the source term $\bm{E}_a$ -- is linearly polarized and breaks the radial symmetry.
We achieve this by decomposing\footnote{For more complicated source terms one can also chose a more general decomposition $\sum_{m\in\mathbb{Z}}\tilde{\bm{E}}^m_a e^{im\phi}$. The steps which we are doing will go through in exact analogy, but one has to do $M$ 2D simulations in the end, where $M$ is the number of $\tilde{\bm{E}}^m_a$'s which are nonzero.} $\bm{E}_a$ as a sum of radially symmetric fields:
\begin{equation}
\bm{E}_a(r,z)=\bm{E}_a^+(r,\phi,z)+\bm{E}_a^-(r,\phi,z),
\label{eq:decompose_linear_source}
\end{equation}
where we have assumed that the $\bm{E}_a=-E_a(r,z)\hat{\bm{e}}_y$ depends only on $r$ and $z$ and that the external $B$-field points in $y$-direction $\bm{B}^{(0)}\sim \hat{\bm{e}}_y$. $r=\sqrt{x^2+y^2}$ is the radial coordinate.
With $m=\pm 1$ we can write
\begin{equation}
\bm{E}_a^{m}=\tilde{\bm{E}}_a^{m}e^{im\phi}=\frac{E_a(r,z)}{2}(-\hat{\bm{e}}_{\phi}+im \hat{\bm{e}}_{r})e^{im\phi}.
\end{equation}
Since we are only interested in a constant external $B$-field over the dielectric haloscope or the dish antenna, it is not important for us that the external magnetic field can only depend on $r$ and $z$.

This ansatz splits the problem into the solution of two independent differential equations, each for one source term $m=\pm 1$, respectively:
\begin{eqnarray}
\nabla\times(\nabla\times \bm{E}^m)-k_0^2\epsilon\bm{E}^m=-k_0^2 \bm{E}_a^m,
\label{eq:vectorized_helmholtz_modeDecomposed}
\end{eqnarray}
where $k_0 = \omega$.
For the corresponding $E$-field solution we make the ansatz $\bm{E}^m=\tilde{\bm{E}}^m(r,\phi,z)e^{im\phi}$ with
\begin{equation}
\tilde{\bm{E}}^m=\tilde{E}^m_{r}(r,z)\hat{\bm{e}}_{r}+\tilde{E}^m_{\phi}(r,z)\hat{\bm{e}}_{\phi}+\tilde{E}^m_{z}(r,z)\hat{\bm{e}}_{z},
\end{equation}
and $e^{im\phi}$ factors out in equation~\eqref{eq:vectorized_helmholtz_modeDecomposed}. Since the $\phi$ dependence is known, we solve the equation for $\phi=0$ in the $r,z$ plane. We use COMSOL for the FEM simulation to compute the unknown functions $\tilde{E}^m_{r}(r,z), \tilde{E}^m_{\phi}(r,z)$ and $\tilde{E}^m_{z}(r,z)$ on the discretized mesh.
The total solution is obtained as a superposition:
\begin{eqnarray}
\bm{E}=\tilde{\bm{E}}^{+}e^{i\phi}+\tilde{\bm{E}}^{-}e^{-i\phi}.
\end{eqnarray}
Therefore, we obtain the full 3D solution for a radial symmetric geometry by solving two ($m=\pm 1$) 2D problems. The strategy of decomposing the linear source term as in equation~\eqref{eq:decompose_linear_source} is used as well for example in the context of the simulation of radial symmetric antennas~\cite{Comsol_horn_antenna}. The decomposition can be interpreted as a decomposition into left and right circular polarized fields~\cite{pedrotti1993introduction}. Here we apply this strategy to axion-electrodynamics for the first time. The $H$-field can be calculated via ${\bm{H}=\frac{\nabla\times\bm{E}}{i\omega \mu}}$ and the time averaged pointing vector in the far field $\bar{\bm{S}}=-\frac{1}{2}\text{Re}\bm{E}\times\bm{H}^*$. A straightforward calculation for the power in $z$-direction which goes through a circular surface at position $z$ yields
\begin{eqnarray}
P_z=-\text{Re}\frac{i\pi}{\omega \mu}\sum_{m=\pm1}\int_0^{R_p}dr \tilde{E}^m_{\phi} \left(r \partial_{z}\tilde{E}^{m*}_{\phi}+i m \tilde{E}^{m*}_{z}\right)
+r \tilde{E}^{m}_{r} \left(\partial_{z}\tilde{E}^{m*}_{r}-\partial_{r}\tilde{E}^{m*}_{z}\right),
\label{eq:Powerz_full}
\end{eqnarray}
where $R_p$ is the radius of the circular surface, where we consider the power flux. 

\subsection{Recursive Fourier Propagation}\label{subsec:recFourierApproach}

When solving equation~\eqref{vectorized_E_wave_equation_lin_media_zerovel_EBharmonic} in geometries with dielectric materials and conductors we encounter solutions with two different dispersion relations~\cite{Millar:2017eoc}. One dispersion relation is axion-like $k=0$ in the zero velocity limit. The second dispersion relation is the usual photon dispersion $k^2=n^2\omega^2$ known from electrodynamics. Rather than trying to detect axion-like mass states, which simply give a stationary $E$-field, a dielectric haloscope or dish antenna uses translation invariance to convert them into propagating photon-like states, which can then exit the system. 
To describe them we exploit the well established scalar diffraction theory of Fourier optics in this section. A scalar diffraction theory is applicable if the optical system is much larger than the photon wavelength~\cite{jackson_classical_electrodynamics_1999} of the propagating fields and the fields are propagating along one preferred axes. The optical size of a dish antenna is given by the dish diameter, while for a dielectric haloscope the optical size is given by the diameter of the dielectric disks. In the proposed dielectric halsocopes MADMAX~\cite{TheMADMAXWorkingGroup:2016hpc,Brun:2019lyf} and LAMPOST~\cite{Baryakhtar:2018doz} as well as in dish antenna experiments such as BRASS~\cite{BRASS:website} this condition will be fulfilled.

As we discuss more explicitly in section~\ref{sec:pec}, the solutions with a photon dispersion relation are target of detection in many axion haloscopes. These can be described by using the classical Maxwell equations, i.e., \eqref{Maxwell_axion_inhomogeneous_matter_gauss_e_1storder}--\eqref{Maxwell_axion_homogeneous_matter_faraday_1storder} with $g_{a\gamma}=0$.
When we combine the Maxwell-Faraday equation and the Ampere law of the classical Maxwell equations, we obtain
\begin{eqnarray}
-\nabla^2\bm{E}-\omega^2n^2\bm{E}=0,
\label{eq:wave_equation_decoupled}
\end{eqnarray}
with the refractive index $n^2=\epsilon$.
In equation \eqref{eq:wave_equation_decoupled} we have neglected the term $\nabla(\nabla\cdot\bm{E})$, i.e., there are no charges $\rho=0$, but $\bm{J}\neq 0$ is still possible. Therefore, this approach neglects near fields generated by any induced charge distributions. We can treat all three components of $\bm{E}$ as scalar fields, since all three components will be independent of each other in equation~\eqref{eq:wave_equation_decoupled}. 
Equation~\eqref{eq:wave_equation_decoupled} is a wave equation and is solved by plane waves fulfilling the photon dispersion relation. The general solution for each component of the electric field $E_i$ can therefore be obtained using a Fourier approach
\begin{equation}
E_i(\bm{x})=\int_{\mathbb{R}^3}\frac{dk^3}{(2\pi)^3}\hat{E}_i(\bm{k})e^{i\bm{k}\cdot\bm{x}},
\label{eq:Mag_E_FT}
\end{equation} 
where $\bm{x} = (x,y,z)$. Let $\bm{E}(x,y)$ be the field at $z=z_S$, then the propagated field is~\cite{GoodmanFourierOptics,jackson_classical_electrodynamics_1999}
\begin{eqnarray}
E_i(\bm{x})=
\int_{\mathbb{R}^2}\frac{dk_xdk_y}{(2\pi)^2}\mathcal{F}(E_i)(k_x,k_y) e^{i|z-z_S|k_z(k_x,k_y)}e^{ik_xx}e^{ik_yy},
\label{Fourier_approach}
\end{eqnarray}
where $\mathcal{F}$ denotes the two dimensional Fourier transformation and where $k_z(k_x,k_y)$ is given by the photon dispersion relation
\begin{eqnarray}
    k_z(k_x,k_y)=\sqrt{(\omega n)^2-k_x^2-k_y^2}.
    \label{eq:photon_dispersion_relation}
\end{eqnarray}
Therefore, in comparison to the 1D case, where the phase evolves as $e^{i \omega n z}$, the phase evolves more slowly as $e^{i k_z z}$ in 3D.
It is evident from equations~\eqref{eq:wave_equation_decoupled} and \eqref{Fourier_approach} that all components of the $E$-field are propagated independently. If one $E$-field component on the surface at $z_S$ is zero, it will be zero in the whole space. 

In section \ref{sec:pec} we will use the presented scalar theory to compute the fields and the power which is emitted by a dish antenna.
Note that there are also other formulations of a scalar diffraction theory such as the one from Kirchhoff and Rayleigh~\cite{jackson_classical_electrodynamics_1999} suitable to obtain far field approximations.
We will furthermore compare the scalar theory to a full vectorial treatment taking into account near fields and boundary charges. We demonstrate that the scalar approach here is sufficient to describe the $E$-field component which is parallel to the polarization direction of the external $B$-field. Recall that this is the component directly coupled to the axion.

\begin{figure}
    \centering
    \includegraphics[width=\textwidth]{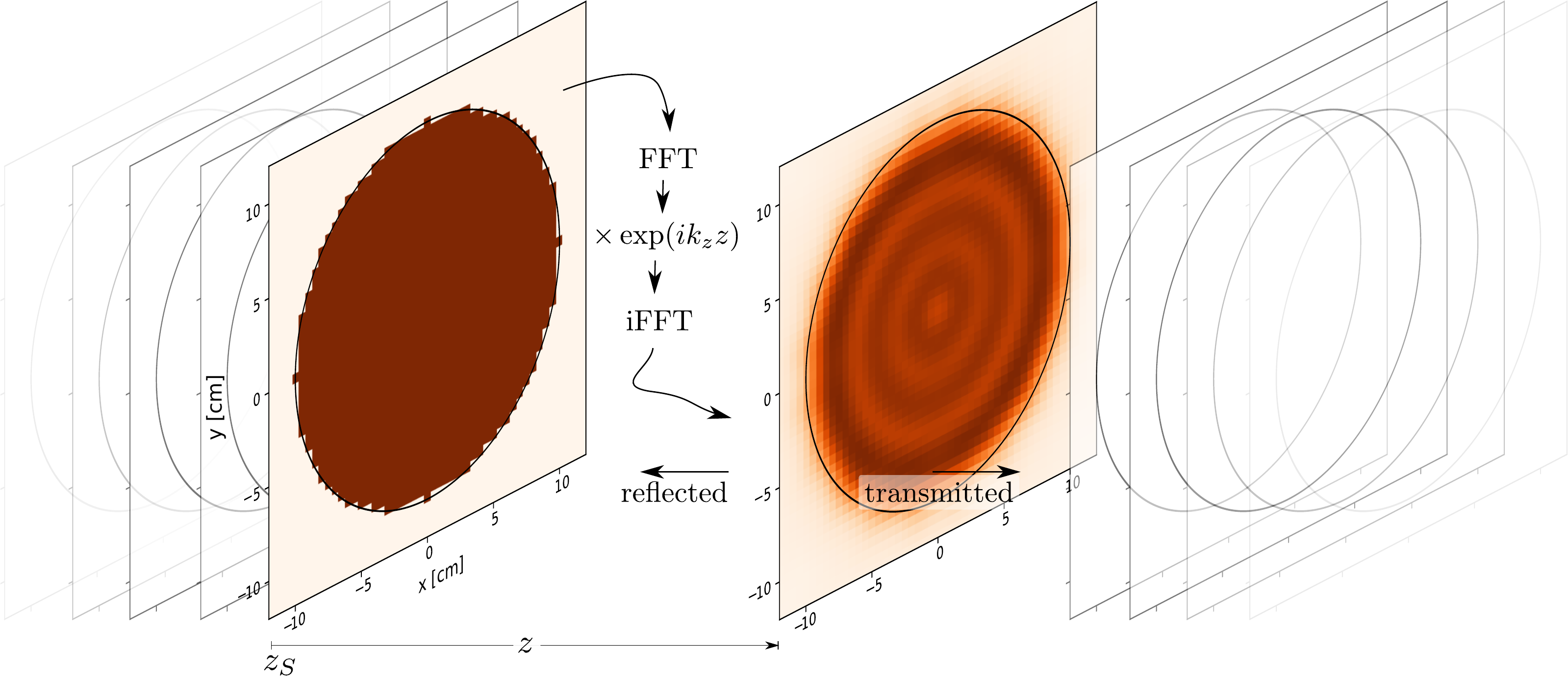}
    \caption{Illustration of the recursive Fourier propagation approach. The field on the left surface in brown is given and propagated to the next surface by applying equation~\eqref{Fourier_approach}. As one can see from \eqref{Fourier_approach} the propagation consists of a Fourier transformation of the initial field and a backtransformation which is taking into account the phase factor $ \exp(i k_z z)$. The implementation can be done with the fast Fourier transformation (FFT) algorithm and the corresponding inverse (iFFT). We also illustrate that the fields are defined on a discretized grid over the different slices. The approach takes into account diffraction phenomena arising from the finite size disks. Repeating the propagation recursively and applying reflection and transmission coefficients at each slice, a 3D multilayer system can be calculated.}
    \label{fig:Recursive_fourier_approach_sketch}
\end{figure}

In sections \ref{sec:dieldisc} and \ref{sec:dieldiscMirror} we demonstrate that one can also use the scalar diffraction theory to compute the 3D fields for more complex systems such as a dielectric disc and a minimal dielectric haloscope. This can be achieved by applying the Fourier propagation approach recursively to propagate the radiation coming from the different interfaces in the system as sketched in figure~\ref{fig:Recursive_fourier_approach_sketch}.
After each step we apply transmission and reflection coefficients when the radiation hits another interface. In this way we are able to recursively construct the $E$-field solution in 3D for systems with more than one interface. 
Note that one can store fields at all layers simultaneously and sum up the reflected and transmitted fields at each layer after each iteration step, such that the numerical complexity grows linear with the number of iteration steps. Note we can use the fast Fourier transformation (FFT) algorithm. In addition, this approach also requires only a discretization in $x$ and $y$ dimensions with grid spacings according to the Nyquist theorem of $\approx \lambda / 2$, and thus makes this method numerically efficient.
A more detailed description of this approach is given for the example of a single dielectric disk in section~\ref{sec:dieldisc}.

%% file: free_space.tex
As a first example we consider a vacuum domain, only containing CDM axions and a static magnetic field. We refer to this in the following as free space.
More precisely we assume that the external magnetic field $\bm{B}^{(0)}$ changes over scales larger than the Compton wavelength of the axion, which suppresses effects due to the inhomogeneity of the $B$-field~\cite{Redondo:2010js}.
It is easy to see that the axion induced field in equation~\eqref{eq:axionmaxwell:axion_induced_field} approximately solves equation~\eqref{vectorized_E_wave_equation_lin_media_zerovel_EBharmonic}, since in the case when $\bm{E}_a$ changes over length scales much larger than a wavelength the derivative terms are negligible.
To avoid the effects from these gradients in our simulations, we let $\bm{B}^{(0)}$ smoothly drop to zero towards the boundaries, such that the first derivative at the boundary is zero and the drop stretches over a $ \text{few}\,\lambda$. 
Because in this case we do not have dominating emitted propagating fields, we will only compare the results of our FEM calculations with each other and with dedicated analytical solutions in this section.

Since we are in free space we have $\mu=1,~\epsilon=1,~\rho_f=0$, and~$\bm{J}_f=0$.
A full analytic solution can be obtained with the theory of retarded potentials for the axion charge and current terms~\cite{Ouellet:2018nfr,Beutter:2018xfx} as in equations~\eqref{eq:maxwell:axion-chargedensity} and  \eqref{eq:maxwell:axion-currentdensity},
because in our approach the axion-Maxwell equations are decoupled.
This gives an electric field of
\begin{eqnarray}
\bm{E}(\bm{x})&=& E_a \Bigg [ \omega^2 \int d^3x' \hat{\bm{B}}^{(0)}(\bm{x}')G - \int d^3x'~~\nabla'G~\nabla'\cdot \hat{\bm{B}}^{(0)}(\bm{x}') \Bigg ], \label{eq:free_space_retarded_potentials}
\end{eqnarray}
with ${E_a=g_{a\gamma}\hat{B}^{(0)} a_0}$ the axion induced field in an idealized 1D calculation~\cite{millar2017dielectric}.
Further, $\bm{B}^{(0)}=\hat{B}^{(0)}\hat{\bm{B}}^{(0)}(\bm{x})$, where $\hat{B}^{(0)}$ is the constant magnitude of the external $B$-field and $\hat{\bm{B}}^{(0)}(\bm{x})$ contains the spatial dependency and is of order one.
$G$ is the Green's function of the scalar wave equation ${G(\bm{x},\bm{x}')=\frac{e^{i\omega |\bm{x}-\bm{x}'|}}{4\pi|\bm{x}-\bm{x}'|}}$.
Note that in the FEM simulations we have to put an external $B$-field by hand which can be non-physical, i.e. ${\nabla\cdot \hat{\bm{B}}^{(0)}\neq 0}$. Such a non-physical $B$-field with small gradients is not problematic as we will see later. However, it leads to the second term in the analytical solution in equation~\eqref{eq:free_space_retarded_potentials}. It comes from an artificial charge density ${\partial_t\rho_{\text{art}}^{\rm (1)}=-\nabla\cdot\bm{J}_a^{(1)}}$ which guarantees that the continuity equation is also fulfilled. If this term is not included the retarded potential solution is not equivalent to the FEM solution of the vectorized Helmholtz equation \eqref{vectorized_E_wave_equation_lin_media_zerovel_EBharmonic}. Note that since the $B$-field changes only on length scales much larger than the photon-wavelength we can express the total $E$-field as the axion-induced field plus a small radiative correction in the following:
\begin{eqnarray}
   \bm{E}(\bm{x})= \bm{E}_a+\bm{E}_{\text{rad}}.
\end{eqnarray}
Again, if the $B$-field gradients are small on scales of the wavelength, then the radiative corrections are small and the total axion-induced $E$-field follows the shape of the external $B$-field $\bm{B}^{(0)}$~\cite{Redondo:2010js}.

\begin{figure}
\centering
  \begin{subfigure}{6cm}
    \includegraphics[width=\textwidth]{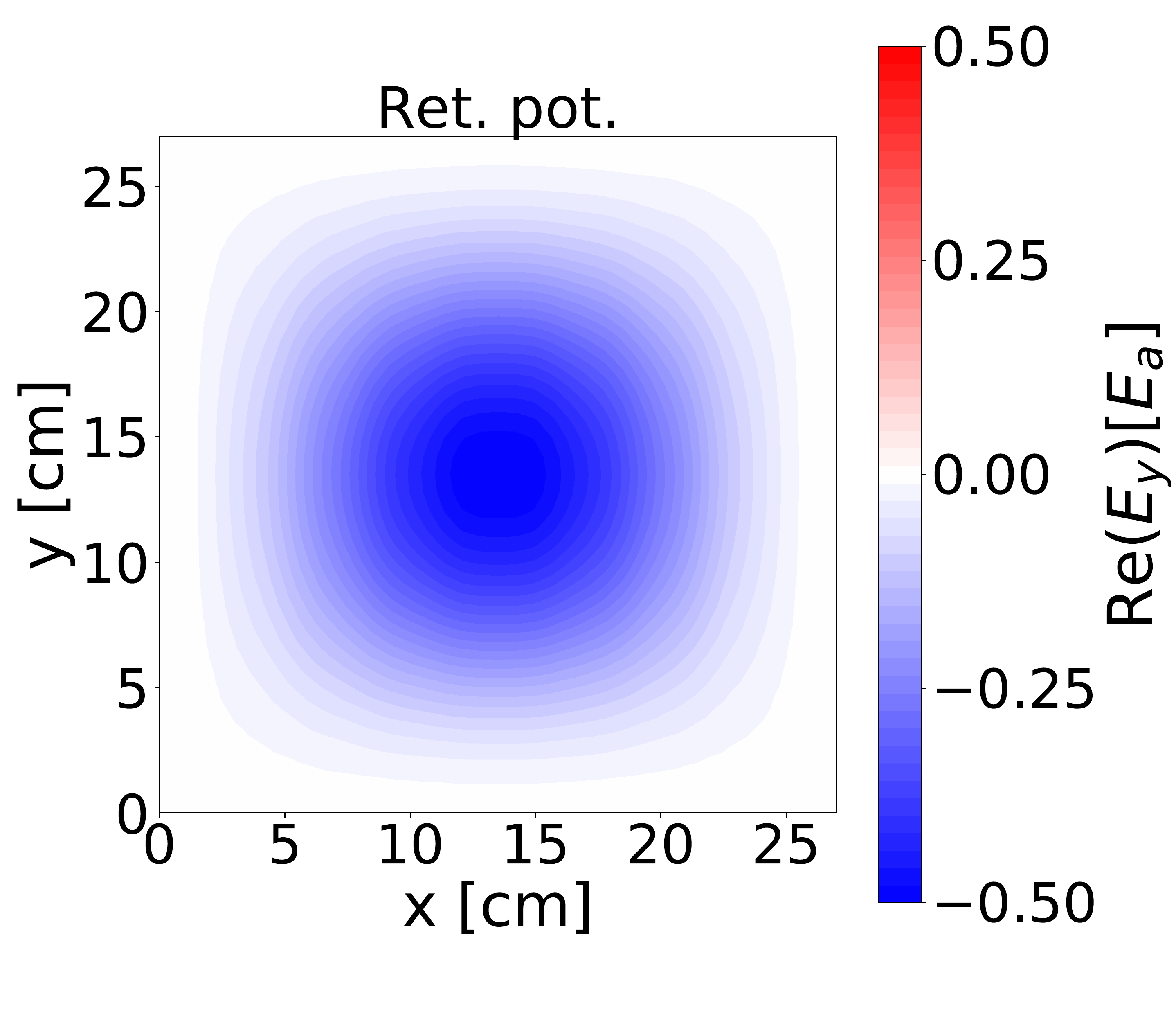}
    \caption{}\label{fig:free_space:RP}
  \end{subfigure}
  \begin{subfigure}{6cm}
    \includegraphics[width=\textwidth]{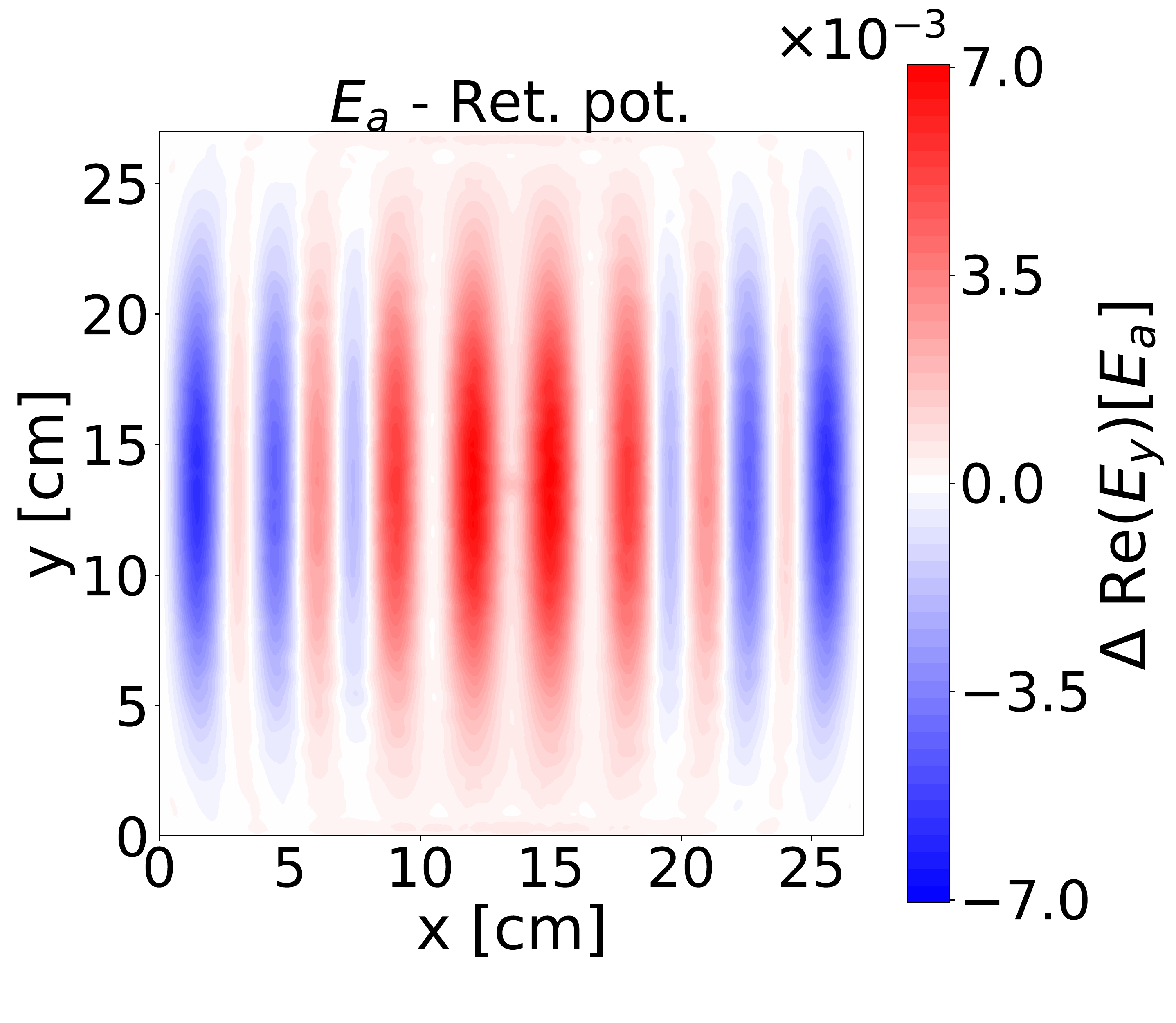}
    \caption{}\label{fig:free_space:RPvsSA}
  \end{subfigure}

  \begin{subfigure}{6cm}
    \includegraphics[width=\textwidth]{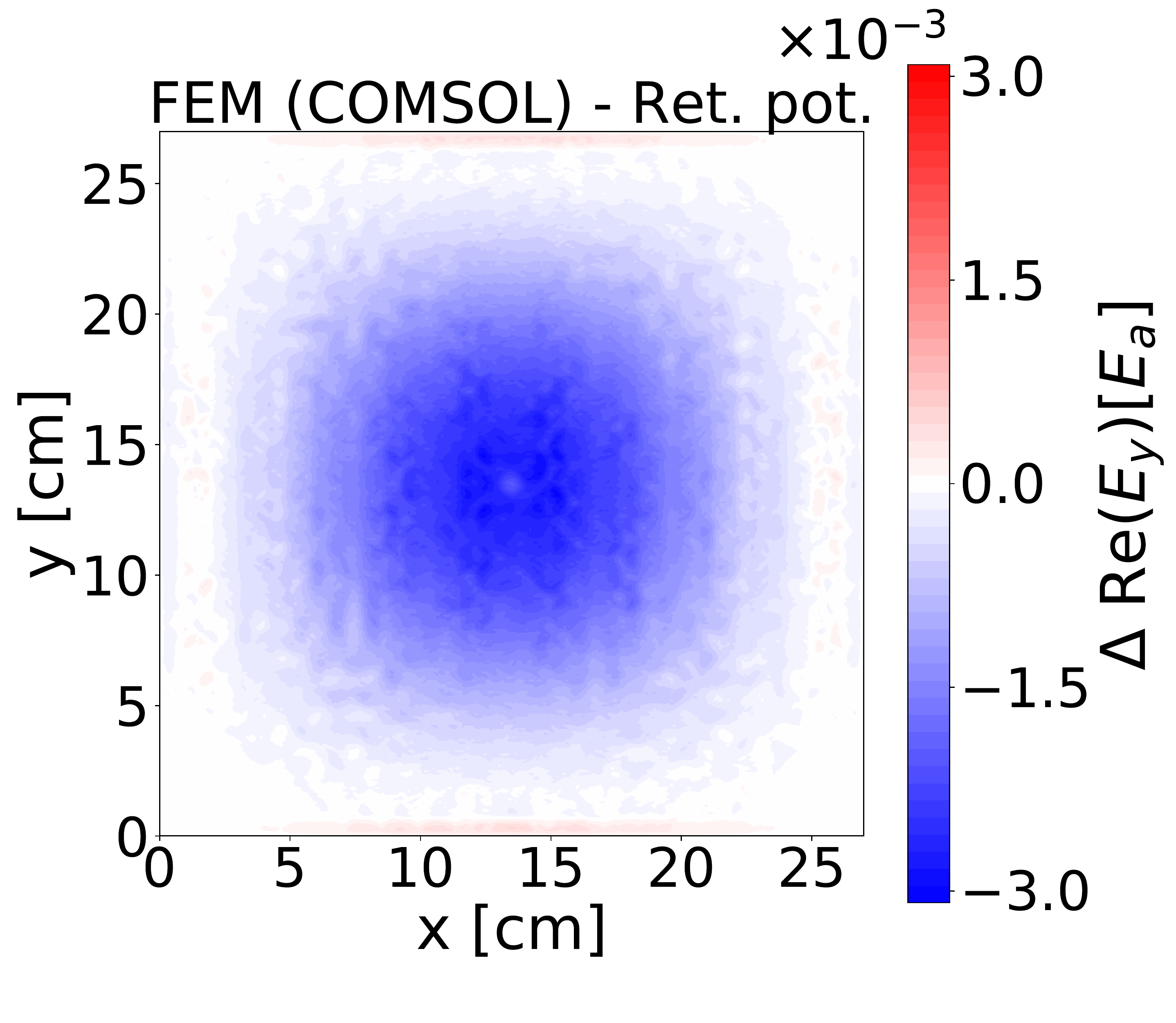}
     \caption{}\label{fig:free_space:RPvsC}
  \end{subfigure}
  \begin{subfigure}{6cm}
   \includegraphics[width=\textwidth]{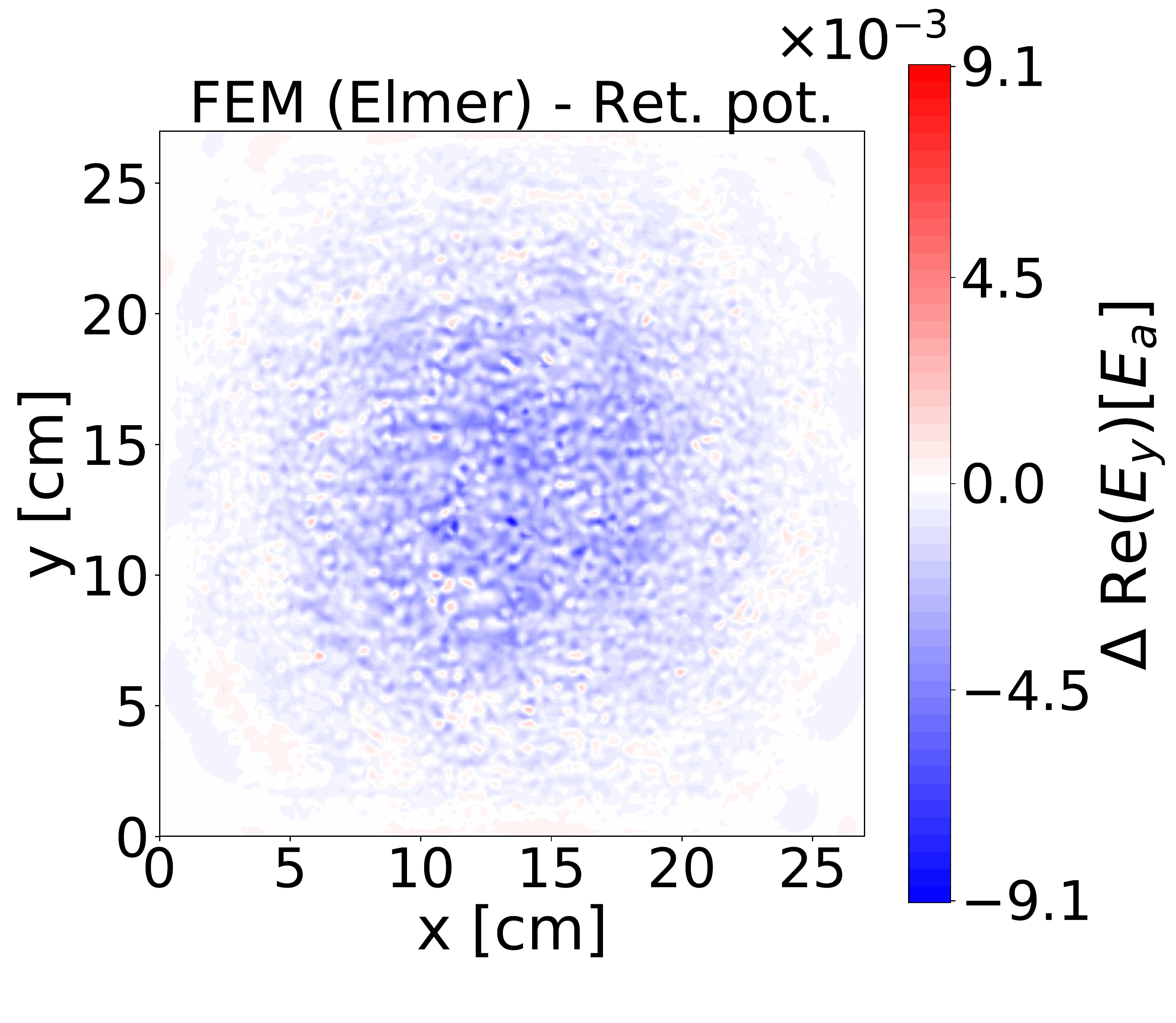}
   \caption{}\label{fig:free_space:RPvsE}
  \end{subfigure}
  \caption{ Assuming a CDM axion background field for $m_a \approx \SI{40}{\micro\electronvolt}$ and an external $B$-field pointing in $y$-direction as given in equation~\eqref{eq:free_space:magnetic_field}, the real part of the resulting $E_y$ field at $z=\SI{15}{\centi\metre}$ is shown in panel (a) as obtained with the method of retarded potentials (ret. pot.), cf.~equation~\eqref{eq:free_space_retarded_potentials}, whereas the other panels show the respective differences to this result when (b)~radiative corrections are neglected, and (c)~COMSOL and (d)~Elmer are used in the computation of ${\rm Re}(E_y)$. The considered simulation domain is a box with $9\lambda \times 9\lambda \times 6.7\lambda$ extend and the frequency of the axion-induced field is \SI{10}{\giga\hertz} as given by the assumed $m_a$ value.}
  \label{fig:freespace} 
\end{figure}

Specifically, let us take for now
\begin{equation}
 \hat{\bm{B}}^{(0)}(\bm{x})=\hat{\bm{e}}_y \sin^2(\frac{\pi x}{L_x})\sin^2(\frac{\pi y}{L_y})\sin^2(\frac{\pi z}{L_z}),
 \label{eq:free_space:magnetic_field}
 \end{equation}
inside the simulation domain $[0,L_x]\times [0,L_y]\times [0,L_z]$, with ${L_x = L_y = 9 \lambda}$, $L_z = \SI{20}{\centi\metre} \approx 6.7\lambda$ and zero outside. For simplicity we have chosen the external $B$-field in equation~\eqref{eq:free_space:magnetic_field} such that it is non-physical, i.e. it has a small divergence $\nabla\cdot\hat{\bm{B}}^{(0)}\neq0$,  since we want a $B$-field which is zero at the boundary of the simulation domain to avoid additional propagating fields coming from the boundaries of the simulation domain. 
Evaluating equation~\eqref{eq:free_space_retarded_potentials} numerically leads to the fields shown in figure~\ref{fig:freespace}\,(a). Figure~\ref{fig:freespace}\,(b) shows the radiative correction, obtained when subtracting the axion-induced field from equation~\eqref{eq:axionmaxwell:axion_induced_field}.
The radiative correction is much smaller than the axion induced field. This explicitly confirms that assuming a constant magnetic field for  haloscope experiments is valid as long as the magnetic field only drops over large length scales.

We show the subtraction of the analytical solutions from the results obtained with our FEM tools in figure~\ref{fig:freespace}\,(c) for COMSOL and in figure~\ref{fig:freespace}\,(d) for Elmer. In addition to COMSOL and Elmer using different meshes, we further verify the independence on the chosen absorbing boundary condition by using different boundary conditions for the two solvers, i.e., impedance boundary conditions in Elmer and PML in COMSOL. Since the radiative correction shown in figure~\ref{fig:freespace}\,(b) does not reappear, we conclude that both solvers calculated the radiative correction $\sim \mathcal{O}(10^{-3}) E_a$ correctly. Even smaller systematics remain, which we think are most likely attributed to boundary conditions not perfectly resembling a free space.
While the numerical noise in COMSOL is smaller than the radiative correction, it can be as large as the radiative correction in Elmer thus dominating over other systematics. As stated above, the order of magnitude of these systematics and numerical deviations is negligible for our purposes. This gives us confidence that our FEM simulation approaches are working as required.

%% file: perfect_mirror.tex
A dish antenna axion haloscope is comprised of the following experimental concept: A magnetized perfect electric conductor (PEC) in the axion CDM background leads to an emission of propagating electromagnetic waves~\cite{Horns:2012jf} with a photon dispersion relation. This emission compensates the axion-induced field $\bm{E}_a$ on the PEC surface such that the total tangential $E$-field is zero as required by  the axion-Maxwell equations. While the dish antenna has already been considered in the idealized 1D setup with a PEC of infinite transverse extend~\cite{Horns:2012jf,millar2017dielectric}, we study a 3D setup with a PEC of finite transverse extend. This allows us to explore finite size effects such as diffraction, boundary charges and near fields.

In section~\ref{sec:pec:diffraction} we will compare the FEM with the Fourier propagation approach and further diffraction theories to study the diffraction problem and to validate the different methods against each other. 
We generalize the analytical result for nonzero axion velocities explicitly in section \ref{sec:pec:velocity}.
In section \ref{sec:pec:nearfields} we see that a scalar diffraction theory is unable to describe near field effects and boundary charges. Therefore, we compare the FEM results also against more complete analytical solutions.

\subsection{Diffraction}
\label{sec:pec:diffraction}

\begin{figure}
\centering
 \begin{subfigure}{0.4\textwidth}
\centering
   \includegraphics[height=0.9\textwidth]{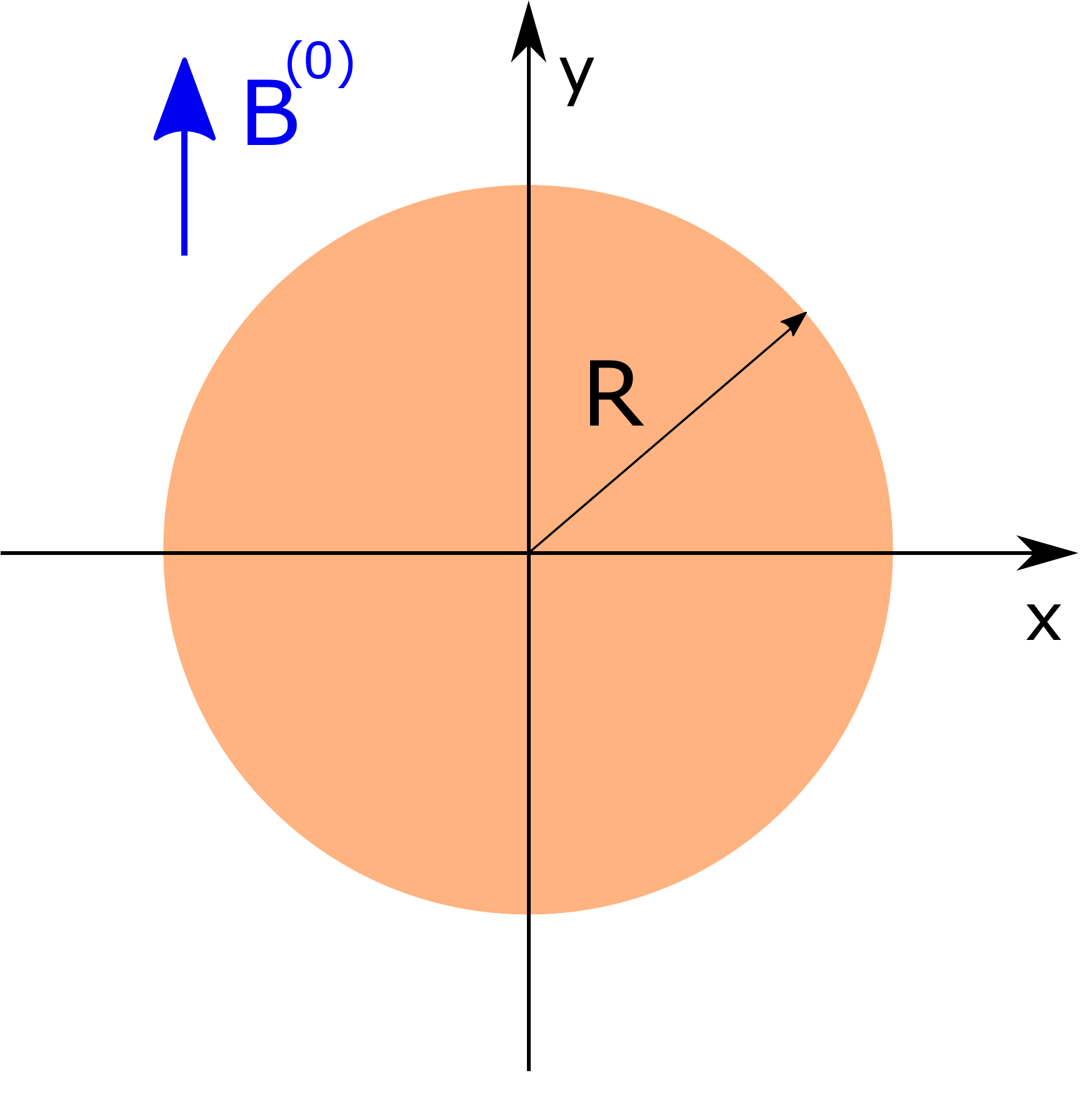}
   \caption{}\label{fig:Circular_Pec_geometry_xy} 
   \end{subfigure}
\begin{subfigure}{0.4\textwidth}
\centering
    \includegraphics[height=0.9\textwidth]{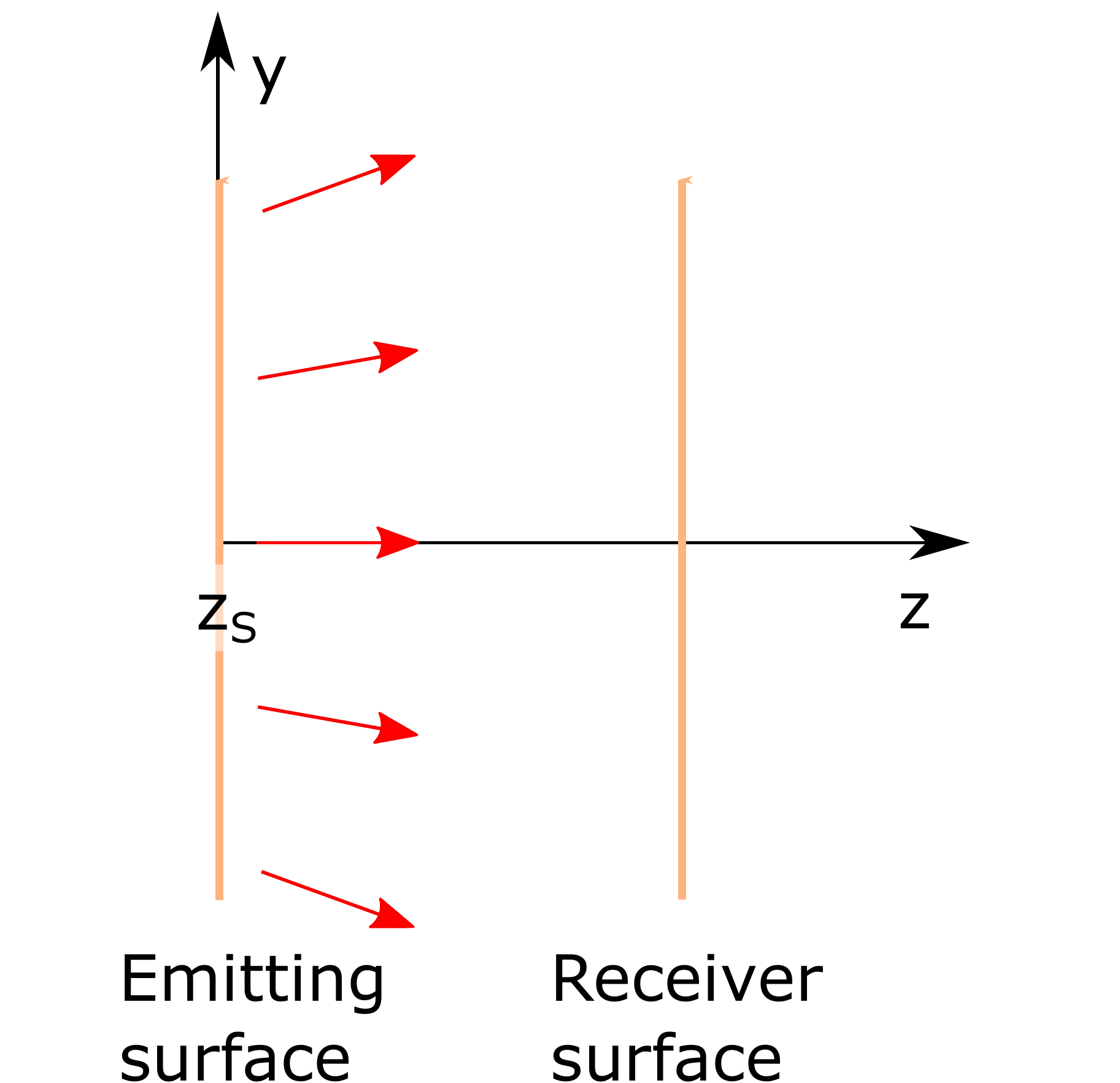}
   \caption{}\label{fig:Circular_Pec_geometry_xz} 
 \end{subfigure}
  \caption{(a) Circular PEC surface of radius $R$ in the $xy$-plane. The external {$B$-field} $\bm{B}^{(0)} = B^{(0)} \hat{\mathbf{e}}_y$ is considered to be constant over the surface and leads to the induced $\bm{E}_a$ field in the presence of the axion CDM background. (b) The circular PEC surface at $z_S=0$ is emitting electromagnetic field as sketched by the red arrows. The receiver surface is a fictitious surface to probe the amount of power that is going through and is not lost due to diffraction. }
  \label{fig:Circular_Pec_geometry} 
\end{figure}

Let us consider now explicitly a circular PEC of radius $R$ at ${z_S=0}$ as shown in figure~\ref{fig:Circular_Pec_geometry} with constant external $B$-field ${\bm{B}^{(0)}=B_0\hat{\bm{e}}_y}$ over the whole PEC.
In this case the Fourier approach~\eqref{Fourier_approach} leads to the following formula for the $E$-field of the emitted electromagnetic wave:
\begin{equation}
   \frac{E_y(\tilde{r},\tilde{z})}{E_{a}}=\int_0^{\infty} d\tilde{k}_{r}  e^{i\sqrt{\tilde{\omega}^2n^2-\tilde{k}_{r}^2}|\tilde{z}|}J_0(\tilde{r}\tilde{k}_{r})J_1(\tilde{k}_{r}),
  \label{FT3d_general_zc8}
\end{equation}
with $r=\sqrt{x^2+y^2}$ the radial coordinate and the normalized variables ${\tilde{k}_{r}=k_{r} R},$ ${\tilde{r}=\frac{r}{R}}$, ${\tilde{z}=\frac{z}{R}}$ and ${\tilde{\omega}=\omega R}$. The $x$ and $z$~components of the emitted fields $E_{x/z}$ are zero, because we assume a constant external $B$-field over the PEC which is polarized in $y$ direction only. Note that the Fourier approach can also be used to calculate the emitted fields of a PEC in an inhomogeneous external magnetic field. In this case, as discussed in section~\ref{sec:free_space}, the axion-induced field $\bm{E}_a$ -- and therefore also the emitted field compensating $\bm{E}_a$ on $S$ -- follow the shape of the external magnetic field on $S$.

\begin{figure}
  \begin{subfigure}{4.3cm}
    \includegraphics[height=4.4cm]{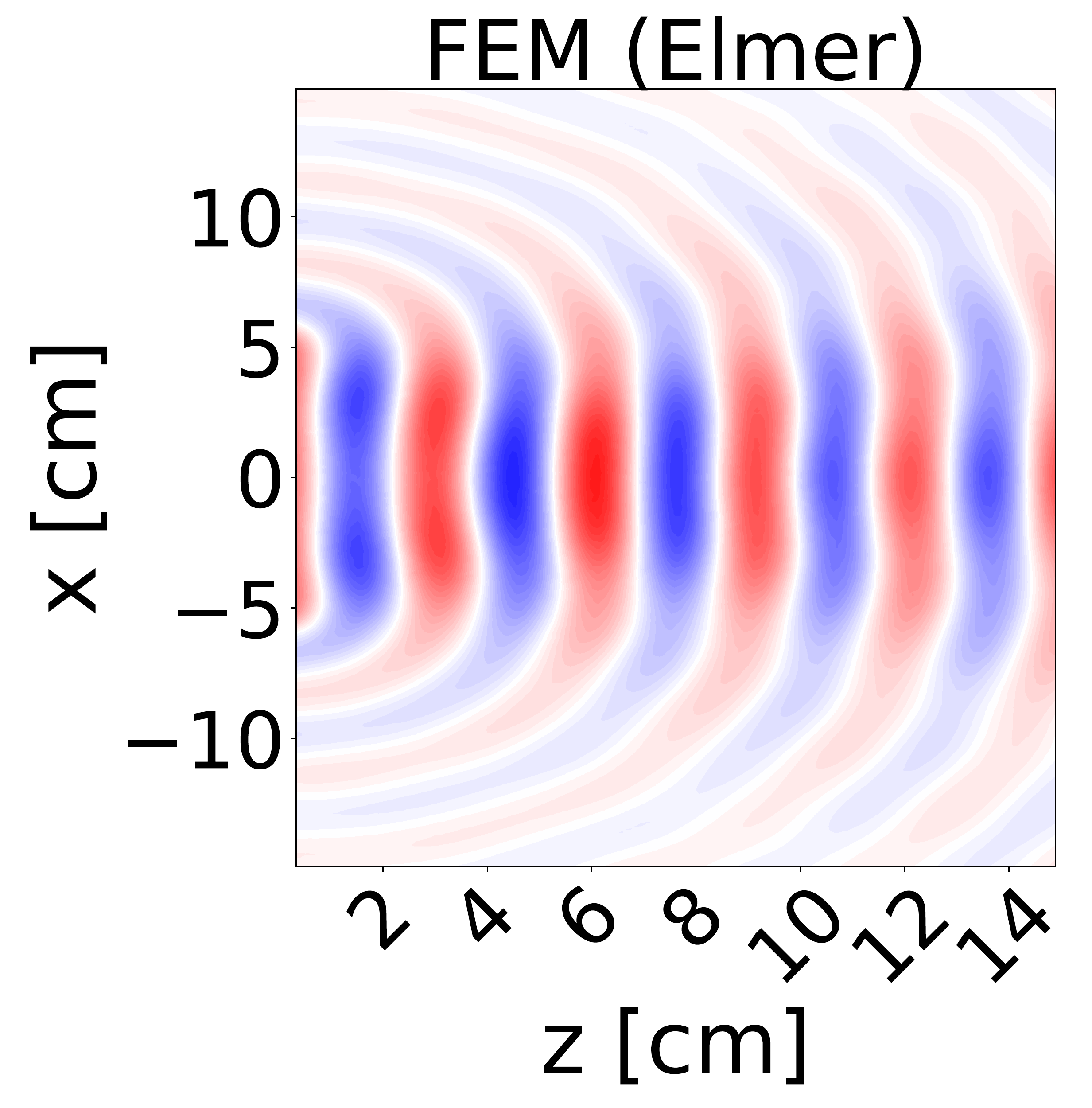}
    \caption{\hspace{-2cm}}\label{fig:PEC:ELMEREy}
  \end{subfigure}
    \begin{subfigure}{4.7cm} \includegraphics[height=4.4cm]{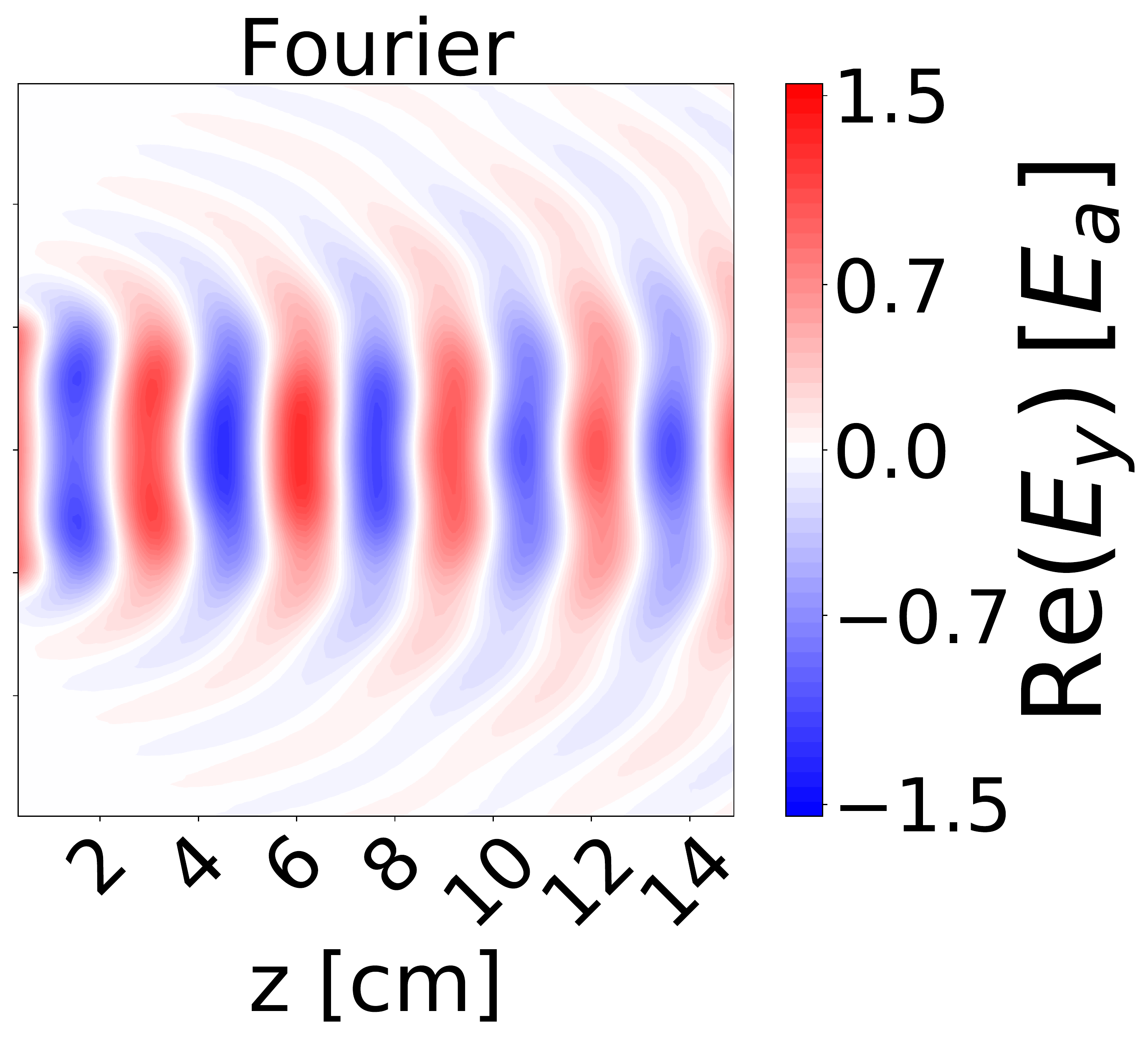}
    \caption{\hspace{2cm}}\label{fig:PEC:fourier_appraoch}
  \end{subfigure}
  \begin{subfigure}{4.3cm}
    \includegraphics[height=4.4cm]{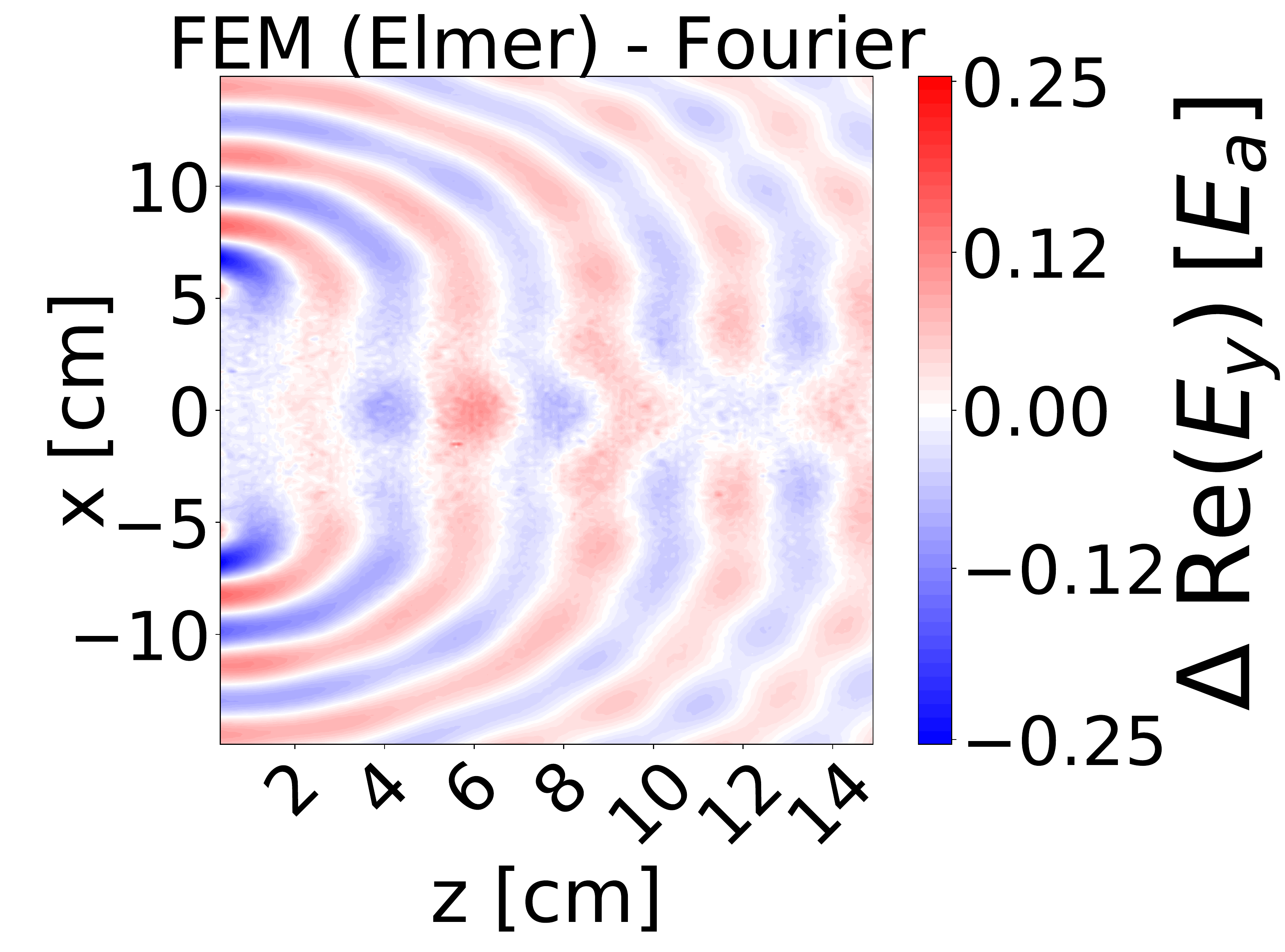}
     \caption{}\label{fig:PEC:Fourier_vs_ELMER}
  \end{subfigure}
  \caption{Single circular PEC located at $z_S=0$ with radius of $\SI{6}{\centi\metre}$ in the CDM axion field and assuming an external magnetic field pointing in $y$-direction. Real part of the $y$-component of the $E$-field in the $xz$-plane at $y=\SI{-2.5}{\centi\meter}$, at a frequency of \SI{10}{\giga\hertz}, i.e., $m_a \approx \SI{40}{\micro\electronvolt}$, corresponding to a wavelength of the propagating electromagnetic fields of $\SI{3}{\centi\metre}$. (a)~shows the FEM solution, where we have subtracted the axion induced field since we are only interested in the emitted fields described by the Fourier approach. The results from the Fourier approach are shown in panel (b). Panel (c) shows the difference between the results shown in panels (a) and (b). The largest difference is observed at the rims of the circular PEC.}
  \label{fig:FEM_vs_Fourier} 
\end{figure}

In figure~\ref{fig:FEM_vs_Fourier}  we compare the result of the Fourier approach to the FEM solution for a PEC with radius $R=\SI{6}{\centi\meter}$. In this paper we chose a circular disk radius which is smaller than in the planned full size haloscopes such as MADMAX, since smaller radii for a given axion mass are more prone to 3D finite size effects, which we investigate here. The results in figure~\ref{fig:FEM_vs_Fourier} show that the Fourier approach can describe the $E$-field of the propagating electromagnetic wave well in the forward direction. The largest differences between the Fourier approach and the FEM solution are at the rims of the PEC. The differences are due to near field effects and boundary charges, which will be discussed in section~\ref{sec:pec:nearfields}.

\begin{figure*}
  \begin{subfigure}{0.49\textwidth}
    \includegraphics[width=\textwidth]{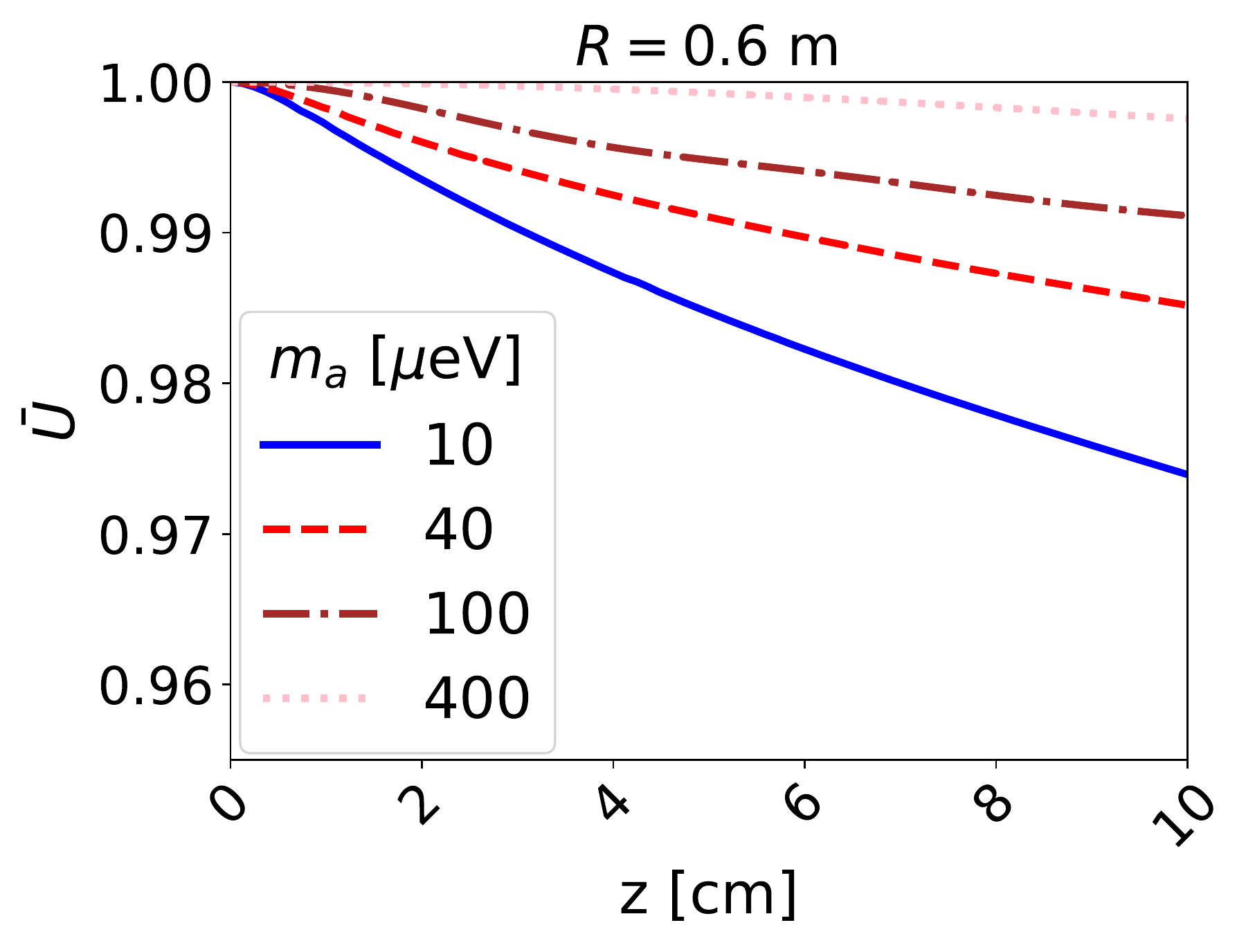}
    \caption{}\label{fig:Different_Frequencies}
  \end{subfigure}
  \begin{subfigure}{0.49\textwidth}
    \includegraphics[width=\textwidth]{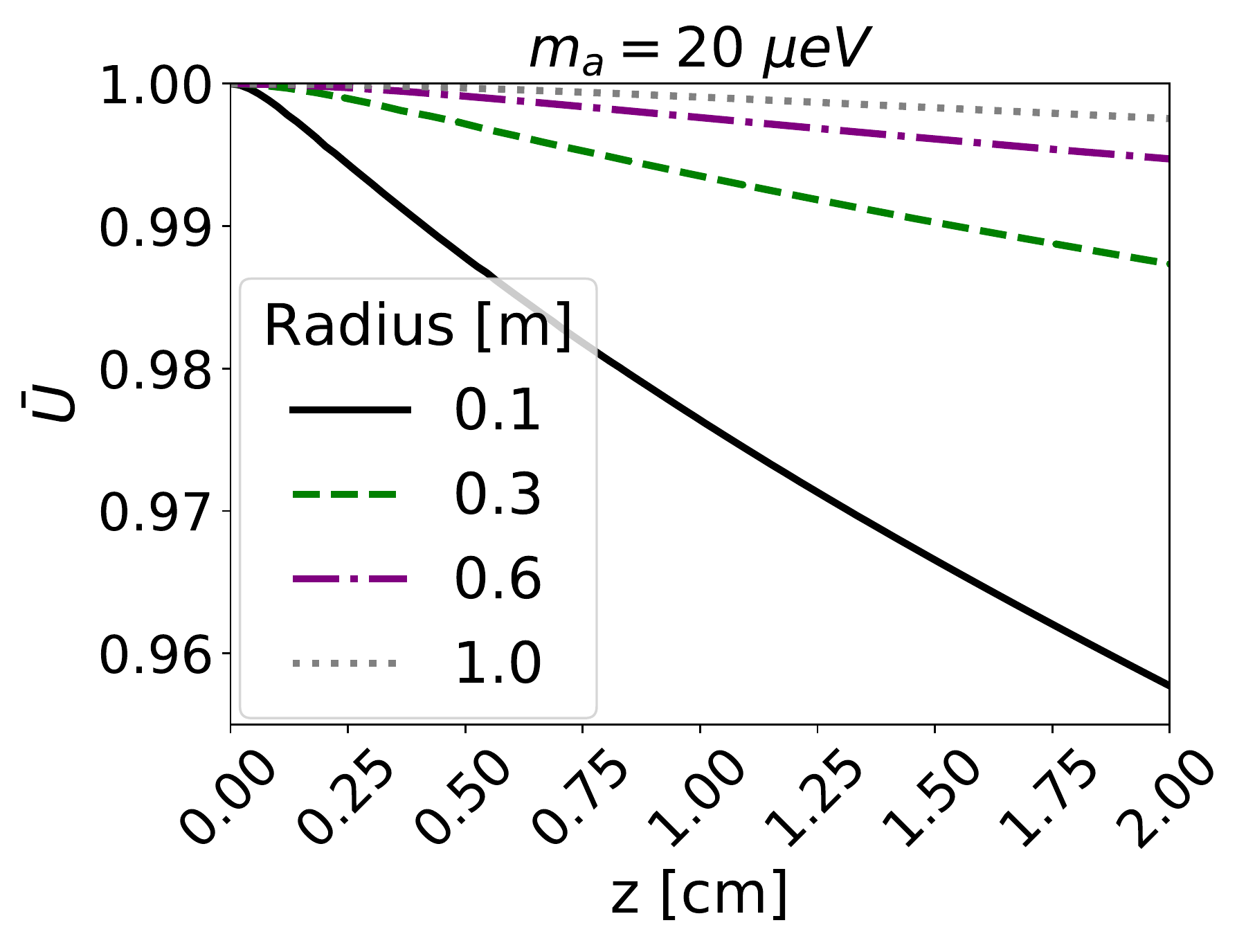}
    \caption{}\label{fig:Fourier_Approach_DifferentRadii}
  \end{subfigure}

  \caption{Received power over emitted power $\bar{U}$ for a circular PEC and an ideal receiver surface of the same size for different distances, obtained with the Fourier approach. Diffraction losses increase with distance. (a) For decreasing frequencies/axion masses, i.e., increasing photon wavelength, the diffraction losses increase. The lowest considered axion mass $m_a=\SI{10}{\micro\electronvolt}$ corresponds to a frequency of $\SI{2.5}{GHz}$ ($\lambda\approx\SI{13}{\centi\metre}$) and the largest one $m_a=\SI{400}{\micro\electronvolt}$ to a frequency of $\SI{100}{GHz}$ ($\lambda\approx \SI{3}{\milli\metre}$). We already presented this data in~\cite{Schutte-Engel:2018mfn}. (b) For increasing PEC radii the diffraction loss is smaller. }
  \label{fig:Fourier_Approach_DifferentRadii_different_Frequencies} 
\end{figure*}

Let us now discuss what diffraction effects imply for dish antennas or dielectric haloscopes.
Figure~\ref{fig:Fourier_Approach_DifferentRadii_different_Frequencies} shows the fraction of received power over emitted power $\bar{U}$ arriving on a surface with the same size as the circular PEC. We obtain $\bar{U}$ at different distances away from the circular PEC with the Fourier approach.
This should already give a first indication of diffraction losses expected between dielectric interfaces in dielectric haloscopes such as MADMAX, where the distance between the interfaces is around $\lambda/2$~\cite{millar2017dielectric}, i.e., at the ${\rm cm}$ level for our calculation.
In figure~\ref{fig:Fourier_Approach_DifferentRadii_different_Frequencies}\,(a) we see that the diffraction loss will increase if one wants to probe smaller axion masses, i.e., larger photon wavelengths. In figure~\ref{fig:Fourier_Approach_DifferentRadii_different_Frequencies}\,(b) $\bar{U}$ is plotted for different radii. If the photon wavelength is much smaller than the diameter of the circular PEC, diffraction effects will not have significant impact on the emitted fields. This statement also applies to dielectric haloscopes which consists of many radiating surfaces, see also section~\ref{sec:dieldiscMirror} below. 

The calculations above illustrate the effects of diffraction few ${\rm cm}$ away from the PEC. In dish antenna experiments the receiver might be placed much farther away from the PEC.
A scalar diffraction theory which is suited to do far field expansions of the emitted $E$-fields was developed by Kirchhoff and Rayleigh. If the $E$-field in the $xy$-plane ($S$) at $z_S=0$ is given, then the fields in whole space are~\cite{jackson_classical_electrodynamics_1999}
\begin{equation}
    E_i(\bm{x})=\frac{k}{2\pi i}\int_SdA'\frac{e^{ikD}}{D}\Big(1+\frac{i}{kD}\Big)\frac{\bm{n}'\cdot\bm{D}}{D}E_{i}(x',y'),
\label{eq:Scalar_Kirchhoff}
\end{equation}
with $\bm{D}=\bm{x}-\bm{x}'$, $\bm{n}'$ is the normal vector on the surface $S$ at point $\bm{x}'$ pointing into the diffraction region.
For large observer distances near the $z$-axis we can expand $D$ in $({x-x'})/{z}$ and~$({y-y'})/{z}$ in equation~\eqref{eq:Scalar_Kirchhoff}~\cite{GoodmanFourierOptics}. We obtain for the propagated field just a single Fourier transformation of the field on the surface $S$
\begin{equation}
    E_i(\bm{x})=\frac{k}{2\pi i}\frac{1}{z}e^{ikz}e^{i\frac{kr^2}{2z}}\int_S dA' e^{-i\frac{k}{z}(xx'+yy')}E_i(x',y'),
    \label{eq:Scalar_Kirchhoff_expanede}
\end{equation}
where we have assumed that $\frac{k(x'^2+y'^2)}{2z}<1$.
In particular, for a circular PEC in a homogeneous external magnetic field as depicted in figure~\ref{fig:Circular_Pec_geometry} this leads the well known Airy disk formula~\cite{GoodmanFourierOptics,Airy}
\begin{equation}
\frac{E_y(\bm{x})}{E_a} = -i kR^2 \frac{e^{ikz}}{z}e^{\frac{ik}{2z}r^2}\frac{J_1(R\,m_a\,\tan\theta)}{R\, m_a\, \tan\theta},    
\end{equation}
where $\theta$ is the polar angle from spherical coordinates.
The opening angle\footnote{When we require all observer coordinates to be large, i.e., $D=|\bm{x}|-\frac{\bm{x\cdot\bm{x}'}}{|\bm{x}|}+\mathcal{O}(1/|\bm{x}|^2)$ \cite{jackson_classical_electrodynamics_1999}, we obtain $\sin\theta=1.22\lambda/(2R)$.} 
of the diffracted $E$-field is defined by the first minimum and is located at
\begin{equation}
     \tan\theta \approx 1.22 \, \frac{\lambda}{2R} \approx \num{38e-3} \, \left(\frac{\SI{1}{\meter}}{2R} \right) \left(\frac{\SI{40}{\micro\electronvolt}}{m_a} \right) .
\end{equation}
This shows again explicitly that diffraction effects decrease with larger radii and axion masses.
Therefore, dish antennas and dielectric haloscopes are limited towards lower axion masses by diffraction effects.

\subsection{Axion Velocity}
\label{sec:pec:velocity}
\begin{figure*}
\centering
  \begin{subfigure}{0.32\textwidth}
    \includegraphics[width=\textwidth]{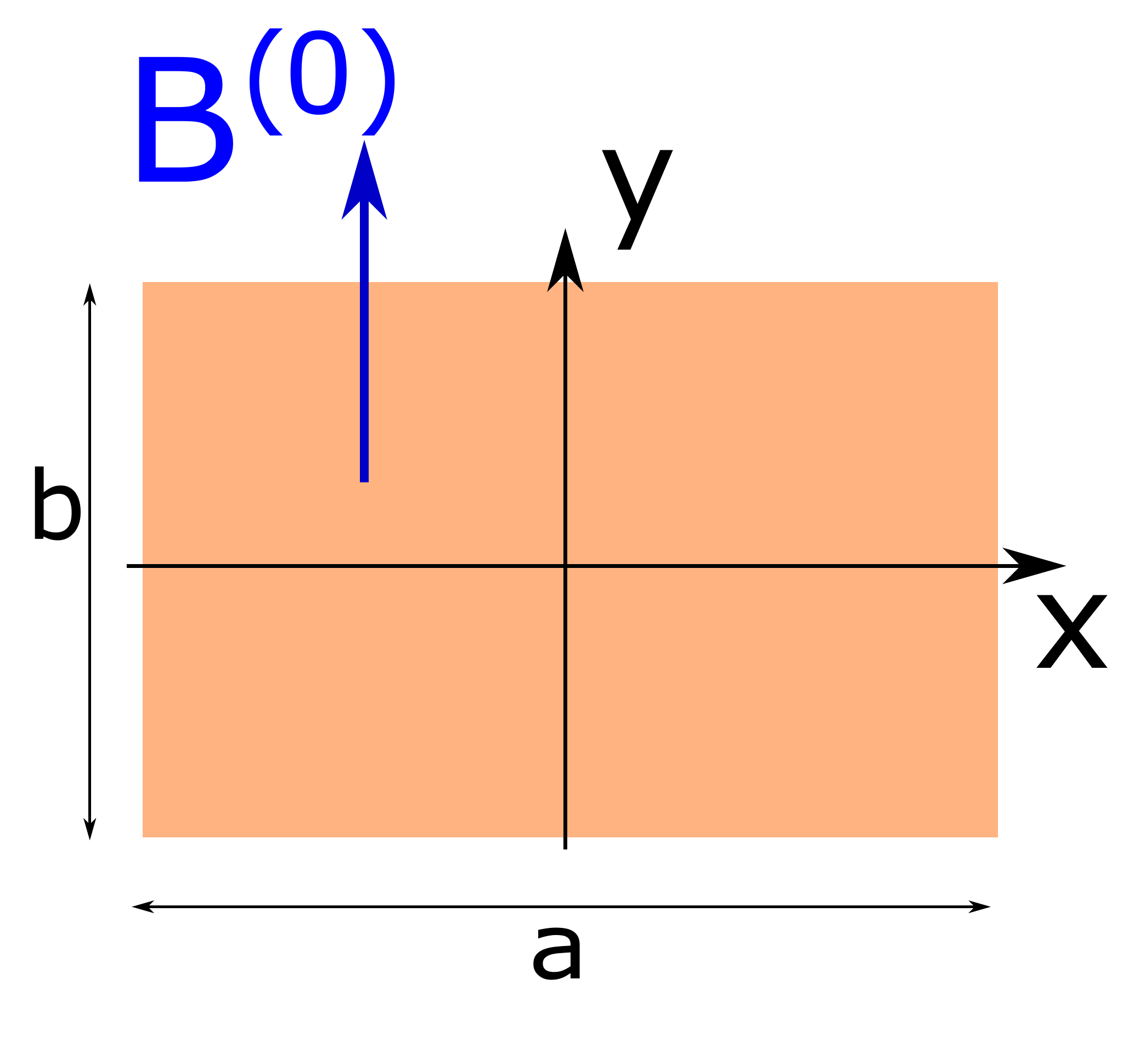}
    \caption{}\label{fig:rect_disk_geometry}
  \end{subfigure}
  \begin{subfigure}{0.32\textwidth}
    \includegraphics[width=\textwidth]{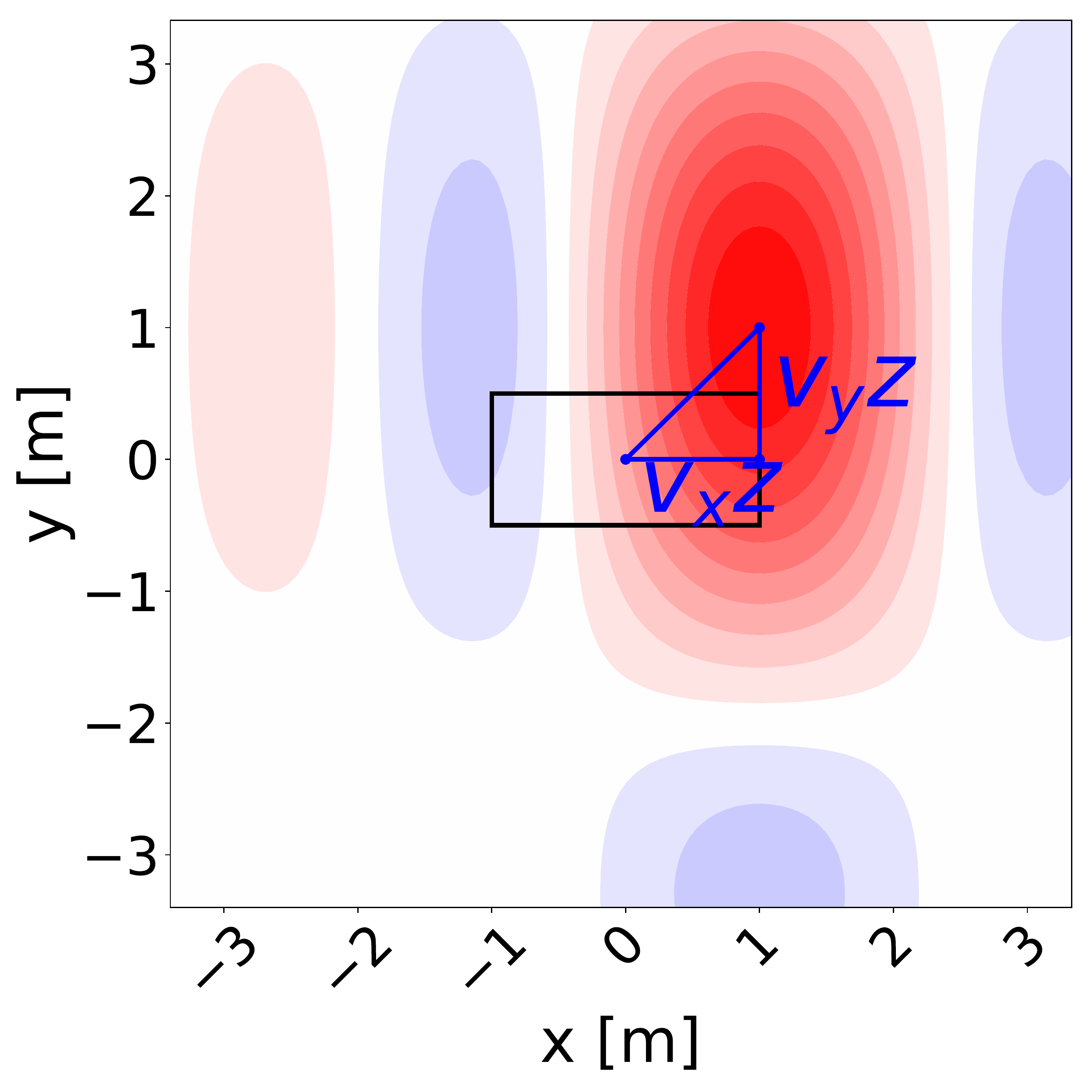}
    \caption{}\label{fig:rect_disk_shift}
  \end{subfigure}
  \caption{(a) Rectangular PEC in the $xy$-plane at $z_S=0$. The external $B$-field is homogeneous over the entire dish antenna and points in $y$-direction. (b) Diffraction pattern of the rectangular disk with $a=\SI{2}{\metre}$ and $b=\SI{1}{\metre}$ at distance $z=\SI{10}{\metre}$ away from the rectangular dish antenna. The axion velocity was set to a large value of $v_{x,y}=0.1$ to illustrate the effect better. The maximum of the diffraction pattern is now shifted from $x=0$ and $y=0$ to $x=v_xz$ and $y=v_yz$. The used wavelength was $\SI{30}{\centi\metre}$.}
  \label{fig:rect_dishantenna}
\end{figure*}

While in the rest of this paper we consider the axion velocity to be zero, in this section we examine axion velocity effects, as also discussed in \cite{Jaeckel:2013sqa,Jaeckel:2015kea,Jaeckel:2017sjb}. We describe the axion field in this case as $a^{(0)}=a_0e^{-i\omega t+i\bm{k}_a\cdot \bm{x}}$, where $\bm{k}_a$ is the axion field wave vector which is only nonzero since we now allow a finite axion velocity. The considered dish antenna is centered around the coordinate center in the $xy$-plane and has dimensions $(a,b)$ as sketched in figure~\ref{fig:rect_dishantenna}\,(a). A rectangular geometry is used in various dish antenna experiments \cite{FUNK:2017icw,Knirck:2018ojz,BRASS:website}.

Furthermore we assume a homogeneous external $B$-field pointing in $y$-direction. To compute the far field of the emitted $E$-field from the dish antenna we use the scalar diffraction theory by Kirchhoff and Rayleigh, which was introduced in the previous section in equation~\eqref{eq:Scalar_Kirchhoff_expanede}. 
The emitted field on the surface $S$, which now represents the rectangular dish antenna, is given by ${E_ae^{i\bm{k}^S\cdot\bm{x}}}$, where $\bm{k}^S$ is the wave vector of the emitted $E$-field. Due to momentum conservation we have $k^{S}_{x,y}=k_{ax,y}=v_{x,y}\omega$, where $\bm{v}$ is the axion group velocity, and $k^{S}_z\approx\omega=k$~\cite{Millar:2017eoc}. The angles ${\tan\alpha\equiv\frac{k_x^S}{k}\approx v_x,\tan\beta\equiv\frac{k_y^S}{k}\approx v_y \lesssim 10^{-3} }$ correspond to the parallel velocities $v_x, v_y$ of the axion to the surface. Below we see that they define the angle of the emitted radiation in the far field.
With the scalar diffraction theory we describe only the emitted $E$-field component parallel to the external $B$-field, which is the leading $E$-field component. The other $E$-field components are proportional to the axion velocity~\cite{Millar:2017eoc}, which is small.
After inserting everything into equation~\eqref{eq:Scalar_Kirchhoff_expanede} we get
\begin{equation}
 \frac{E_i(\bm{x})}{E_a}=\frac{e^{ikz}}{i\lambda z}e^{i\frac{kr^2}{2z}}ab ~\text{sinc}\left[\dfrac{ka}{2} \left(v_x-\frac{x}{z}\right)\right]\text{sinc}\left[\dfrac{kb}{2} \left(v_y-\frac{y}{z}\right)\right],
\label{eq:Rectangular_disc_farfield}
\end{equation}
with $\text{sinc}(x)\equiv\frac{\sin(x)}{x}$.
It is evident that the velocity effect leads to a shift of the diffraction maximum to $(v_x z, v_y z)$ after distance $z$. To illustrate this we plot the fields of a rectangular PEC with dimensions $\SI{2}{\metre} \times \SI{1}{\metre}$ at distance $z=\SI{10}{\metre}$ for a relatively high axion velocity $v_{x,y}=0.1$ in figure~\ref{fig:rect_dishantenna}\,(b).

In dielectric haloscopes the emitted fields are propagated many times between the dielectric interfaces when resonant. Therefore, the shift of the diffraction maximum grows with propagation distance $z$ and therefore might also be a potential loss mechanism. Assuming the interfaces have distance $\lambda/2$, an internal resonance with quality factor $\tilde{Q}$ causes a virtual propagation distance $\tilde{Q} \, \lambda/2$. Requiring the shift over this distance to be much smaller than the haloscope radius $R$, very roughly limits $\tilde{Q} \ll {2R}/{\lambda} \times 10^3$. Dielectric haloscope designs aim to operate in more broadband configurations and so naturally avoid the high $Q$ limit.
In addition, since the shift of the diffraction maximum depends on the direction of the axion ``CDM wind'', this may be exploited to infer quantities of the CDM velocity distribution similarly as discussed in~\cite{Knirck:2018knd}.

\subsection{Near Fields}\label{sec:pec:nearfields}

In figure~\ref{fig:FEM_vs_Fourier}  we observe that the Fourier approach describes well the far field behaviour of the $y$-component of the emitted electromagnetic waves.
The difference between FEM and Fourier approach is below $10\%$ in the far field. The largest differences between Fourier approach and FEM are observed at the rims of the circular PEC. Here additional radiation appears on the sides of the PEC. Therefore, in the course of this paper we will refer to the fields not described by the Fourier approach as near fields.
For a more complete analytical understanding we have to take into account the vectorial nature of the $E$-field and boundary charges at the rims of the dish antenna. We discuss both effects in the following.

\begin{figure}
  \begin{subfigure}{0.49\textwidth}
    \includegraphics[width=\textwidth]{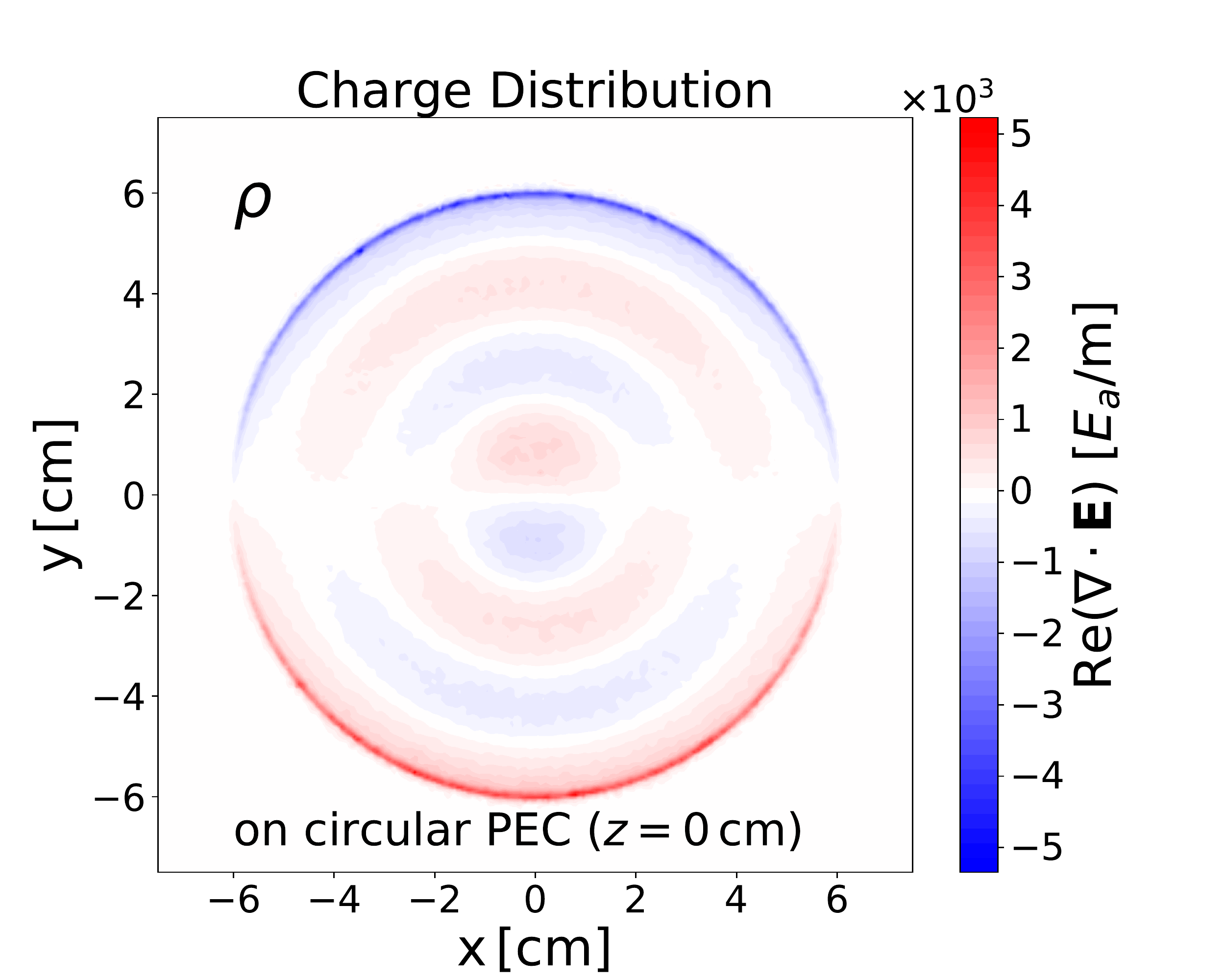}
    \caption{}\label{fig:PEC:NearFields:ChargeDist}
  \end{subfigure}%
  \begin{subfigure}{0.49\textwidth}
    \includegraphics[width=\textwidth]{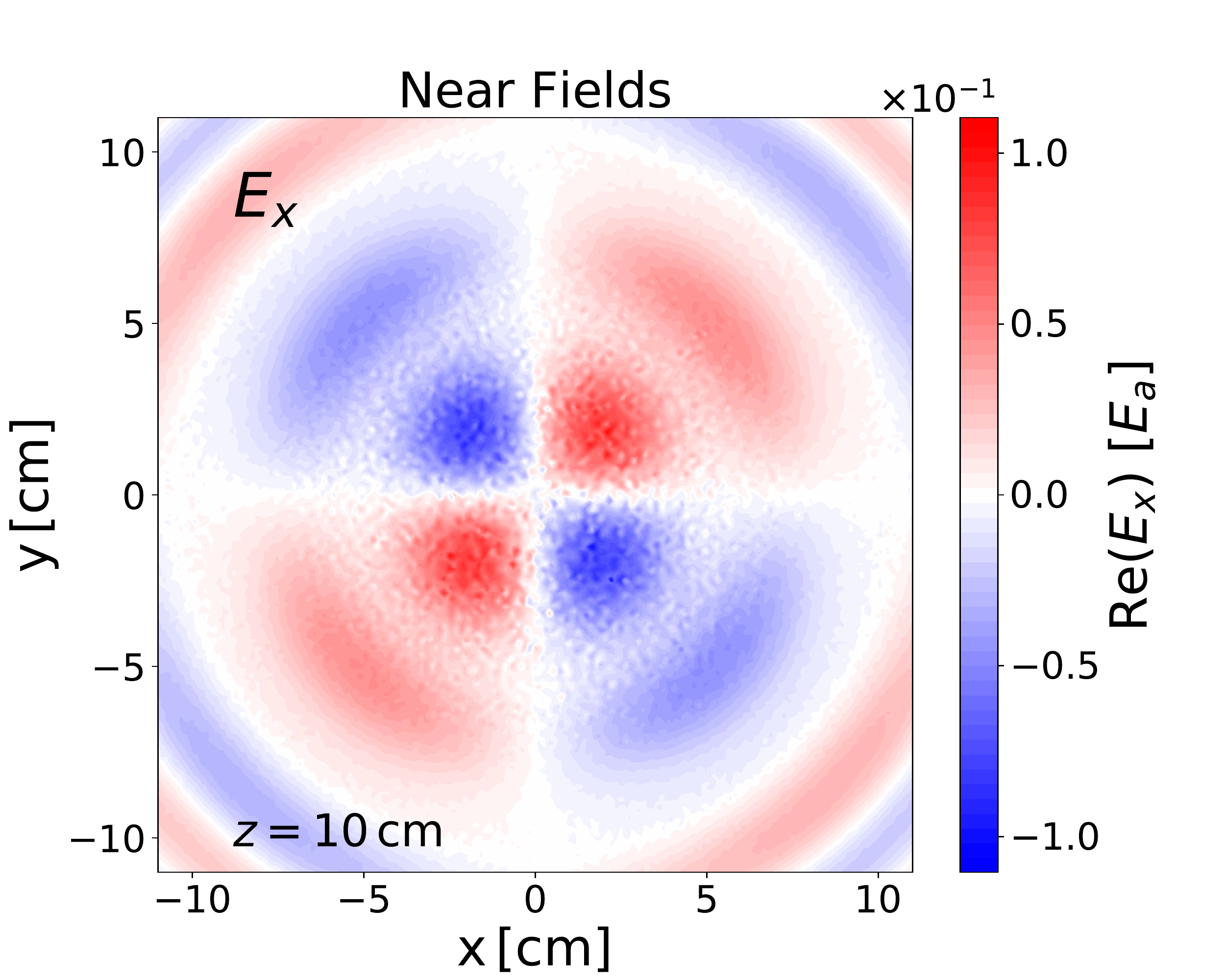}
    \caption{}\label{fig:PEC:NearFields:Ex}
  \end{subfigure}
  \caption{Near field effects of a single circular PEC with radius of $ \SI{6}{\centi\metre}$ obtained with a FEM simulation (Elmer), at  \SI{10}{\giga\hertz}, i.e., $m_a \approx \SI{40}{\micro\electronvolt}$, and external magnetic field pointing in $y$-direction. (a)~Charge distribution on the circular PEC. (b)~$E_x$ fields $\SI{10}{\centi\metre} \approx 3.3\lambda$ from the circular PEC away. The field pattern can be understood phenomenologically when one imagines an $E$-field in (a) pointing from the positive charges towards the negative charges. For example, the $x$-component of the electric field is $0$ along the axis $x = 0$ since the charge distribution is symmetric when mirrored at this axis.}
  \label{fig:PEC:NearFields} 
\end{figure}

The vectorial nature of the emitted $E$-field was neglected in the scalar Fourier approach by setting $\nabla(\nabla\cdot\bm{E})=0$ in deriving equation~\eqref{eq:wave_equation_decoupled}. A vectorial description of diffraction which includes near field effects is given by the \textit{vector Kirchhoff diffraction formula}~\cite{jackson_classical_electrodynamics_1999}
\begin{eqnarray}
\bm{E}_k(\bm{x})=
\int_{S}dA'\Big[i\omega(\bm{n}'\times \bm{B})G+(\bm{n}'\times\bm{E})\times\nabla' G+(\bm{n}'\cdot\bm{E})\nabla' G\Big],
\label{eq:Kirchhoff}
\end{eqnarray}
where $\bm{n}'$ is defined as in the previous section as well as the surface $S$, i.e., $S$ is the $xy$-plane at $z_S=0$. The $E$ and $B$-fields which appear in equation~\eqref{eq:Kirchhoff} depend on the primed variables. For a circular PEC the fields on $S$ are only nonzero for $r={\sqrt{x^2+y^2}}<R$. Here in particular we choose constant fields $\bm{E}=E_a\hat{\bm{e}}_y$ and $\bm{B}=-E_a\hat{\bm{e}}_x$ over the PEC and zero fields outside. 
In the Fourier approach only the $y$-component of the $E$-field is nonzero, but in the FEM solutions we observe that all three $E$-field components of the emitted $E$-field are nonzero. 
Using the vector Kirchhoff diffraction formula~\eqref{eq:Kirchhoff} we obtain an additional $z$-component, but the $x$-component of the $E$-fields is still zero, which is in contrast to the FEM solution for the $x$-component shown in figure~\ref{fig:PEC:NearFields}\,(b).

To also describe the $x$-component of the emitted field we have to take into account boundary charges which are present at the rims of a finite sized PEC. 
Fields induced by the boundary charges are not taken into account by the Kirchhoff formula.
The axion induced field drives the electrons in the PEC up and down such that at the boundary of the PEC the charges accumulate.
This physical picture can be confirmed with the FEM solution which we show in figure \ref{fig:PEC:NearFields}. The free charge density which is maximal at the PEC rims is shown in figure~\ref{fig:PEC:NearFields}\,(a) and we plot the $x$-component of the FEM $E$-field in figure \ref{fig:PEC:NearFields}\,(b). 
We can describe the boundary charges as a line charge density $\sigma_L\sim \sin\phi$ which also comes with a line current density $\bm{K}_L\sim\cos\phi\hat{\bm{e}}_{\phi}$, where $\phi$ is the azimuth angle of cylinder coordinates. The line charge and current density lead to additional terms for the emitted $E$-field.
We therefore have to add the following two \textit{boundary} terms to the Kirchhoff terms
\begin{eqnarray}
    \bm{E}_b(\bm{x})=R\int_0^{2\pi}d\phi' \sigma_L (\nabla G)_L+i\omega R \int_0^{2\pi}d\phi'\bm{K}_LG_L,
\label{eq:Elinechargedensity}
\end{eqnarray}
where the subscript $L$ at $G$ and $\nabla G$ means that we evaluate the corresponding term at $\bm{x}'=R\cos\phi'\hat{\bm{e}}_x+ R\sin\phi'\hat{\bm{e}}_y$. The total field emitted by the PEC is then given by $\bm{E}=\bm{E}_k+\bm{E}_b$.
The charge density term in \eqref{eq:Elinechargedensity} also naturally arises when one takes into account that the $E$ and $B$-fields in the Kirchhoff formula are actually not allowed to be discontinuous. In the case of discontinuous fields one gets a correction term which is exactly the charge density term in \eqref{eq:Elinechargedensity}. The Kirchhoff terms in combination with the charge density term in \eqref{eq:Elinechargedensity} is known also as the Stratton-Chu formula~\cite{stratton1941}.

\begin{figure}
\includegraphics[width=0.49\textwidth]{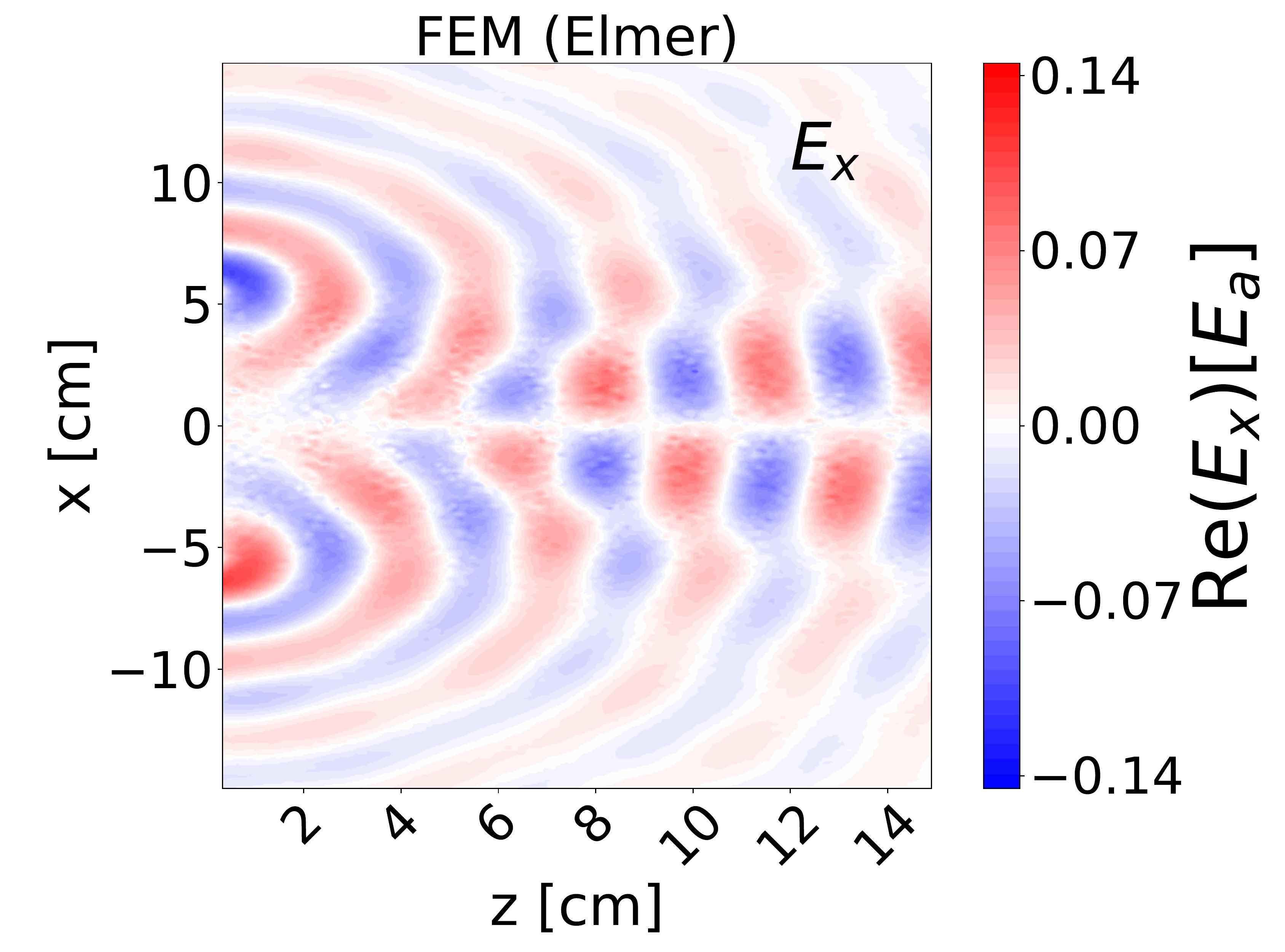}
\includegraphics[width=0.49\textwidth]{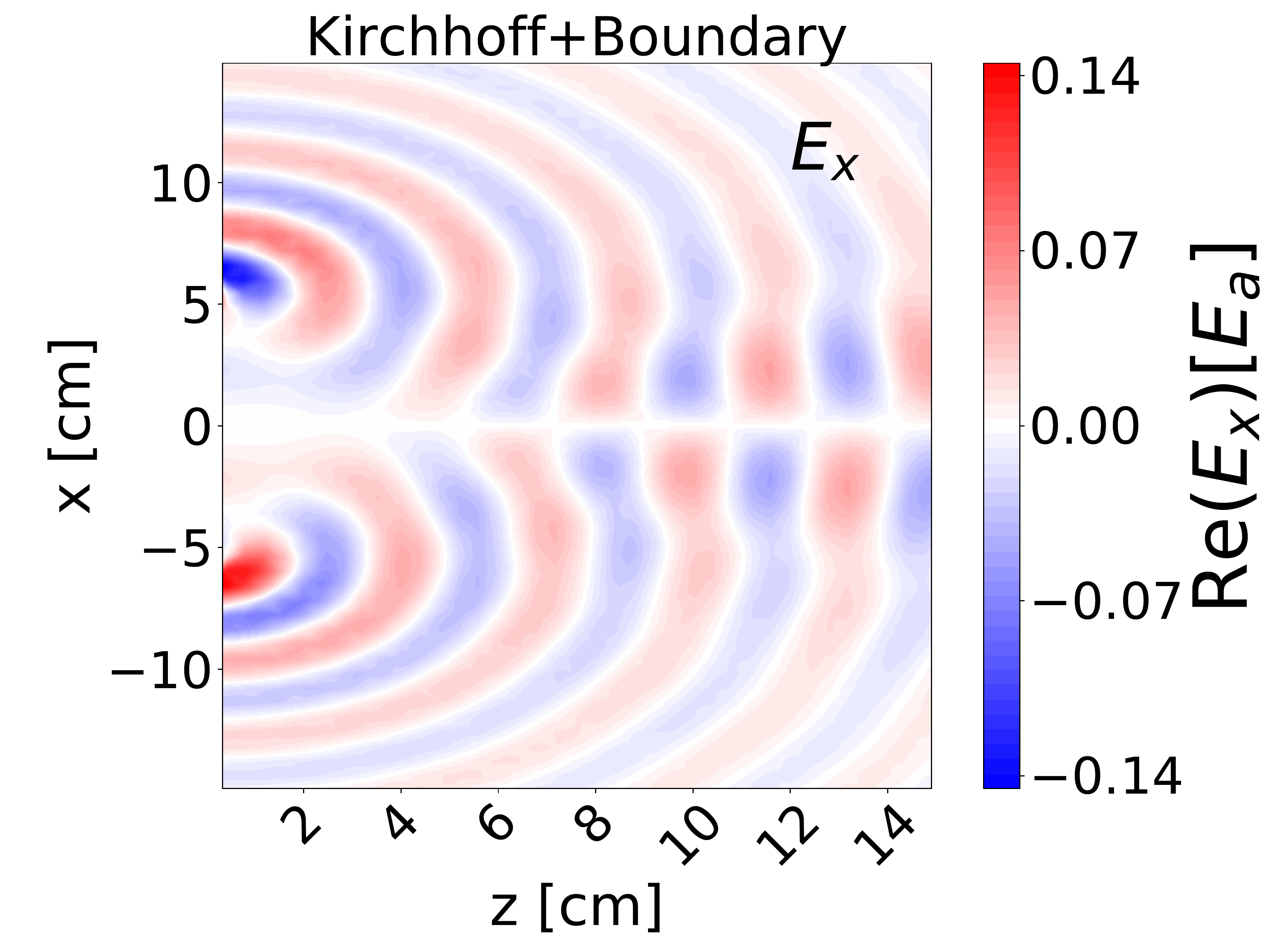}
\includegraphics[width=0.49\textwidth]{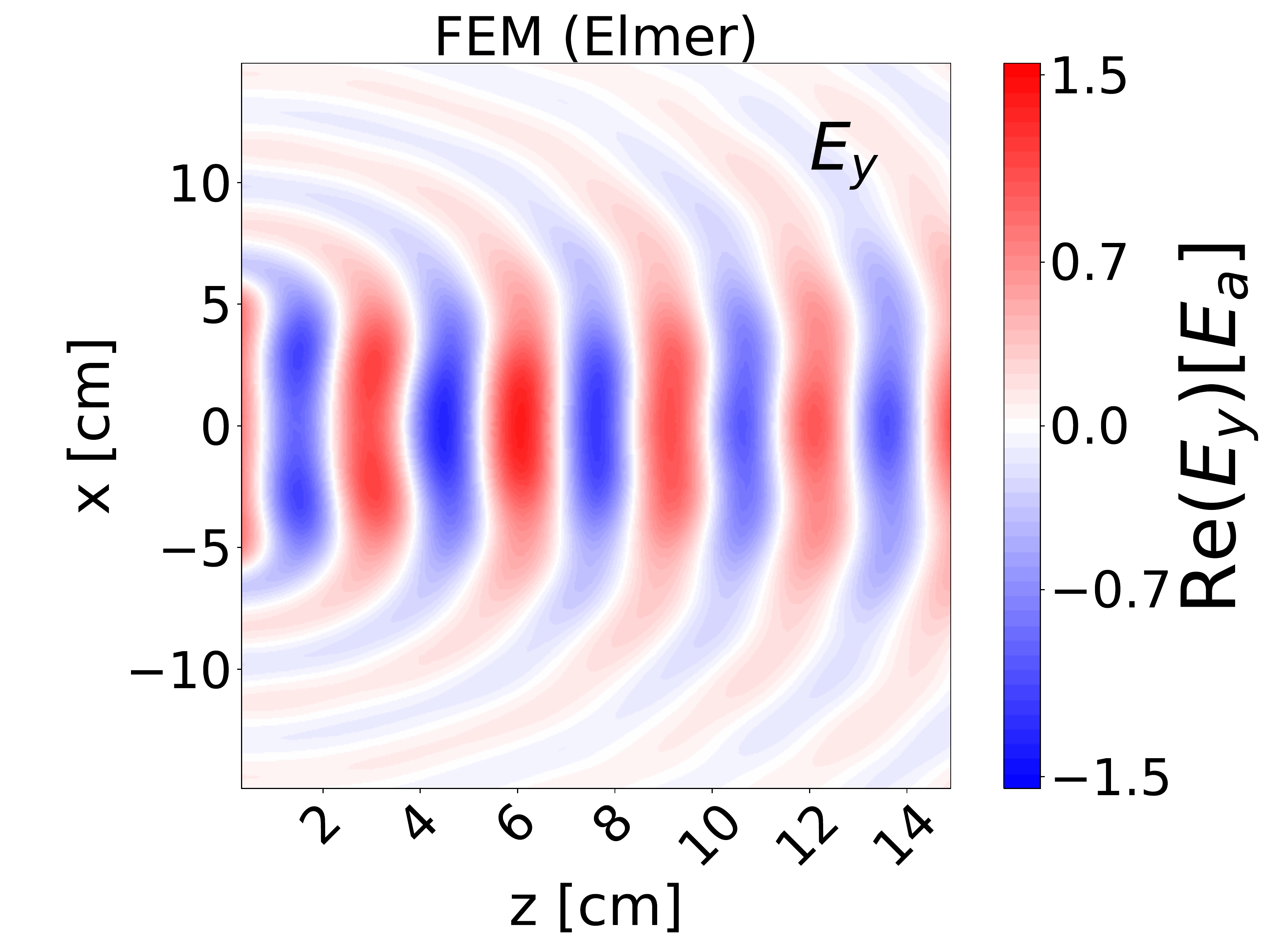}
\includegraphics[width=0.49\textwidth]{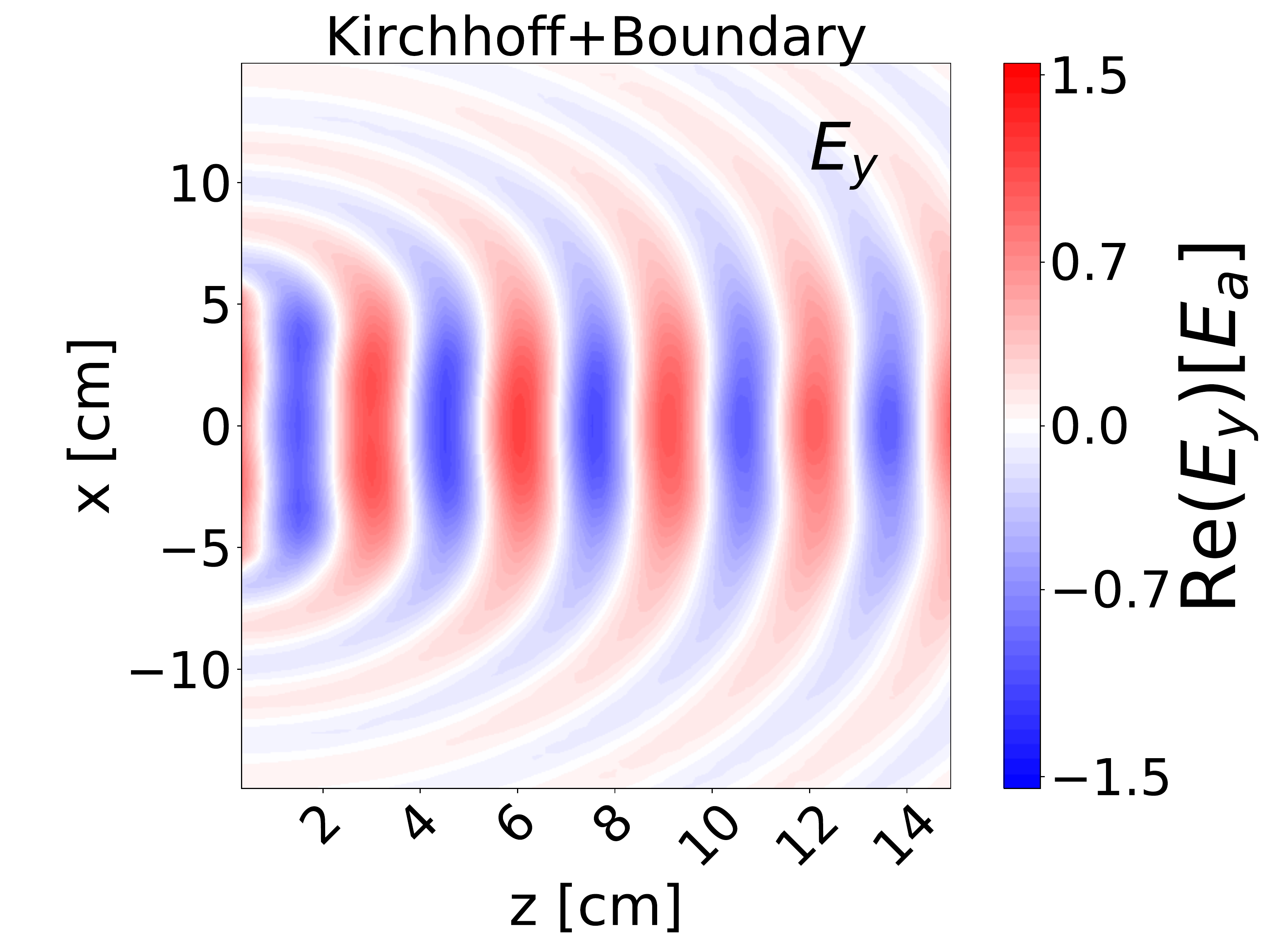}
\includegraphics[width=0.49\textwidth]{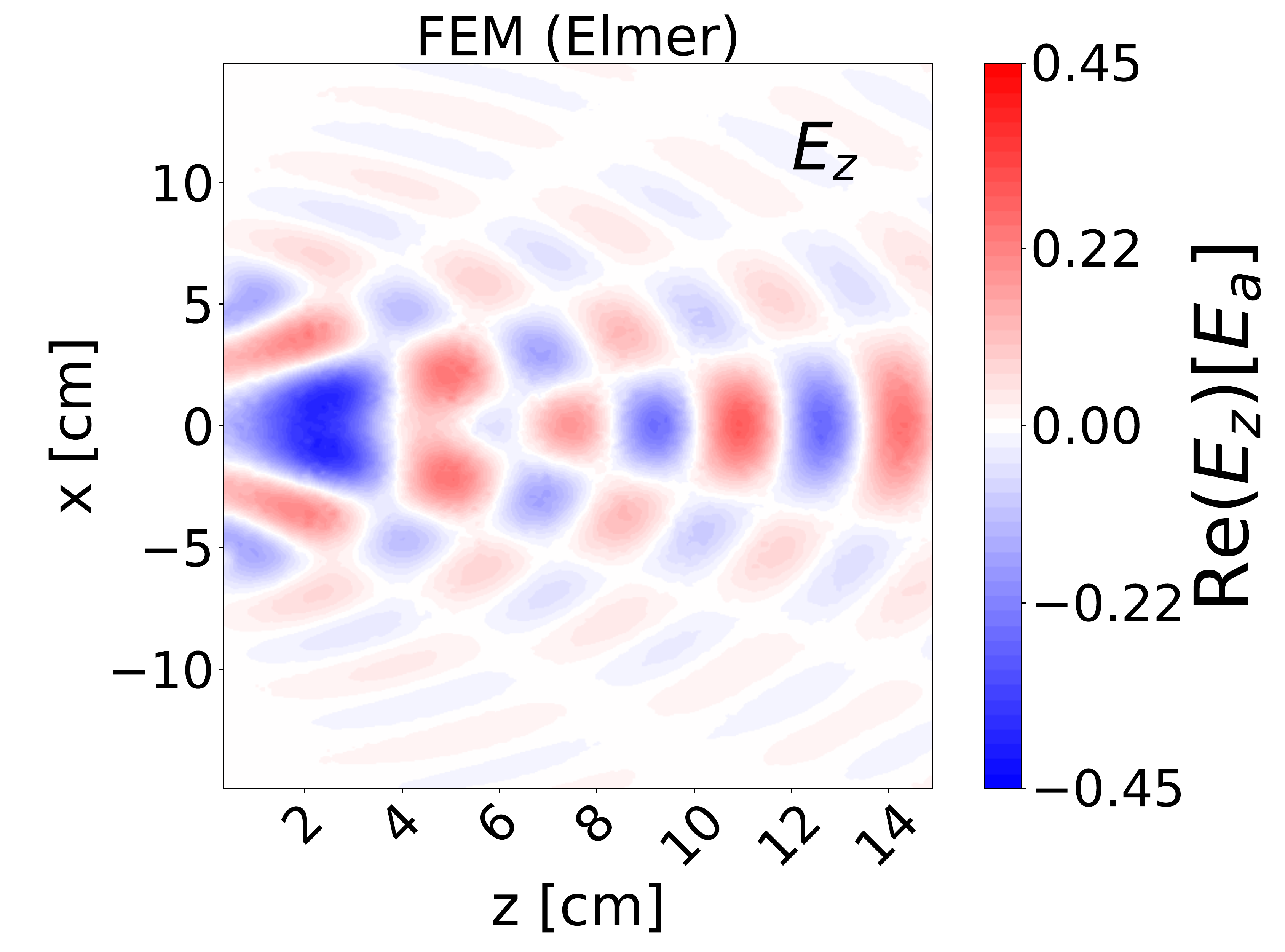}
\includegraphics[width=0.49\textwidth]{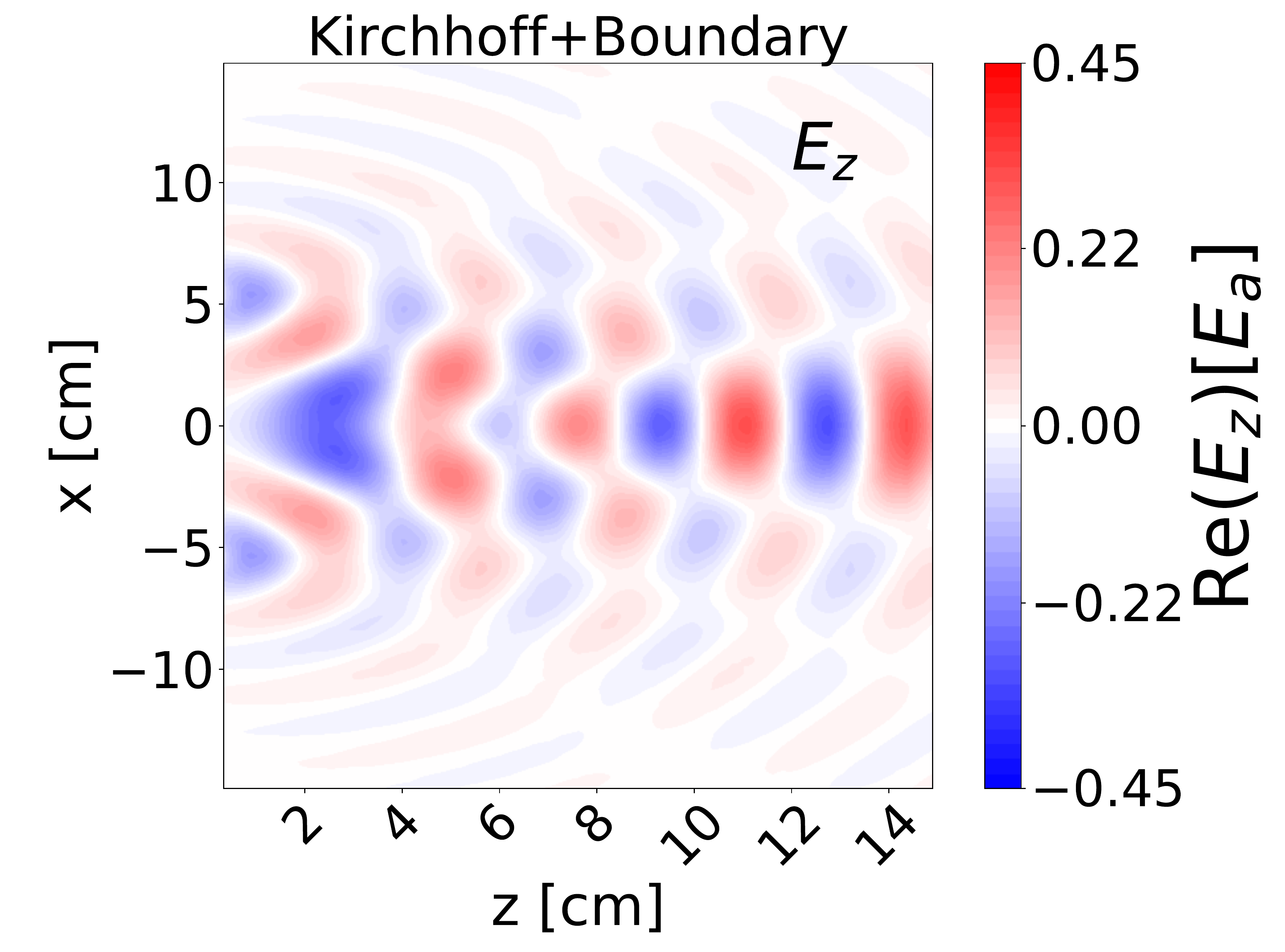}
\caption{\label{fig:FEM_vs_KirchhoffPlusLinecharge}
All components of the emitted fields from a single circular PEC located at $z_S=0$ with a radius of $\SI{6}{\centi\metre}$, at \SI{10}{\giga\hertz}, i.e., $m_a \approx \SI{40}{\micro\electronvolt}$, assuming an external magnetic field pointing in $y$-direction. All panels show a $xz$-slice at $y=\SI{2.5}{\centi\metre}$. Each row represents one $E$-field component, $E_x$ (top), $E_y$ (middle), $E_z$ (bottom). In the left column we show the result from the FEM solution while the right column shows the analytical result based on the Kirchhoff fields, cf.\ equation~\eqref{eq:Kirchhoff}, and boundary line fields, cf.\ equation~\eqref{eq:Elinechargedensity}. We show the results obtained with Elmer. The results obtained with Comsol are the same up to differences dominated by numerical noise.}
\end{figure}

In figure~\ref{fig:FEM_vs_KirchhoffPlusLinecharge} we compare the FEM solution to the analytical formula, i.e., the Kirchhoff formula plus the boundary fields. The exact magnitude of the line charge and current density are not known in general and therefore they have been scaled. However, as we are only trying to identify the underlying physical processes, which lead to the shown $E$-fields, needing such a scaling is not problematic. After the scaling all three components in figure~\ref{fig:FEM_vs_KirchhoffPlusLinecharge} of the $E$-fields agree. This gives us confidence that we have understood the physics behind the FEM solution.

In dielectric haloscopes near field effects and boundary charges may be important, because the dielectric disks are typically separated by a distance of around half a wavelength~\cite{millar2017dielectric}.
The influence of the near fields and boundary charges among diffraction in the context of a minimal dielectric haloscope is also discussed more closely in section~\ref{sec:dieldiscMirror}.

%% file: dielectric_disc.tex
The most simple setup containing more than one boundary between media with different dielectric constants $\epsilon$ is a dielectric disk. In this section we assume a circular disk of a radius of $R = 2 \lambda = \SI{6}{\centi\metre}$, dielectric constant $\epsilon = 9$ (sapphire) for varying thicknesses $d_\epsilon$. 
First, we compare the total emitted power, reflectivity and transmissivity of a dielectric disk in FEM directly with the 1D model in section~\ref{sec:dieldisc:bf}. Afterwards, we compare its diffraction pattern to predictions from the Fourier propagation approach in section~\ref{sec:dieldisc:diffraction}.

\subsection{Boost Factor and Reflectivity}
\label{sec:dieldisc:bf}
For a single dielectric disk the emitted axion-induced power, reflectivity and transmissivity are the primary output of the 1D model, so comparing them will allow us to most directly test the effects of going to 3D and disks with finite sized transverse extend.
For the reflection and transmission coefficients, a Gaussian beam~\cite{goldsmith1998quasioptical,Lee:2008:PTS:1525539} with a beam waist of $w_0 = 5\, {\rm cm}$ was focused on the front surface of the dielectric disk. In Elmer the Gaussian beam was forced to propagate into the system by using the Robin boundary conditions from equation~\eqref{eq:robin-bc} and setting $g$ such that the Gaussian beam fulfills the boundary condition. The respective power is obtained by integrating the flux $\int \bar{\bm{S}}\cdot d\mathbf{A}$ at the front and back simulation domain boundaries, where $\bar{\bm{S}}$ is the time averaged Poynting vector. Numerical errors can be evaluated by varying the integration surface and are below one percent of the maximal output power.\footnote{A few percent of power is radiated to the outside of the finite integration surface in the 3D FEM simulations. We estimate this non-captured power with the Fourier approach and correct the obtained values respectively.}

\begin{figure}
	\centering
    \includegraphics[width=0.95\textwidth]{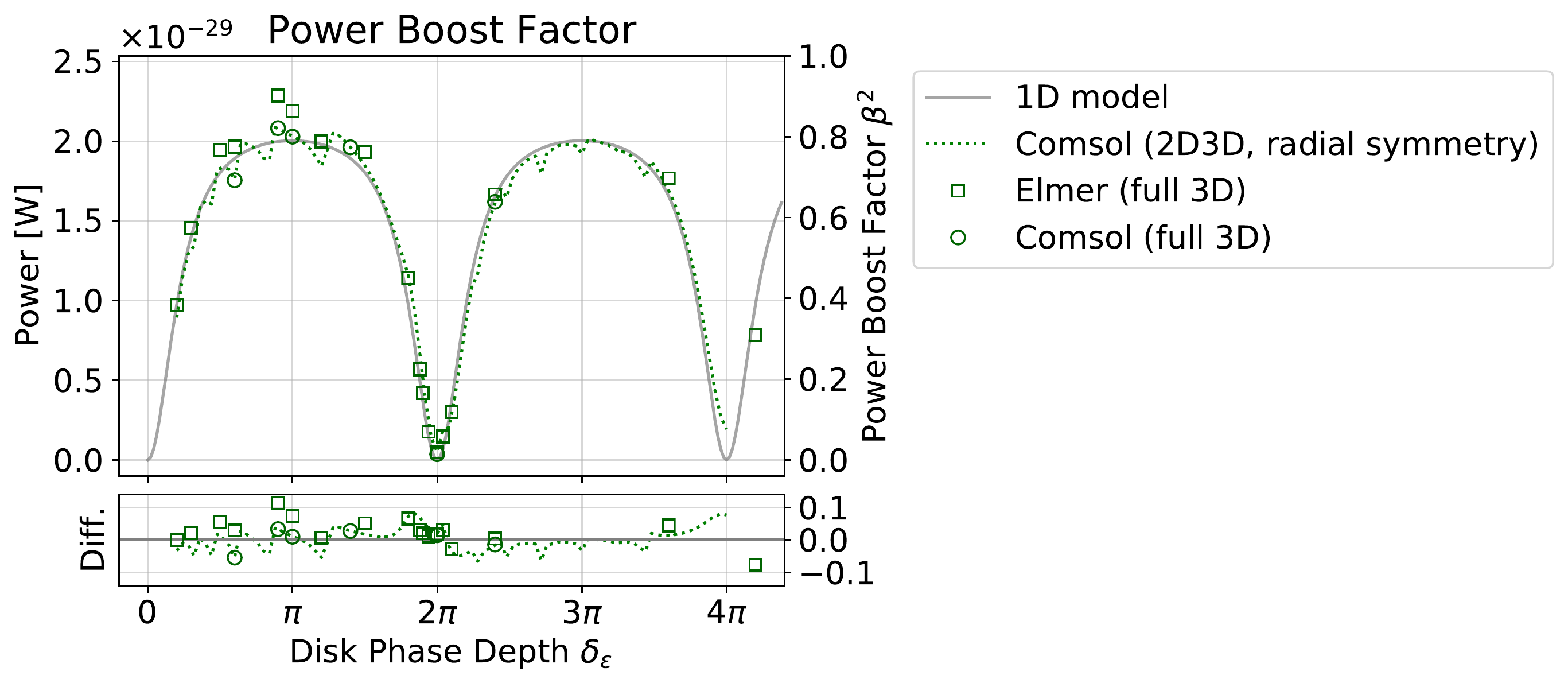}
	\caption{\label{fig:dieldisk:bf} Power boost factor $\beta^2$ of a single dielectric disk against disk phase depth ${\delta_\epsilon = n \omega d_\epsilon}$ where $d_\epsilon$ is the disk thickness. We consider a single circular dielectric disk with radius ${R=2\lambda}$, ${\epsilon = 9}$, at \SI{10}{\giga\hertz}, i.e., $m_a \approx \SI{40}{\micro\electronvolt}$, assuming ${|C_{{\rm a}\gamma}| = 1}$ and ${B^{(0)} = \SI{10}{\tesla}}$. We show the power obtained in Elmer (squares) and COMSOL (circles) in 3D and by using 2D radial symmetry in COMSOL (green dotted line), compared with the result from the 1D model (solid gray line). The lower panel shows the difference with respect to the 1D model result in terms of power boost factor $\beta^2$.}
\end{figure}

\begin{figure}
	\centering
    \includegraphics[width=0.95\textwidth]{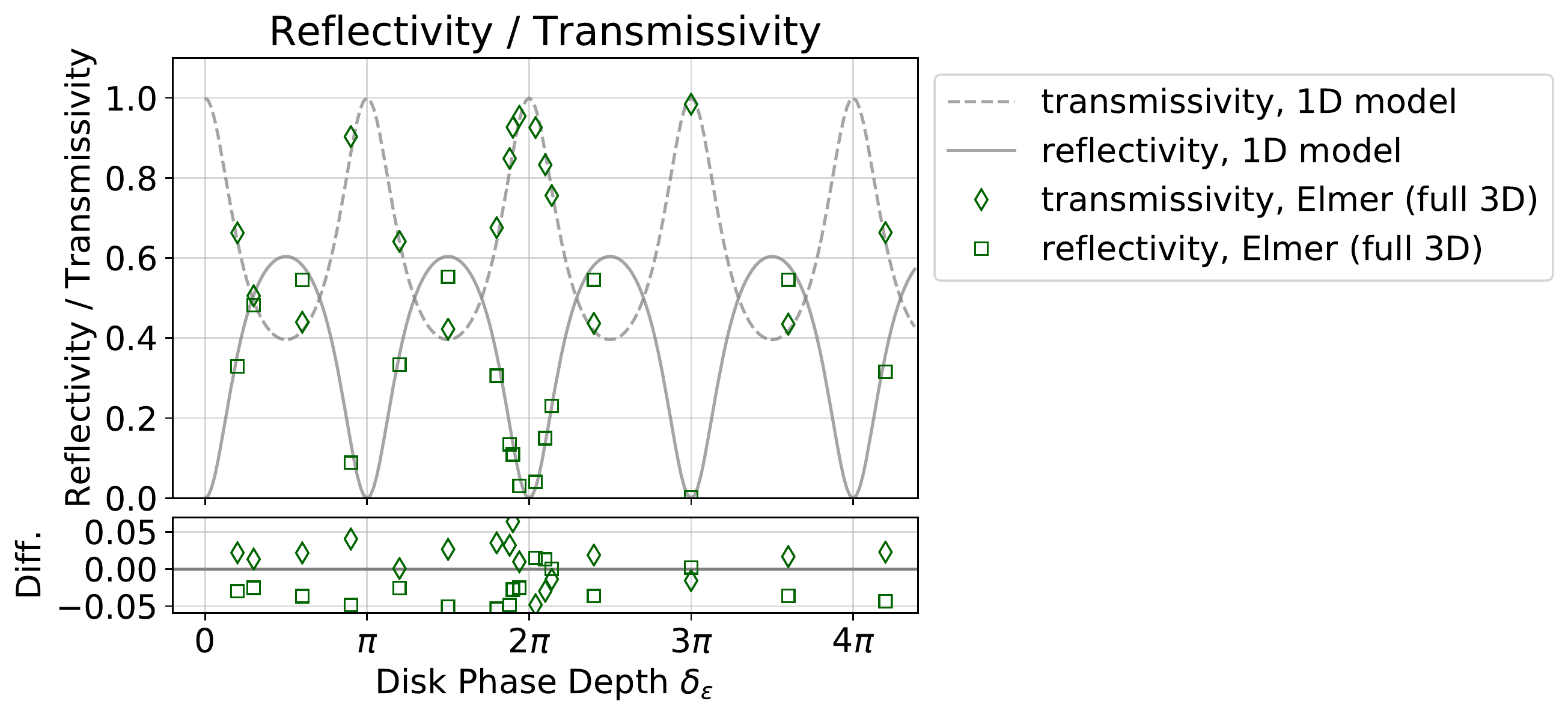}
	\caption{\label{fig:dieldisk:reflectivity} Reflectivity and transmissivity of a single dielectric disk against disk phase depth ${\delta_\epsilon = n \omega d_\epsilon}$ where $d_\epsilon$ is the disk thickness, analogous to figure~\ref{fig:dieldisk:bf}.
	We show reflectivity (squares) and transitivity (diamonds) obtained with Elmer for a Gaussian beam with waist $w_0 = 5~{\rm cm}$ at the disk. The reflectivity from the 1D model is indicated by solid gray line and the transmissivity from the 1D model by the dashed gray line. The disk is transparent for phase depths of integer multiples of $\pi$. The lower panel shows the respective differences with respect to the 1D model result.}
\end{figure}

We compute the emitted axion-induced power (power boost factor $\beta^2$) with both COMSOL and Elmer in full 3D. While Elmer solves the vectorized Helmholtz equation in 3D, we also exploit the radial symmetry in COMSOL (2D3D approach), see section~\ref{subsec:fem-tools}. Figure~\ref{fig:dieldisk:bf} shows the different emitted powers against disk phase depth $\delta_\epsilon = n \omega d_\epsilon$ compared with the 1D model prediction. The results are within $10\%$ of the 1D model predictions. Figure~\ref{fig:dieldisk:reflectivity} shows reflectivity and transmissivity which are within $ 5\%$ of the 1D model predictions.
Similar results are obtained for the phase of the emitted fields. 

As we deliberately choose a small radius of only two wavelengths, if deviations from the 1D model are to be found at all, they would be found here.
Diffraction will cause phase shifts inside the disk compared to the 1D case and power loss to the sides, while the near fields of each surface may directly affect the emission from the other surface.
The approximate match even for a disk with $2\lambda$ radius is encouraging, since haloscope experiments like MADMAX aim for much larger disks. For such disks one expects these effects to be less dominant as demonstrated in the previous section.

\subsection{Diffraction}
\label{sec:dieldisc:diffraction}
We use the recursive Fourier propagation approach to predict the diffraction pattern of a single dielectric disk as introduced in section~\ref{subsec:recFourierApproach}. To this end we have to consider the emissions from both dielectric disk surfaces, their propagation through the disk and eventually their interference outside of the disk. Explicitly, with the disk surfaces at $z_1$ and $z_2$, the diffraction pattern outside of the disk can be approximated by iterating the following steps for all emitted fields:
\begin{itemize}
    \item Considering the fields $E(z_1)$ emitted at $z_1$, the fields at the opposite interface $E(z_2)$ are obtained using the Fourier approach with~equation~\eqref{Fourier_approach}.
    \item The fields $E(z_2)$ outside of the disk surfaces at, i.e., at $r > R$, are set to zero.
    \item At the next surface ${\mathcal{T} E(z_2)}$ is transmitted outside and $\mathcal{R} E(z_2)$ is reflected inside the disk, where $\mathcal{R}$ is the complex reflectivity and $\mathcal{T} = (1+\mathcal{R})$ the complex transmissivity of the surface from inside.
    \item Repeating the above for $N$ iterations between the two interfaces and adding up all fields outside, gives a prediction for the diffraction pattern of a single disk.
\end{itemize}
Note that in the third step we take ${\mathcal{R} = (n_i - n_j)/(n_i + n_j)}$, which holds for plane waves under a normal incident angle in {medium~$i$} on the boundary to {medium~$j$}. Other angles could be accounted for by making $\mathcal{R}(k_x, k_y)$ dependent on the transverse momenta. However, $\mathcal{R}$ only depends at second order on the incident angle (see ``Fresnel equations,'' e.g., in \cite{jackson_classical_electrodynamics_1999}), so assuming normal incidence is a good approximation in our case.

\begin{figure}
  \begin{subfigure}{4.3cm}
    \includegraphics[height=4.2cm]{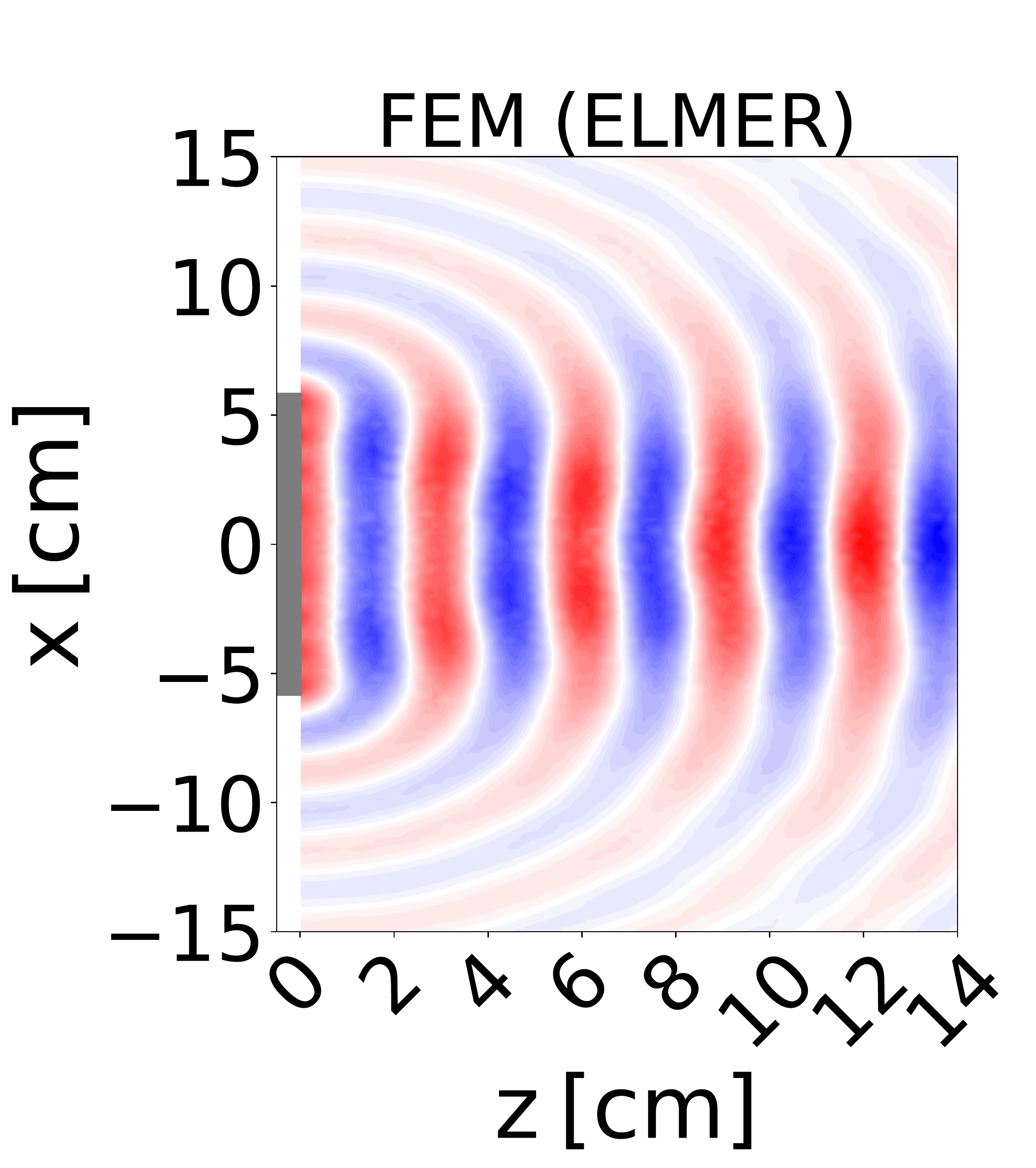}
     \caption{}\label{fig:dielectric_disk:Elmer-xz}
  \end{subfigure}
  \begin{subfigure}{4.7cm}
    \includegraphics[height=4.2cm]{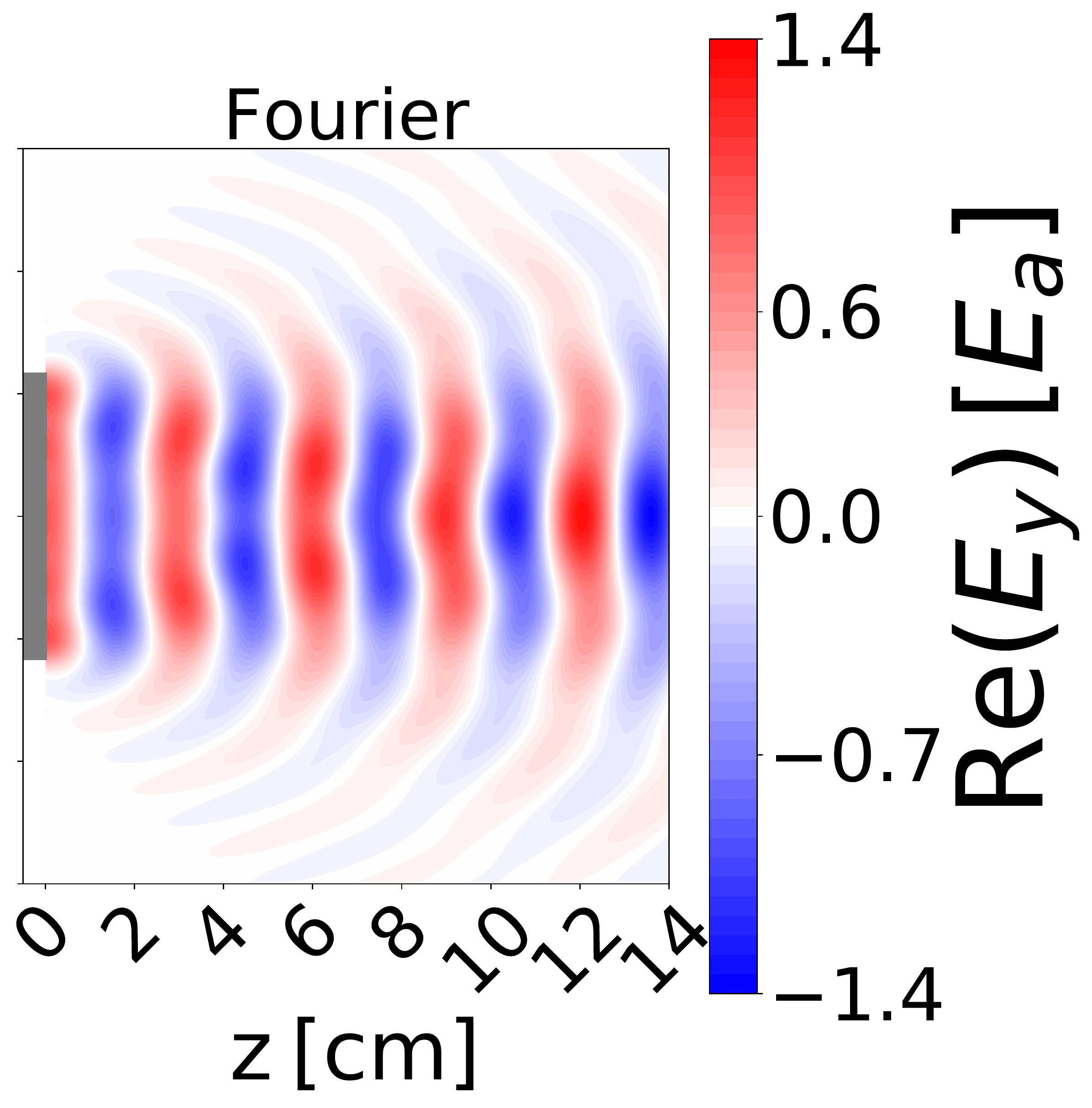}
     \caption{}\label{fig:dielectric_disk:Fourier-xz}
  \end{subfigure}
  \begin{subfigure}{4.5cm}
    \includegraphics[height=4.2cm]{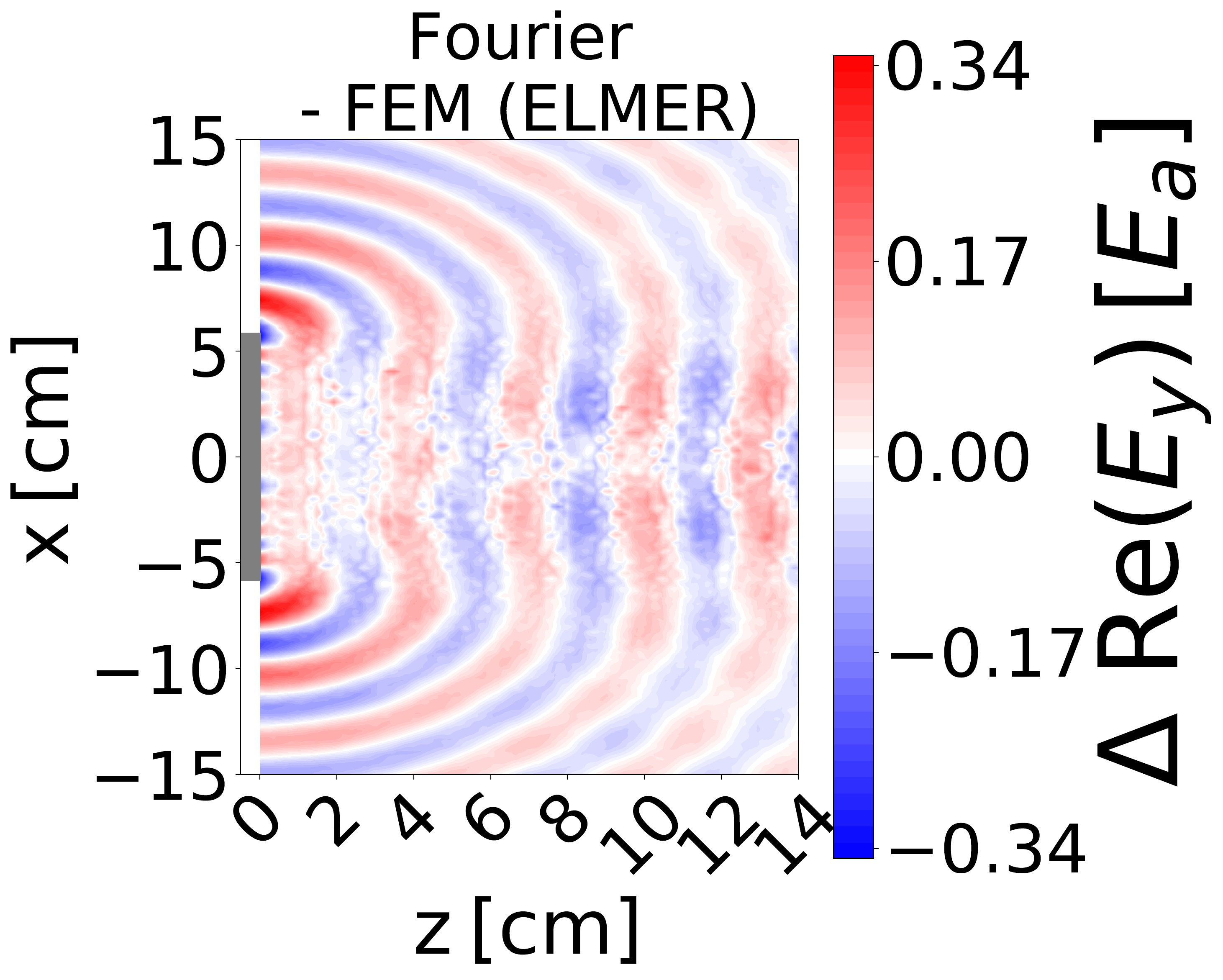}
     \caption{}\label{fig:dielectric_disk:Elmer-vs-Fourier}
  \end{subfigure}
\caption{\label{fig:circular_disc_max_field}
Real part of the $E_y$-field for a circular dielectric disk with $\epsilon = 9$ and phase depth $\delta_\epsilon=\pi$, i.e., with thickness $d_\epsilon=\SI{5}{\milli\meter}$ at \SI{10}{\giga\hertz}, i.e., $m_a \approx \SI{40}{\micro\electronvolt}$ for an external magnetic field pointing in $y$-direction. We show the fields in the $xz$-plane, analogous to figure~\ref{fig:FEM_vs_Fourier}. Panel~(a) shows the FEM solution after subtraction of the axion-induced field, panel~(b) the result obtained with the recursive Fourier propagation approach for $N=25$ iterations, and panel~(c) the difference between~(a) and~(b).}
\end{figure}

Figure~\ref{fig:circular_disc_max_field} shows the diffraction patterns at 10~GHz ($m_a\approx 40~\mu\mathrm{eV}$) for a circular dielectric disk with a radius of $R = 4\lambda = \SI{6}{\centi\metre}$, $\epsilon=9$ and a thickness of $d_\epsilon=\SI{5}{\milli\metre}$, i.e., a phase depth of $\delta_\epsilon=\pi$. We compare the result for the real part of the $E_y$-field obtained with the full 3D FEM with the one from the recursive Fourier 
propagation approach for $N=25$ iterations.
We find that the simple propagation approach matches the fields far away from the disk well. In the region around the rims of the disk we find the largest discrepancies. As we have seen already in section~\ref{sec:pec}, this is due to the fact that the recursive Fourier propagation approach does not include the charge distributions at the rims of the disks which lead to an additional emission. 
In addition, the disk has an interface to vacuum also at its rims. Therefore, it emits axion-induced radiation in radial directions at angles where the interface of the rim gets parallel to the external magnetic field.
In a realistic experimental setup we expect these boundary effects to be small, because the diameter of the disk will be larger and we are not going to detect the radiation to the sides.

%% file: tilted_disc.tex
\begin{figure}
    \centering
    \includegraphics[width=0.55\textwidth]{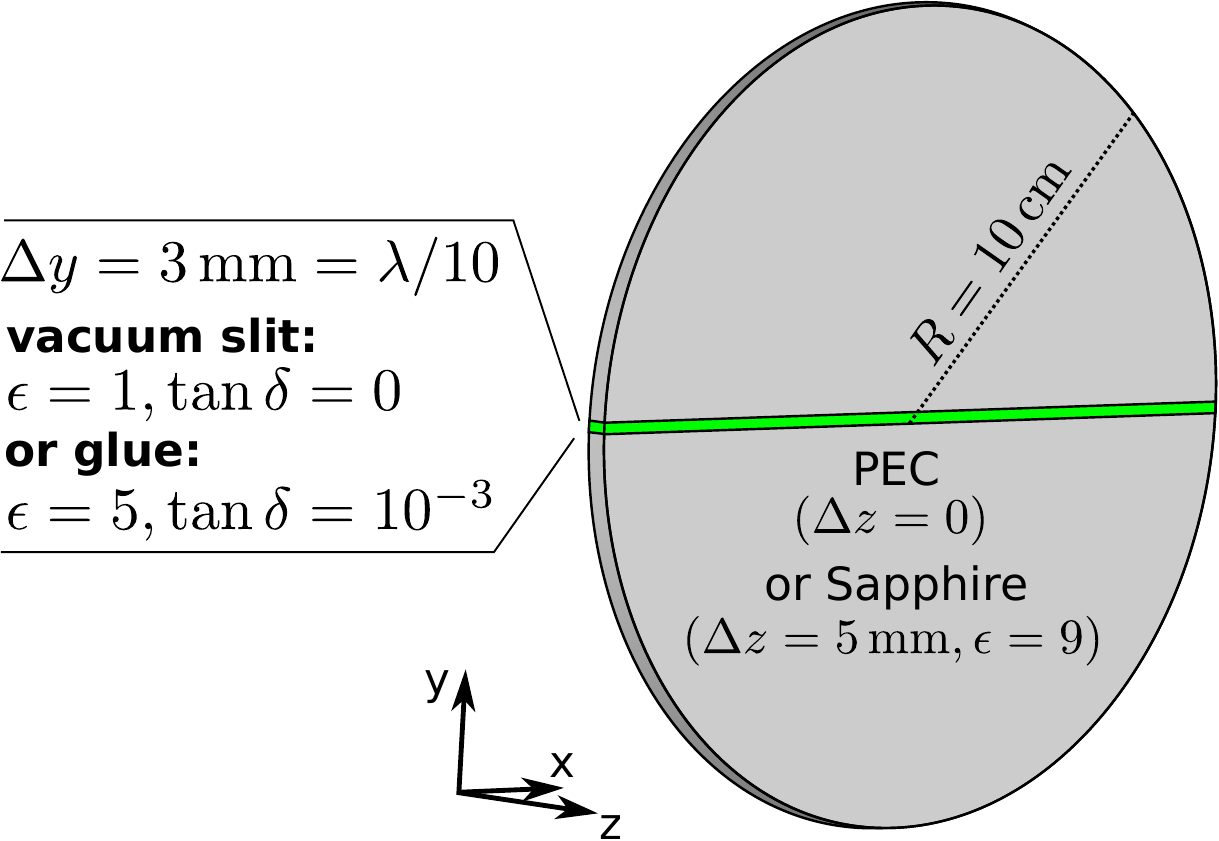}
    \caption{Case definition for simulations with tiled disks. We consider either an infinitesimal thin PEC or a $\lambda / (2 n) =  \SI{5}{\milli\meter}$ thick sapphire disk, each with a vertical/horizontal slit filled with either vacuum or glue with $\epsilon = 5$ and $\tan \delta = 10^{-3}$ inside an external magnetic field pointing in $y$-direction.}
    \label{fig:tiled_disk__powers:cases}
\end{figure}

In the previous sections we have established FEMs for simple cases and have outlined central effects from diffraction and near fields. We now apply this method to settings with geometrical imperfections that may impact the performance of large volume haloscopes.
In MADMAX to get sufficiently large disks many smaller patches of dielectric material will need to be glued together. Due to the non-trivial geometry we omit the application of the Fourier propagation approach in this section and obtain all results in this section with a 3D FEM simulation.
We consider circular disks with a radius of $R \approx 3.3 \lambda \approx \SI{10}{\centi\metre}$ and a slit with an exaggerated width of $\lambda/10 = \SI{3}{\milli\metre}$, cf.~figure~\ref{fig:tiled_disk__powers:cases}. We explore the cases of a PEC and a sapphire disk with a vacuum slit, as well as a sapphire disk with a slit filled with a glue. The glue has $\epsilon = 5$, which is around the expected value for example for \emph{Stycast 2850FT}~\cite{Halpern:86}, and ${\rm tan}\,\delta = 10^{-3}$. For each of these cases we consider slits parallel and orthogonal to the polarization of the axion induced field $\bm{E}_a$.
We choose to place the horizontal (vertical) slit not exactly in the center of the disk but slightly displaced such that its lower edge (its left edge) lies at the center of the disk. This means that the disk is not separated symmetrically, as in a realistic setup also the tiles will not be geometrically perfect.

\subsection{Boost Factor}
\begin{figure}
\centering
    \includegraphics[width=\textwidth]{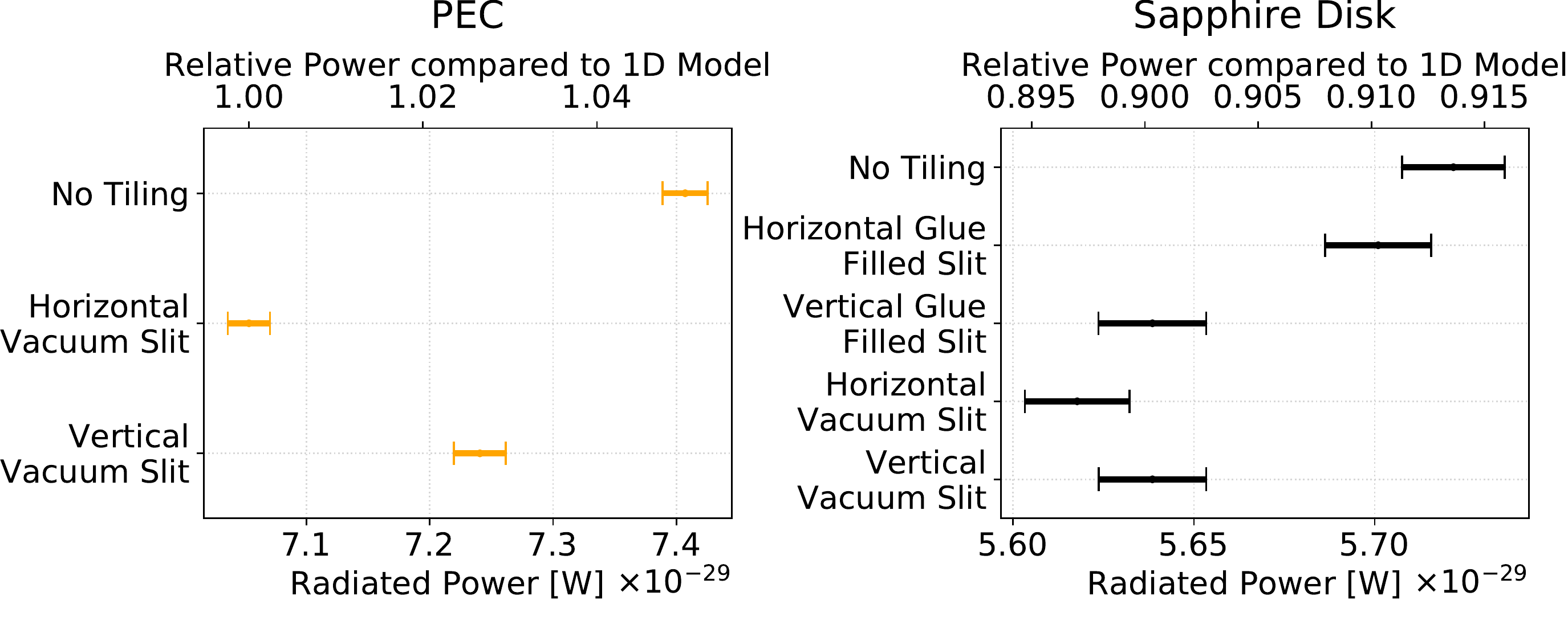}
   \caption{\label{fig:tiled_disk__powers} Power emitted from untiled disks compared to various different tiled disks, at \SI{10}{\giga\hertz}, i.e., $m_a \approx \SI{40}{\micro\electronvolt}$, assuming ${|C_{{\rm a}\gamma}| = 1}$ and an external magnetic field ${B^{(0)} = \SI{10}{\tesla}}$ pointing in the $y$-direction (vertical). The power deviation from the untiled disks roughly matches the naive expectation from the reduction of the size of the emitting surface area of around $2\%$.}
\end{figure}

Figure~\ref{fig:tiled_disk__powers} shows the power emitted by the various tiled disks mentioned before compared with the powers emitted by an untiled PEC and by an untiled sapphire disk, as obtained with Elmer in 3D.
We first compare the Elmer results for the power output of the untiled PEC/disk to corresponding results from the 1D model and observe similar deviations as in the previous section in figure~\ref{fig:dieldisk:bf} (i.e., $10\%$ level). As discussed above, we believe them to arise from diffraction and near field systematics. Since we use the same mesh and solver for all tiled disk cases considered here, also numerical systematics are expected to be the same for all compared tiled disks. Therefore, comparing them to each other is valid although the absolute deviation from the 1D model result is larger.

Now turning our attention to the comparison between the various tiled disks, note that the slit considered here reduces the surface area emitting axion-induced electromagnetic waves by $\approx 2\%$.
This seems to lead to a power-reduction of around the same order compared to the untiled disks as can be seen in figure~\ref{fig:tiled_disk__powers}.
For the sapphire disk with a horizontal glue filled slit the power is not significantly reduced because the surface of the dielectric glue ($\epsilon = 5$) emits electromagnetic waves as well. However, if the slit with the glue is placed parallel to the applied external magnetic field, the boundary condition on the electric field between the glue and the sapphire may constrain possible propagation modes, inhibiting the emission again.
Nevertheless, the overall result is encouraging, since a reduction of emitted power at the order of the relative area covered by gluing slits would not significantly affect the experimental sensitivity, even for multiple slits.

\subsection{Diffraction}
\begin{figure}[tbp]
    \centering
	\includegraphics[width=\textwidth]{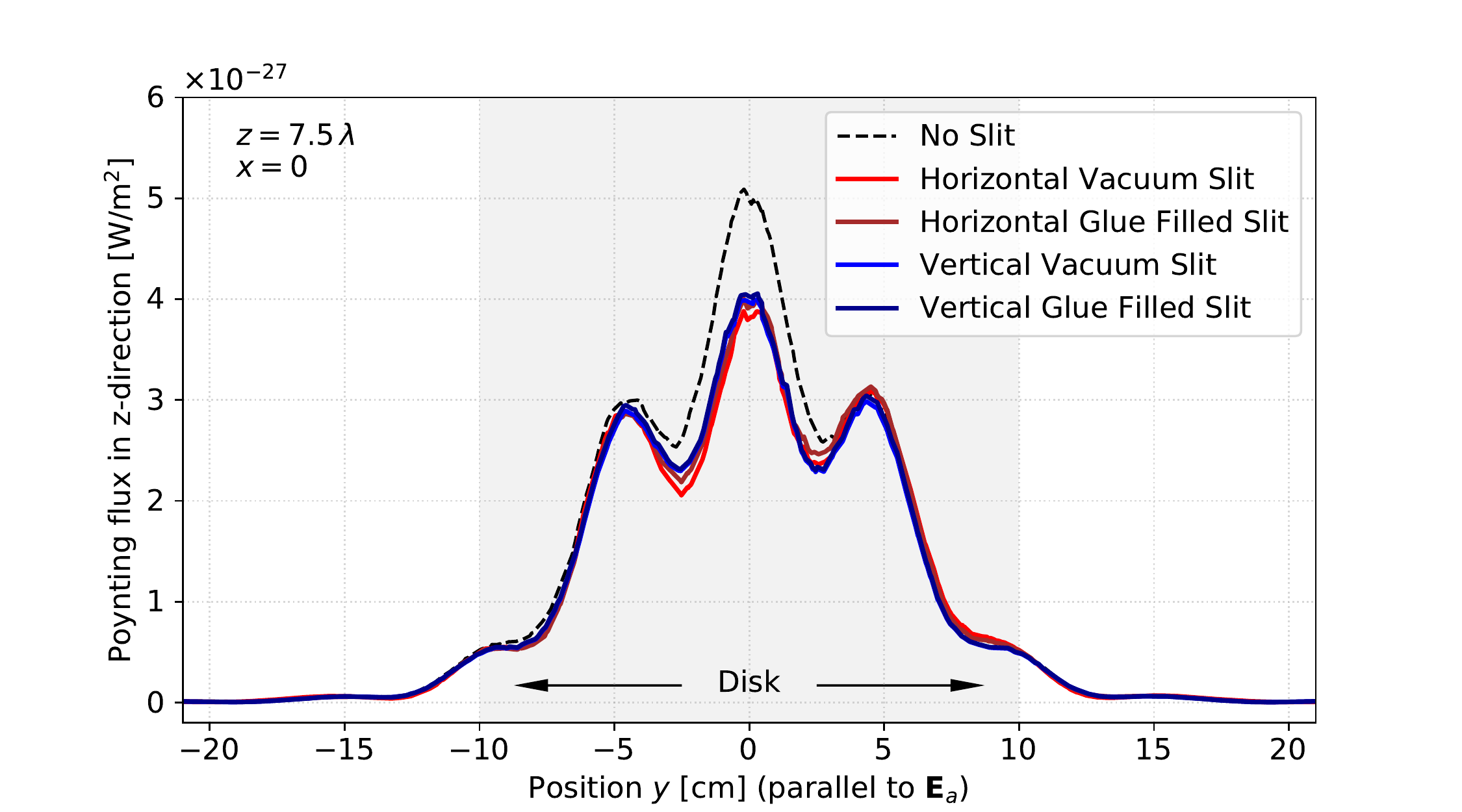}
	\caption{\label{fig:tiled_disk:diffraction}Diffraction pattern of different tiled sapphire disks compared to an untiled sapphire disk at a distance $z=7.5\lambda = \SI{22.5}{\centi\meter}$, at \SI{10}{\giga\hertz}, i.e., $m_a \approx \SI{40}{\micro\electronvolt}$, assuming ${|C_{{\rm a}\gamma}| = 1}$ and an external magnetic field ${B^{(0)} = \SI{10}{\tesla}}$ pointing in the $y$-direction (vertical) 
	as obtained with Elmer. The gray area indicates the extend of the dielectric disk ($D \approx 6.7 \lambda \approx \SI{20}{\centi\meter}$).}
\end{figure}
Figure~\ref{fig:tiled_disk:diffraction} shows the diffraction pattern of the studied tiled dielectric disks in comparison to the one of an untiled disk. The patterns are shown at a distance of ${7.5\,\lambda = \SI{22.5}{\centi\meter}}$ away from the disk in the polarization direction of the axion-induced field $\bm{E}_a$.
The slight asymmetries reflect that the gluing position is not exactly in the center of the disk as mentioned above. More importantly, the diffraction pattern is suppressed with respect to the untiled case by $\approx 20 \%$ at low radii. 
At large distances momenta separate spatially, since they propagate at different angles away from the disk, as discussed above in equation~\eqref{eq:Scalar_Kirchhoff_expanede}.
Therefore, the reduction of the diffraction pattern at low radii indicates that low transverse momenta $k_x, k_y$ are suppressed by the tiling. It is consistent with the naive expectation that a gap or gluing spot in the disk inhibits large transverse wavelengths, corresponding to aforementioned low momentum modes.
Note that it is not trivial to implement these effects in the recursive Fourier propagation approach for the diffraction patterns presented above, since now the 3 regions of the disk (upper half, lower half and glue)  have to be treated separately with appropriate boundary conditions.
The diffraction pattern determines the momentum in $x$- and $y$-direction and the dispersion relation then the momentum in $z$-direction, as discussed in section~\ref{sec:pec}. Therefore, a change due to tiling might cause an additional phase shift compared to the 1D model, which may affect the boost factor. In addition, if the power is radiated at higher transverse momenta, this increases the diffraction loss, as discussed e.g. in section~\ref{sec:pec:diffraction}. Lastly, it obviously changes the beam shape within the dielectric haloscope, which has implications on antenna design.

\subsection{Near Fields}
\begin{figure}
    \centering
	\includegraphics[width=0.78\textwidth]{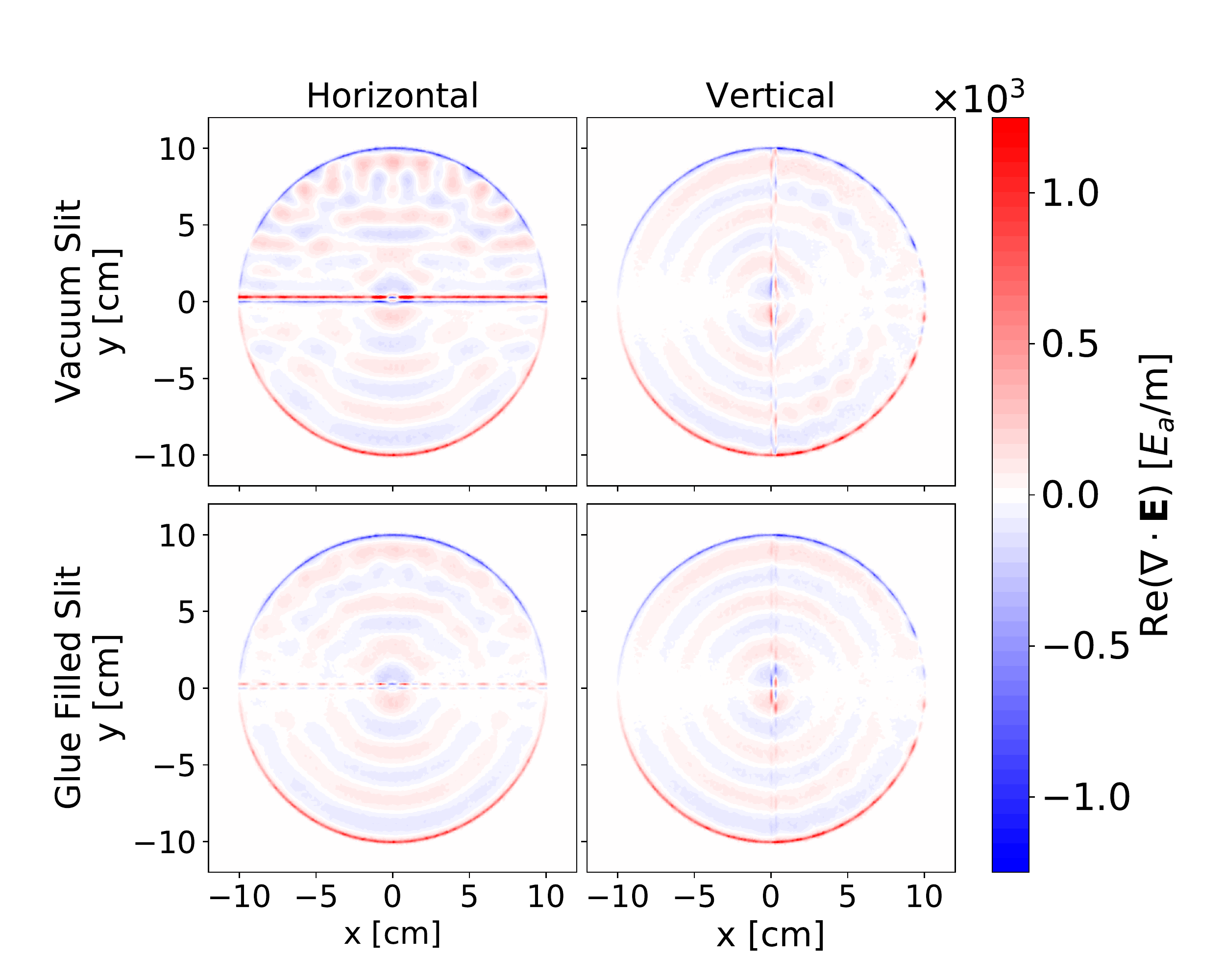}
	\caption{\label{fig:tiled_disks:charges} Polarization charge distributions $\rho = \nabla \cdot \mathbf{E}$ on various tiled sapphire disks (horizontal/vertical slit filled with vacuum/glue) with radius $\approx 3.3 \lambda \approx \SI{10}{\centi\meter}$, at \SI{10}{\giga\hertz}, i.e., $m_a \approx \SI{40}{\micro\electronvolt}$, for an external magnetic field pointing in $y$-direction. The small asymmetry induced by the slit affects mainly surface wave propagation.}
\end{figure}

Figure~\ref{fig:tiled_disks:charges} shows the effect of the tiling on the polarization charges of a dielectric disk. It is apparent that the polarization charge distributions over the whole disk become asymmetric just due to the small asymmetry of the tiling. These asymmetries mainly manifest in different surface wave patterns on both sides of the disks. 
This again may impact the emitted wave from a tiled disk in momentum space, as discussed in the previous section.
In addition, we see additional polarization charge accumulations when the disk is horizontally separated, just as naively expected. The effects of these additional charge accumulations cancel out in the far field but may contribute to the near fields of the disk. 
In the case of a glue filled slit these changes are more moderate due to the smaller difference in dielectric constant. The effect of tiling is expected to be even less relevant for larger disks with larger tiles.

%% file: disk_mirror.tex
In the previous sections~\ref{sec:pec} and~\ref{sec:dieldisc} we have studied the axion-induced electromagnetic emission from a PEC and a circular dielectric disk separately in the presence of a strong external $B$-field. Combining these two objects, we study 3D effects of a minimal dielectric haloscope in this section for the first time. To this end we apply the recursive Fourier propagation approach and FEMs as described in section~\ref{sec:toolbox}.

\subsection{Boost Factor}\label{subsec:boostfactor}

\begin{figure}
 \begin{subfigure}{0.47\textwidth}
   \includegraphics[width=\textwidth]{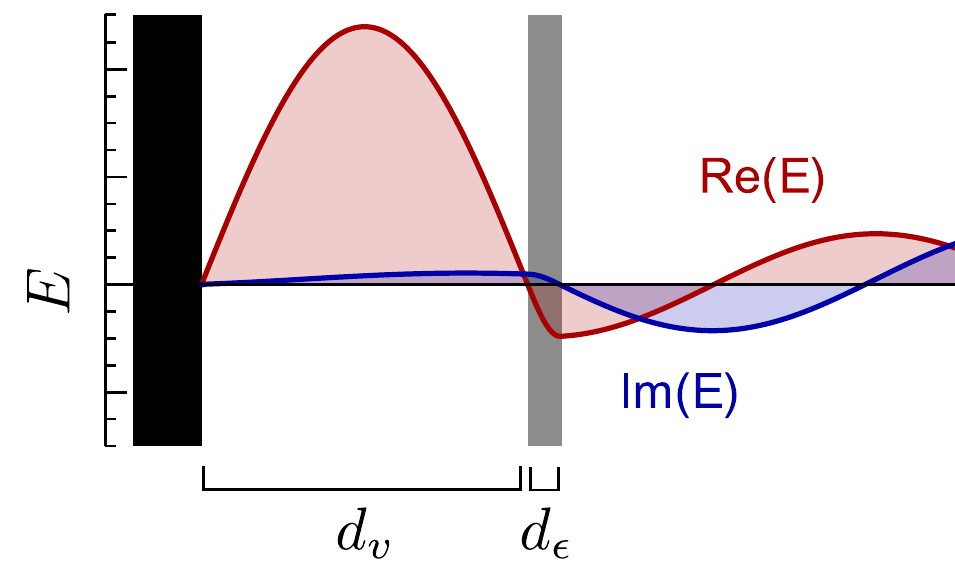}
    \caption{}\label{Fig:Disk_Mirror_Geometry_1D}
  \end{subfigure}\hfill%
  \begin{subfigure}{0.5\textwidth}
    \includegraphics[width=\textwidth]{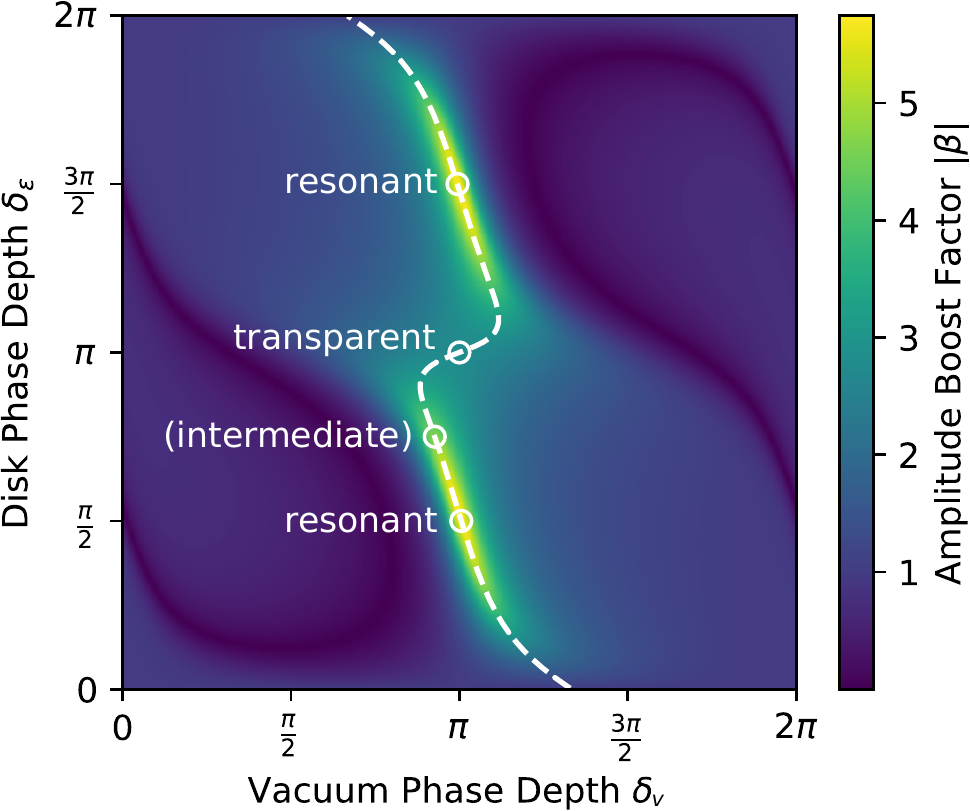}
    \caption{}\label{fig:Disk_Mirror_Boost_1D}
  \end{subfigure}
  \caption{(a) Real and imaginary part of the emitted $E$-field in a minimal dielectric haloscope setup, which consists vacuum gap with phase~depth ${\delta_v= d_v\omega}$ between a PEC mirror and a dielectric disk of phase~depth~${\delta_{\epsilon}=nd_{\epsilon}\omega}$~\cite{millar2017dielectric}.
  (b)~Amplitude boost factor $|\beta|$ for the minimal dielectric haloscope as obtained in the 1D~model. We indicate maximally resonant, transparent and intermediate cases as discussed in the main text. The considered 3D simulations sweep along the parameter space indicated by the dashed line, i.e., for each disk phase depth the vacuum phase depth is optimized to obtain the maximal boost factor.}
  \label{fig:Disk_Mirror_1D} 
\end{figure}

Both PEC (dish antenna) and a single dielectric disk are limited in the amount of generated power. In terms of the boost factor $\beta^2$, which describes the emitted power compared to the power emitted by a single PEC in the 1D model, both are limited to be below one. We now turn our attention to dielectric haloscopes with a single dielectric disk and a PEC (``minimal dielectric haloscope setup''), which can provide already boost factors greater than one.

First let us recap the basic properties of such a setup in 1D \cite{millar2017dielectric} in order to study how they change in 3D. 
In figure~\ref{fig:Disk_Mirror_1D}\,(a) we show a sketch of such a minimal dielectric haloscope. 
For the further discussion we will use the phase depths $\delta_v=\omega d_v$ and ${\delta_{\epsilon}=n\omega d_{\epsilon}}$, where $d_v$ is the distance between disk and mirror and $d_{\epsilon}$ the thickness of the disk and $n=\sqrt{\epsilon}$ its refractive index. In figure \ref{fig:Disk_Mirror_1D}\,(b) we show the boost amplitude from the 1D calculation, where the white dashed line marks for each $\delta_{\epsilon}$ the optimal $\delta_v$ such that the boost factor is maximized. As depicted in the figures the optimal phase depth for this setup is around $\delta_v\approx \pi$. Putting the dielectric disk one wavelength further away from the PEC does not change the situation in the 1D model, hence the boost factor is in general periodic and maximal at around $\delta_v\approx \pi+2\pi n,n\in\mathbb{N}$.  $\delta_v\approx \pi$ means physically that the distance between disk and mirror is $\frac{\lambda}{2}$.
In the resonant case ($\delta_{\epsilon}=\frac{\pi}{2}$ and $\delta_v\approx\pi$) the boost factor reaches its global maximum and the minimal dielectric haloscope is most resonant. This can be understood physically, since for $\delta_{\epsilon}=\frac{\pi}{2}$ the reflectivity of the dielectric disk is maximal (see figure~\ref{fig:dieldisk:reflectivity}). In the transparent case (${\delta_{\epsilon}=\pi}$) the reflectivity of the dielectric disk is zero in 1D, but we still get a boost factor which is larger than one  due to constructive interference of the axion-induced emissions from the PEC mirror and dielectric disk. The case at $\delta_{\epsilon} = \frac{3\pi}{4}$ between resonant and transparent case is denoted as intermediate case in the following.

In the following we will apply the two techniques introduced in section~\ref{sec:toolbox} to explicitly compute the $E$-fields and the boost factors for the minimal dielectric haloscope in 3D. We want to point out that we present here the full 3D field solutions in axion-electrodynamics, since other axion-electrodynamics studies for cavity experiments typically do not compute explicitly the full outpropagating power of the system, but just the modes of the cavity in 3D.
The disk and PEC in the considered minimal dielectric haloscope have both a radius of $R \approx 3.3\lambda \approx \SI{10}{\centi\metre}$ and we consider dielectric disks with $\epsilon=4$ and $\epsilon=9$. Note that sapphire disks ($\epsilon = 9$) with this size are actually currently used in a MADMAX proof of principle study~\cite{Brun:2019lyf}. In the final MADMAX experiment~\cite{TheMADMAXWorkingGroup:2016hpc} the disk diameter will be around~$\SI{1}{\metre}$. 
In addition, we only consider simulations at \SI{10}{\giga\hertz} ($m_a \approx \SI{40}{\micro\electronvolt}$), while the envisioned search range of MADMAX reaches from \SI{10}{\giga\hertz} up to \SI{100}{\giga\hertz} ($m_a \approx \SI{400}{\micro\electronvolt}$).
Therefore, the small radius and frequency considered here are conservative and, as discussed for the dish antenna in section~\ref{sec:pec}, diffraction effects are expected to be less dominant for larger radii and frequencies as envisioned in the final MADMAX setup.

\begin{figure}
\centering
\large{Mirror and Single Dielectric Disk Haloscope -- Power Boost Factor} \\
  \includegraphics[width=\textwidth]{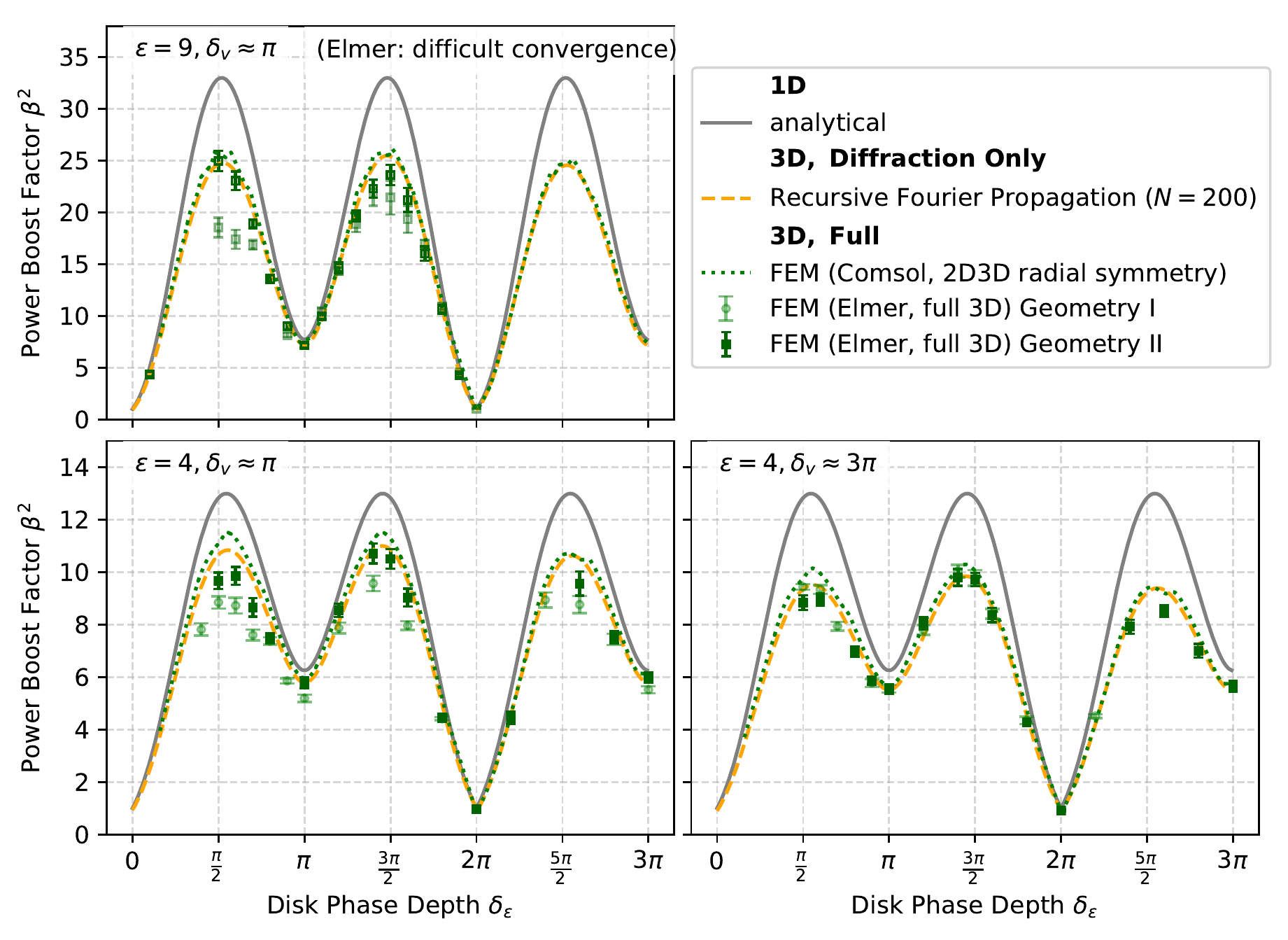}
  \caption{Power boost factor $\beta^2$ as a function of the disk phase depths $\delta_{\epsilon}=\omega n d_{\epsilon}$ for the minimal dielectric haloscope shown in figure~\ref{fig:Disk_Mirror_1D} at \SI{10}{\giga\hertz} ($m_a \approx \SI{40}{\micro\electronvolt}$) and for $\epsilon=9$ (top) and $4$ (bottom). 
   The disk is considered at vacuum phase distances to the PEC mirror of $\delta_v \approx \pi$ (left) and $3\pi$ (right) at which $\beta^2$ is maximized in the 1D model. Both disk and PEC mirror have an identical radius of $R \approx 3.3\lambda \approx \SI{10}{\centi\metre}$. We show our 3D results obtained with various methods introduced in section~\ref{sec:toolbox} in comparison to 1D~results~\cite{millar2017dielectric}. The power boost factor is consistently reduced compared to the 1D model for resonant configurations, i.e., around $\delta_{\epsilon}=\pi/2$, $3\pi/2$, and $5\pi/2$. When we get good convergence of the full 3D FEM calculations, i.e., self-consistent results, all 3D methods give results consistent to within a few percent.}
  \label{fig:Disk_Mirror} 
\end{figure}

In figure~\ref{fig:Disk_Mirror} we show the boost factor from the 1D model and compare it to the 3D calculations. The comparisons are done for three different cases, two different refractive indices and for two different vacuum phase depths $\delta_v\approx\pi,~3\pi$. In all cases the 3D calculations show a boost factor reduction with respect to the idealized 1D calculation. Thus, 3D effects play a role in dielectric haloscopes. We find the largest boost factor reduction in the resonant case, i.e., at $\delta_{\epsilon}=\frac{\pi}{2},\frac{3\pi}{2}\cdots$ to be around 25\% for a sapphire disk. 
In these cases the reflectivity of the dielectric disk is maximal (see figure~\ref{fig:dieldisk:reflectivity}) and the electromagnetic waves which are emitted from the surfaces stay inside the system the longest. 
Therefore, our result follows the naive expectation to find the largest diffraction losses in the resonant case.
Comparing the simulations for $\epsilon = 9$ and $\epsilon = 4$, we see that in the latter case the power reduction is less dominant, which is expected since a dielectric disk with $\epsilon = 4$ has a smaller reflectivity than a disk with $\epsilon = 9$ and makes the system therefore less resonant.
Comparing further the $\epsilon = 4$ cases with $\delta_v\approx\pi$ and $3\pi$, we see that the power boost is reduced for the larger vacuum gap. This is only evident from the 3D solutions and not from the 1D solution, where the power boost is the same. The larger gap between disk and mirror leads to more diffraction losses in the resonant system.
Note that the planned dielectric haloscopes MADMAX and LAMPOST will not operate in the most resonant case.

Let us now discuss the comparison between the various 3D methods in figure~\ref{fig:Disk_Mirror}.
We first consider the diffraction-only calculation using the recursive Fourier propagation approach for $N=200$ iterations (yellow dashed lines) and the full 3D COMSOL result which is obtained by using the radial symmetry (2D3D approach, dotted green lines).
The comparison shows that the recursive Fourier propagation approach can reliably describe the boost factor. This is a surprising result, since the recursive Fourier propagation approach is based on a scalar diffraction theory and neglects boundary and near field effects, which are taken into account by the FEM solutions. Since the distance of the disk and PEC is around $\frac{\lambda}{2}$ and $\frac{3}{2}\lambda$, one may expect that near field effects can play a major role. Nevertheless, our study explicitly shows that these effects only cause small deviations in terms of the emitted power of the minimal dielectric haloscope.

Comparing the full 3D Elmer simulation and the 3D COMSOL simulation making use of the radial symmetry (2D3D approach) shows significant deviations for the first time in this paper. Nevertheless note that the boost factor is still most reduced in the resonant case for the full 3D solution obtained by Elmer.
The error bars in figure~\ref{fig:Disk_Mirror} are obtained by integrating the outgoing power over different $xy$-slices and different quadrants of the simulation domain, i.e., show the self-consistency of the result.
To further check consistency and convergence, we show the results for two different geometries of the outer simulation domain, geometry I and a smaller geometry II.
For a vacuum phase depth of about $\pi$ the Elmer calculations have trouble to converge\footnote{While in previous chapters the convergence parameter in Elmer was always around or below $10^{-7}$, for a sapphire disk haloscope the convergence parameter did not reach below $10^{-4}$, for a $\epsilon = 4$ disk it varied between $10^{-1} - 10^{-9}$ depending on geometry and disk phase depth, more details c.f.~\cite{elmer-solver-manual}.}, do not lead to consistent results and have large numerical errors. When increasing the distance of the dielectric disk and mirror by one wavelength (figure~\ref{fig:Disk_Mirror}\,(right)), the results from both geometries become consistent and confirm the other two 3D results.
On the one hand this shows the limitations of a full 3D FEM simulation in Elmer for large geometries, which can be overcome by assuming radial symmetry (2D3D approach) and reducing the problem by one dimension, as described in section~\ref{subsec:fem-tools}. On the other hand, when the full 3D FEM calculations converge, our results explicitly confirm the results obtained assuming radial symmetry (2D3D approach) in COMSOL or just taking diffraction (recursive Fourier propagation approach) into account.
In addition, even considering most conservatively the full 3D Elmer simulations one expects at most a reduction of at most $50\%$ compared to the 1D model for the power boost factor. Therefore, all our calculations explicitly confirm the feasibility of the dielectric haloscope concept in 3D.

\subsection{Diffraction and Radiation Pattern}\label{subsec:beam_shape}

In this section we explicitly look at the radiation pattern generated by the minimal dielectric haloscope in order to study diffraction and near field effects of the system more thoroughly. This is important also in order to aid antenna design to receive the emitted power from such a system in an optimal way.

\begin{figure}[H]
   \includegraphics[width=0.49\textwidth]{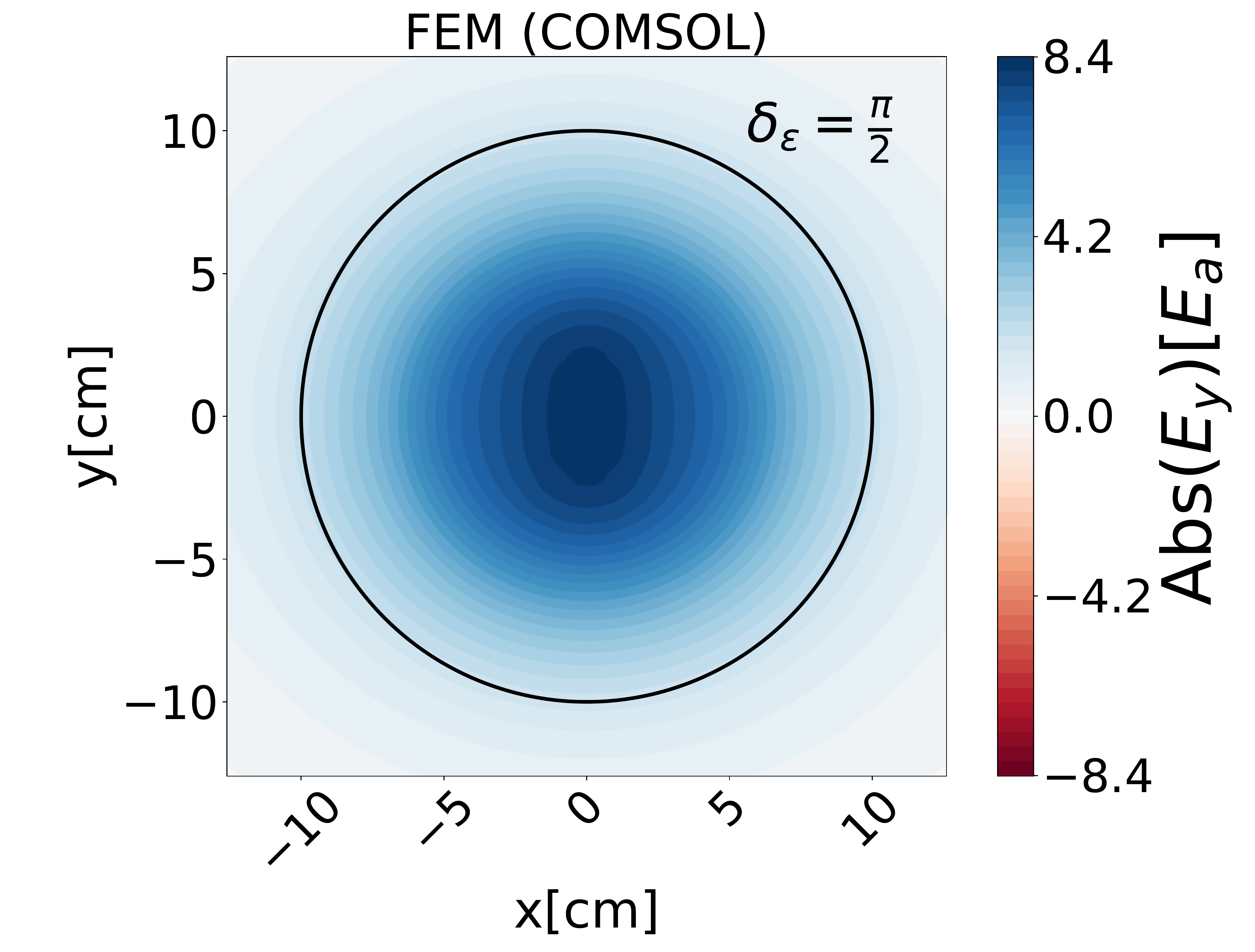}
   \includegraphics[width=0.49\textwidth]{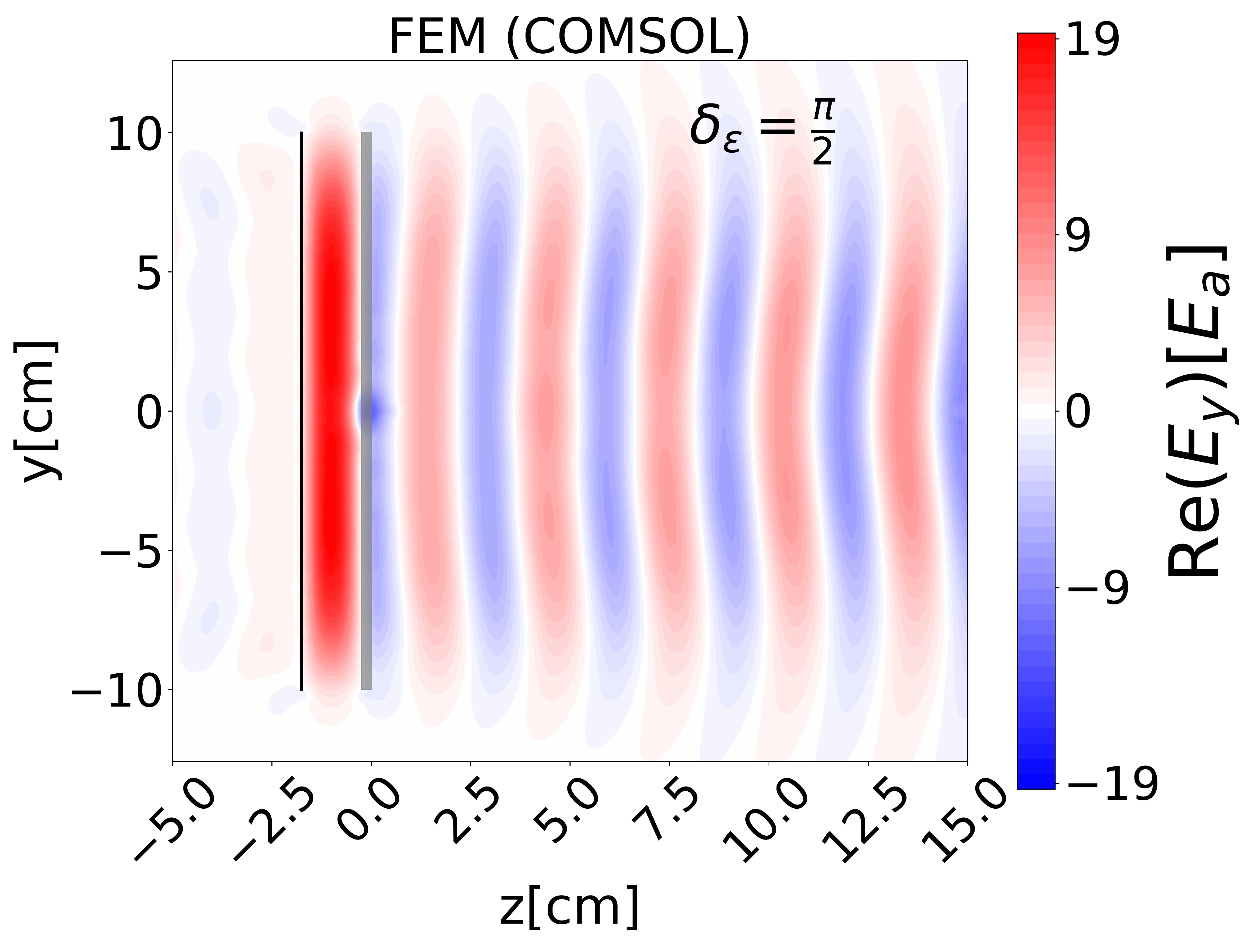}
   \includegraphics[width=0.49\textwidth]{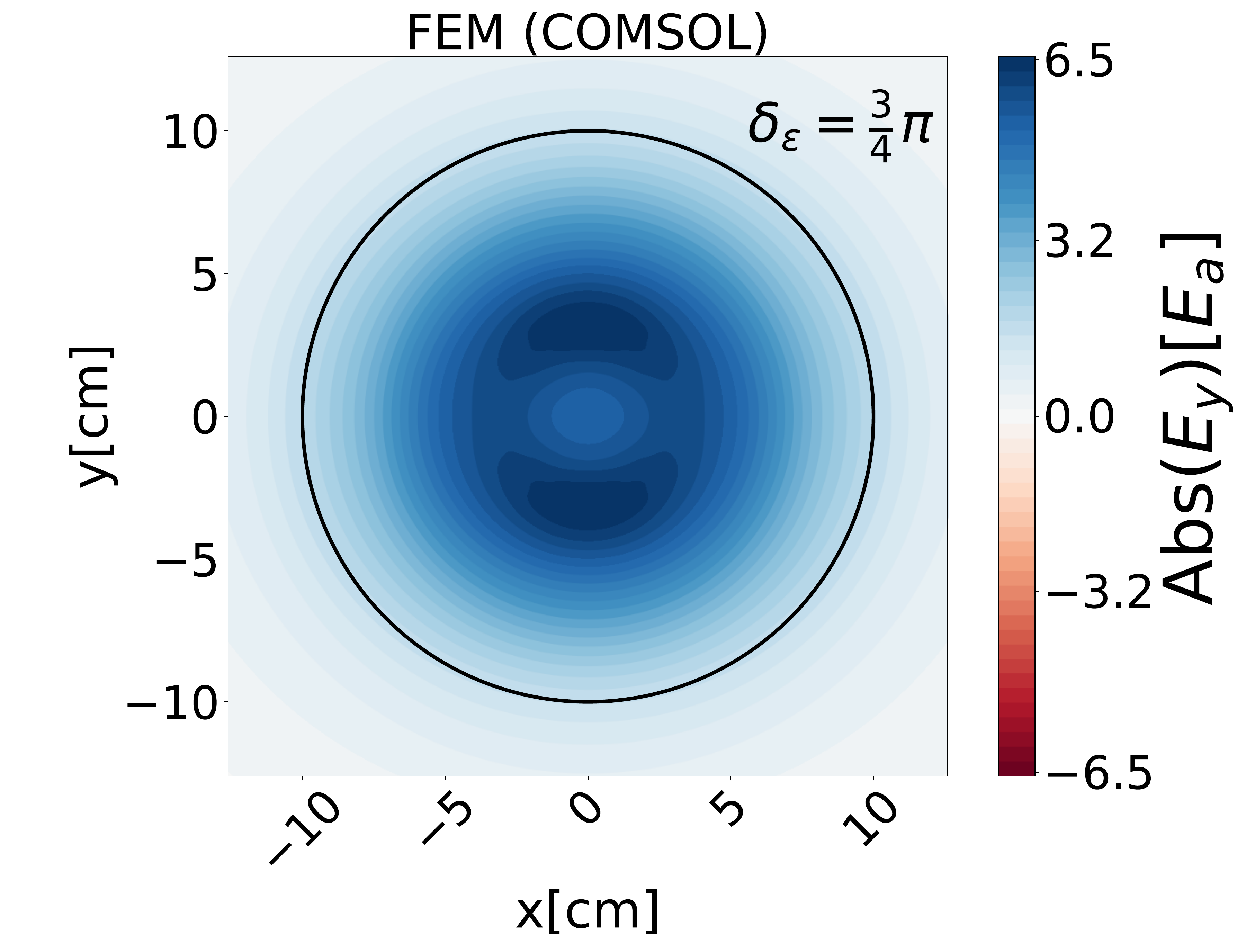}
   \includegraphics[width=0.49\textwidth]{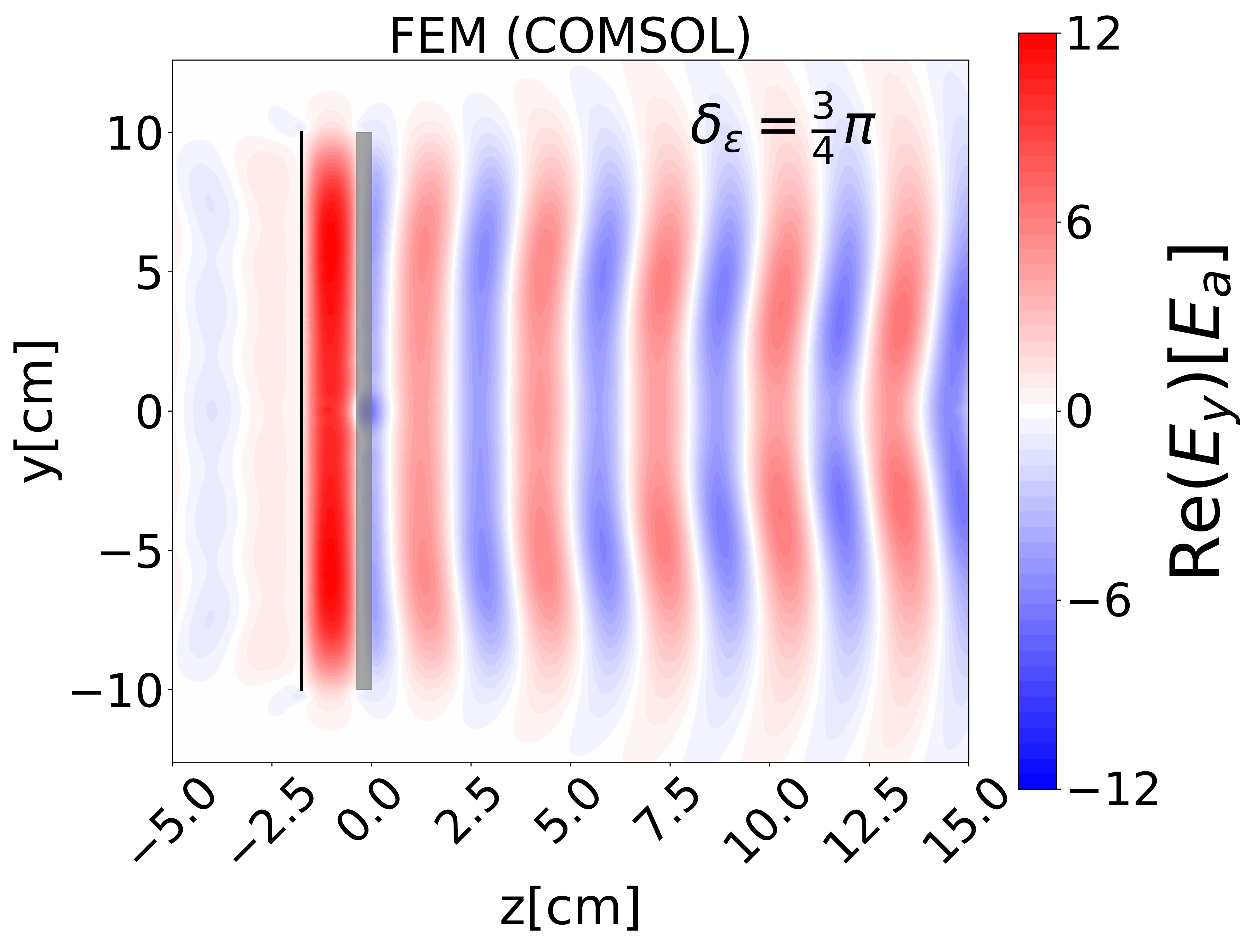}
   \includegraphics[width=0.49\textwidth]{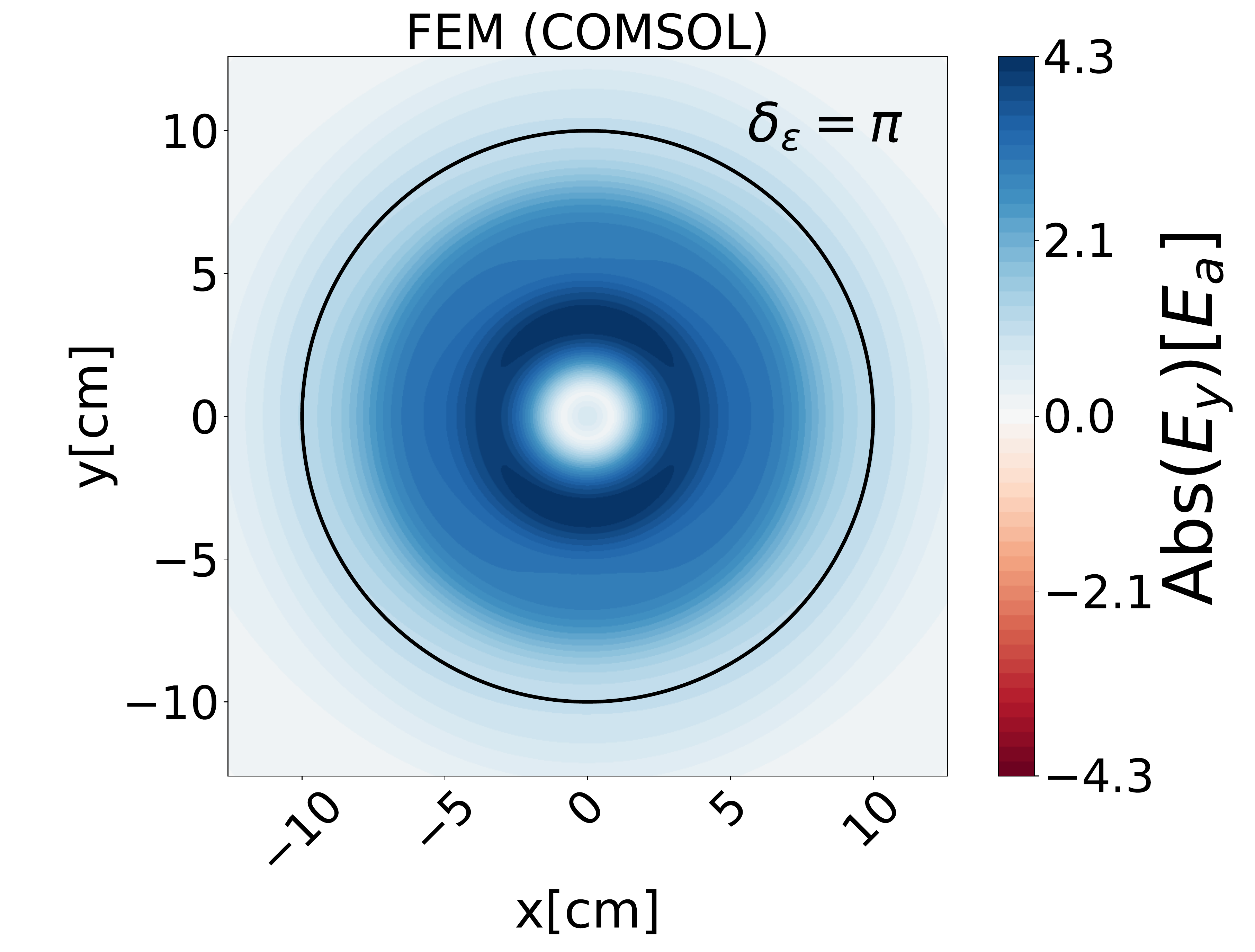}
   \includegraphics[width=0.49\textwidth]{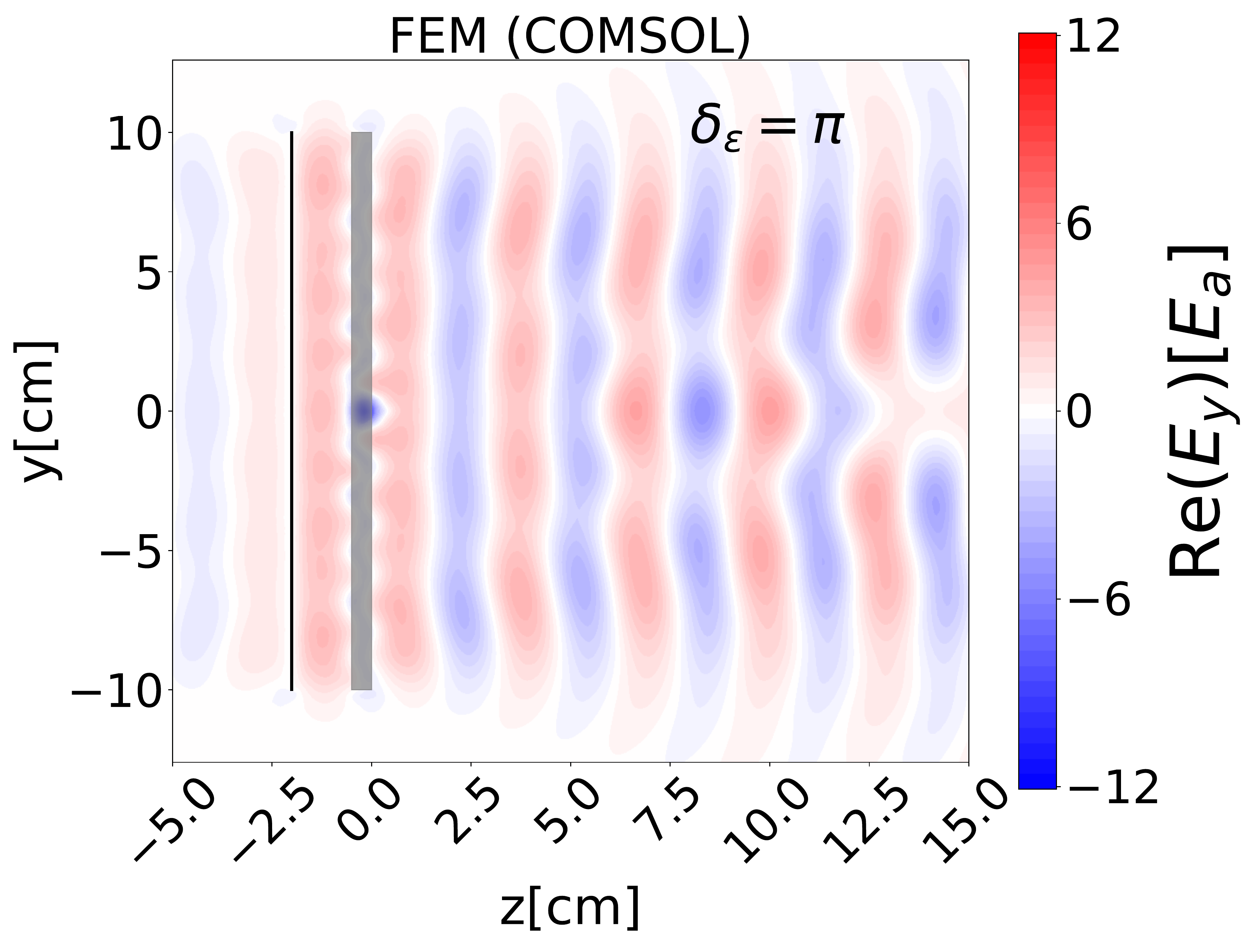}
  \caption{Radiation patterns for a minimal dielectric haloscope. We show the $y$-component of the $E$-field in the $xy$-slices (left) and $zy$-slices (right). The $xy$-slices are evaluated $\SI{14}{\centi\metre}$ away from the minimal dielectric haloscope and the $zy$-slices are evaluated at $y=\SI{0}{\centi\metre}$. The considered frequency is \SI{10}{\giga\hertz}, i.e., $m_a \approx \SI{40}{\micro\electronvolt}$, and the external magnetic field points in $y$-direction.  We show the slices for different phase depths: resonant case {$\delta_{\epsilon} = \pi/2$} (top), intermediate case {$\delta_{\epsilon} = 3\pi/4$} (middle), and transparent case {$\delta_{\epsilon} = \pi$} (bottom). The field pattern acquires more substructure when going from the resonant to the transparent case.}
  \label{fig:Disk_Mirror:shapes_Ey} 
\end{figure}

In figure~\ref{fig:Disk_Mirror:shapes_Ey} we show the $y$-component of the full solution of equation~\eqref{vectorized_E_wave_equation_lin_media_zerovel_EBharmonic} for the radial symmetric minimal dielectric haloscope obtained with the 2D3D FEM approach.
We again consider the resonant case ($\delta_{\epsilon}=\frac{\pi}{2}$), the transparent case ($\delta_{\epsilon}=\pi$) and an intermediate case  ($\delta_{\epsilon}=\frac{3}{4}\pi$). In the right column we observe that the field between the PEC and the disk decreases when going from the resonant case (top) to the transparent case (bottom). As expected from the discussion of the boost factor in figure~\ref{fig:Disk_Mirror} we observe that the emitted field is maximal in the resonant case.  We clearly observe 3D effects, like diffraction of the emitted waves. The shown $E$-field acquires more substructure when going from resonant to the transparent case. This can be also seen in the left column of figure~\ref{fig:Disk_Mirror:shapes_Ey}, where we plot the fields in the $xy$-slice. Going from the resonant case (top) to the transparent case (bottom) we observe again that the magnitude of the $E$-field decreases. Furthermore we see that for the resonant case the beam is concentrated more in the center, while we observe more substructure for the transparent case.
This is as expected, since in the transparent case the emissions from both disk surfaces essentially interfere without being reflected on the disk surface. In the resonant case, however, the system is aligned to be on resonance for only one particular wavelength in $z$-direction. When aligned to be on resonance in the 1D model, the photon dispersion relation fixes the transverse wavelength of such a resonantly enhanced signal to be very large, suppressing substructure.

\begin{figure}[H]
   \includegraphics[width=0.49\textwidth]{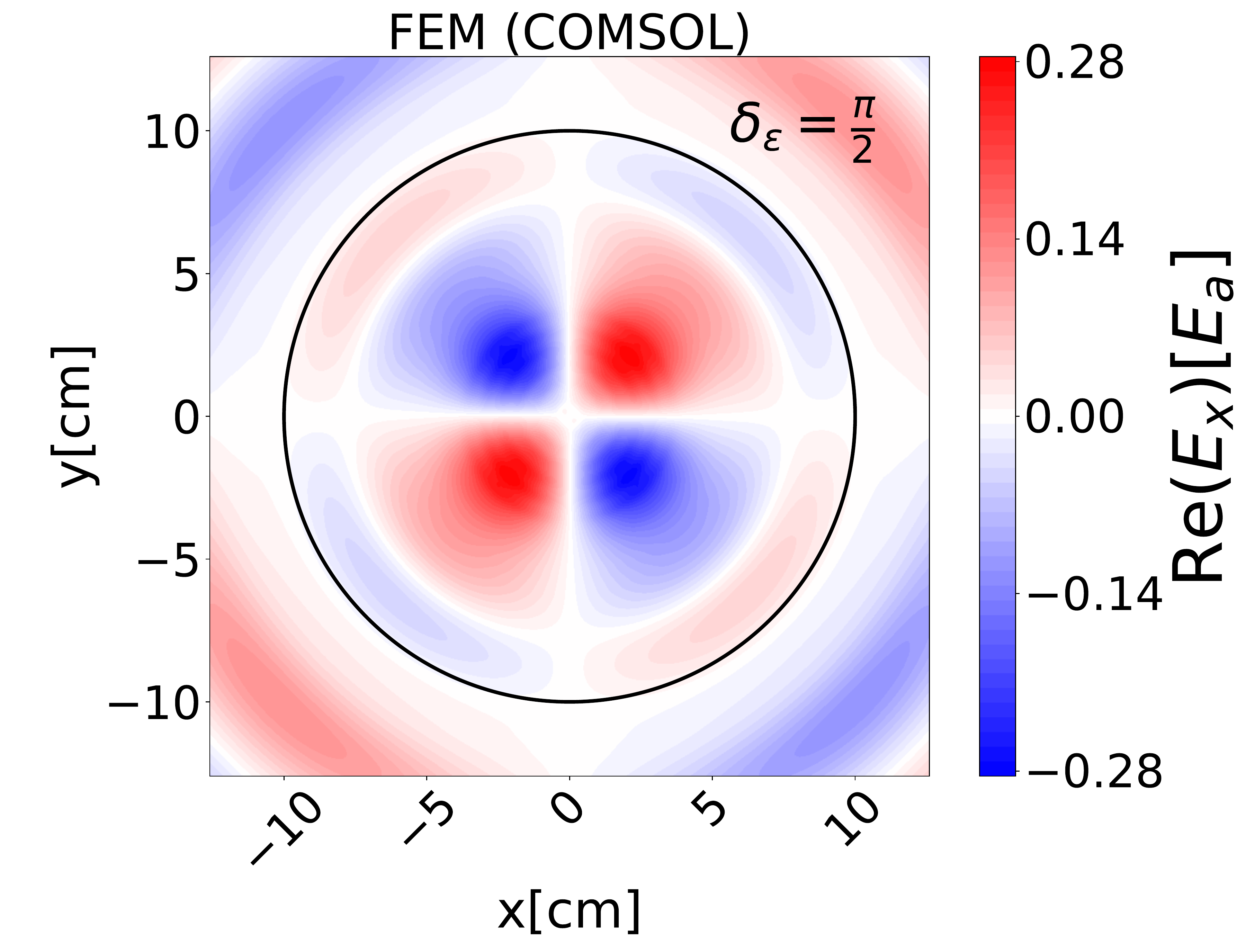}
   \includegraphics[width=0.49\textwidth]{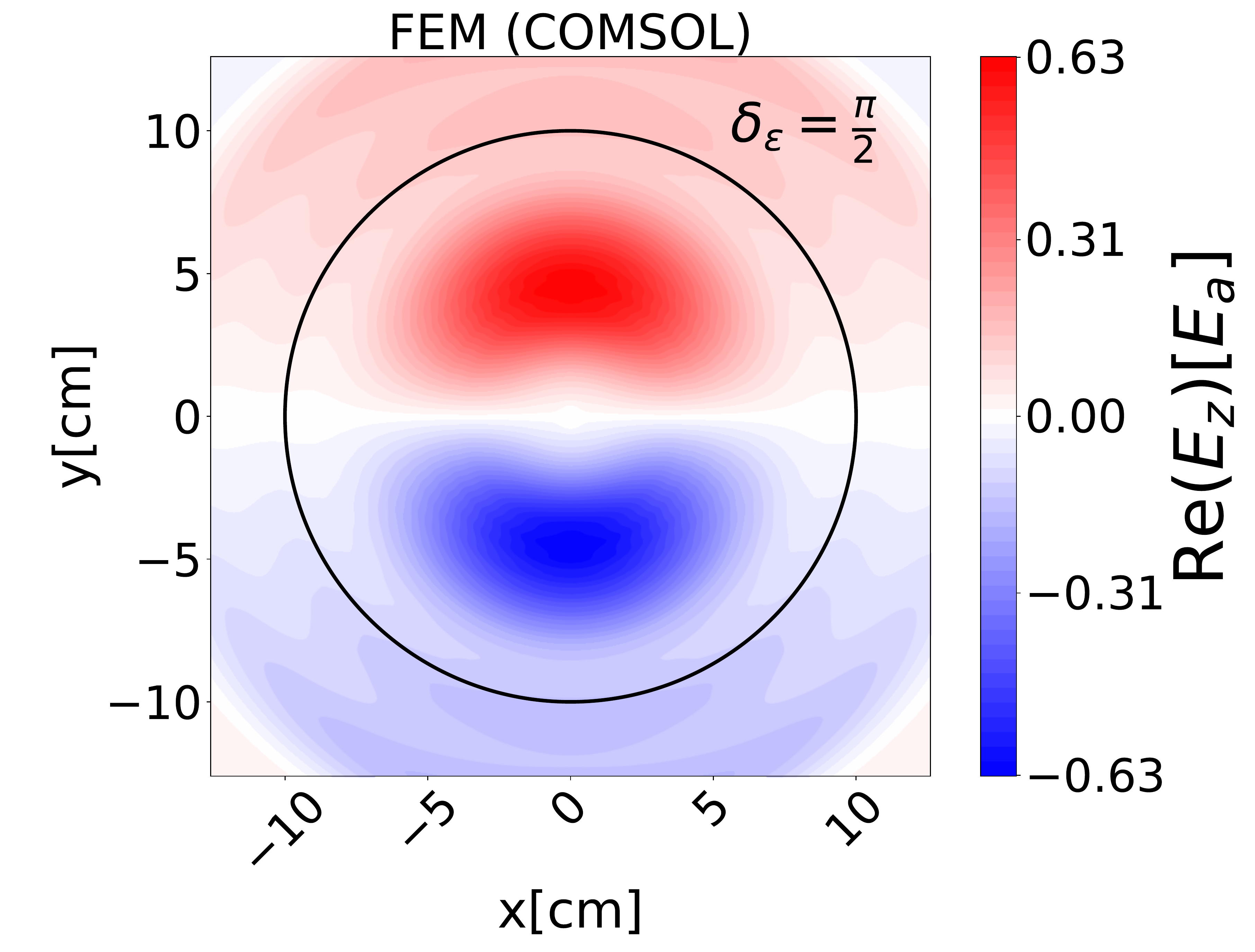}
   \includegraphics[width=0.49\textwidth]{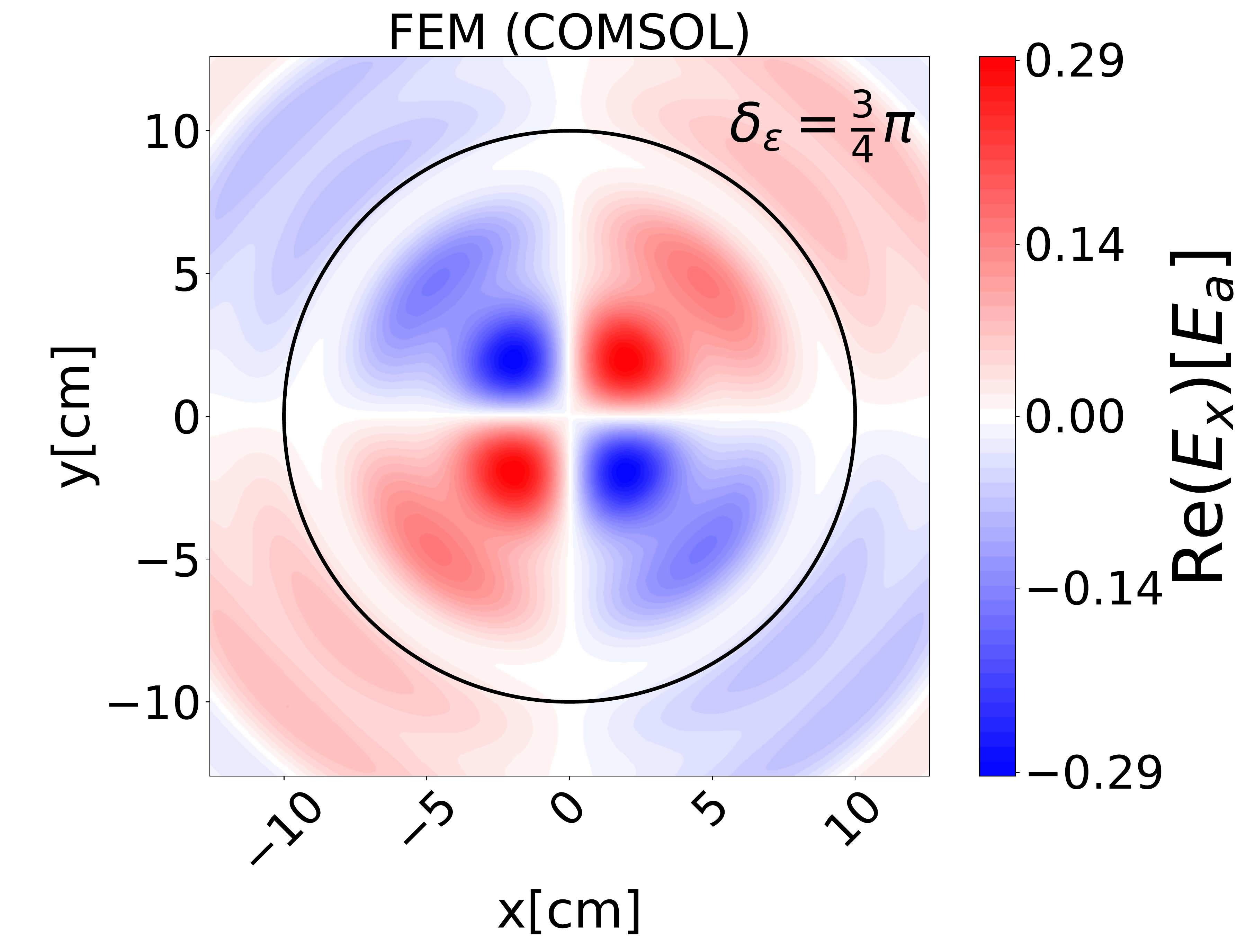}
   \includegraphics[width=0.49\textwidth]{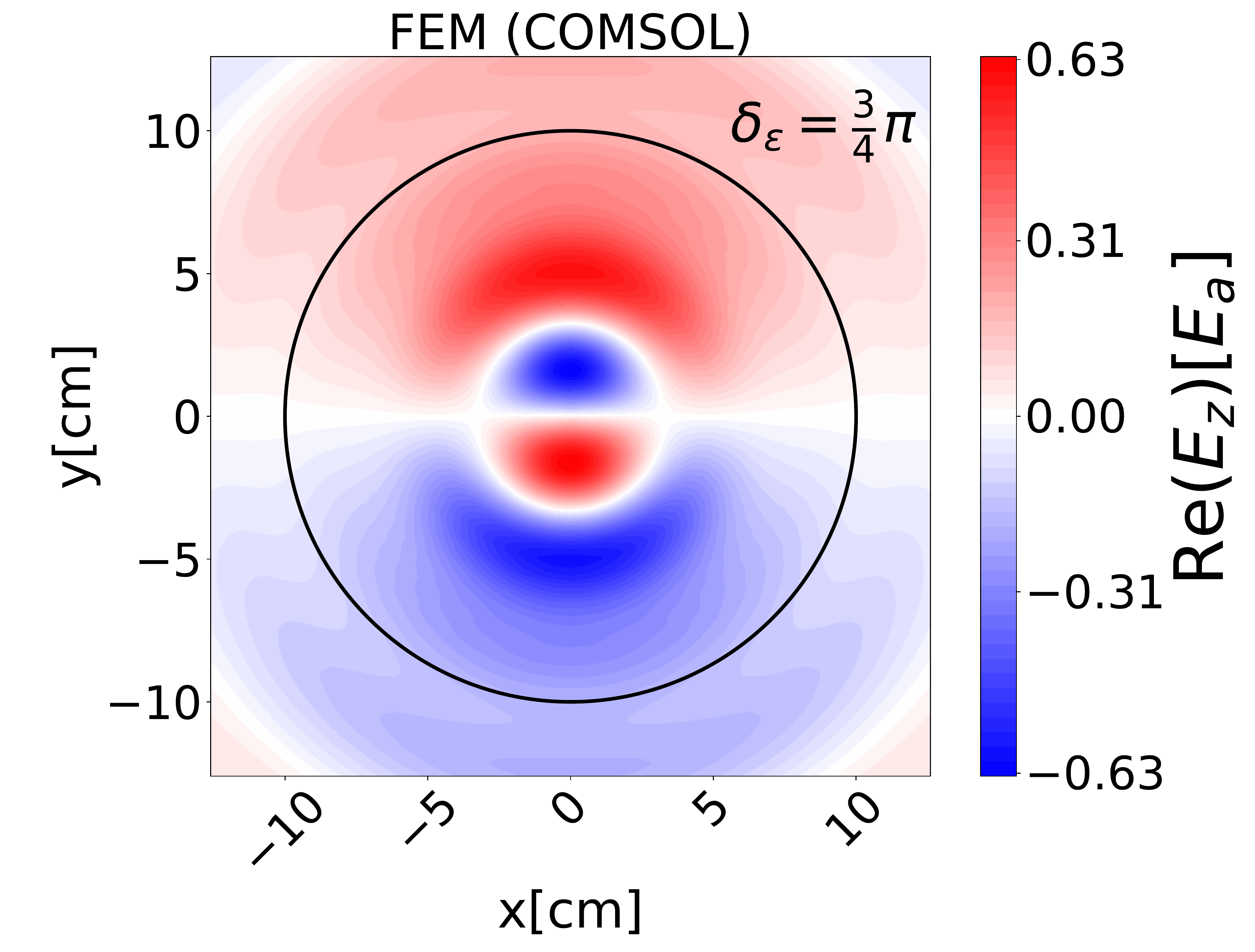}
   \includegraphics[width=0.49\textwidth]{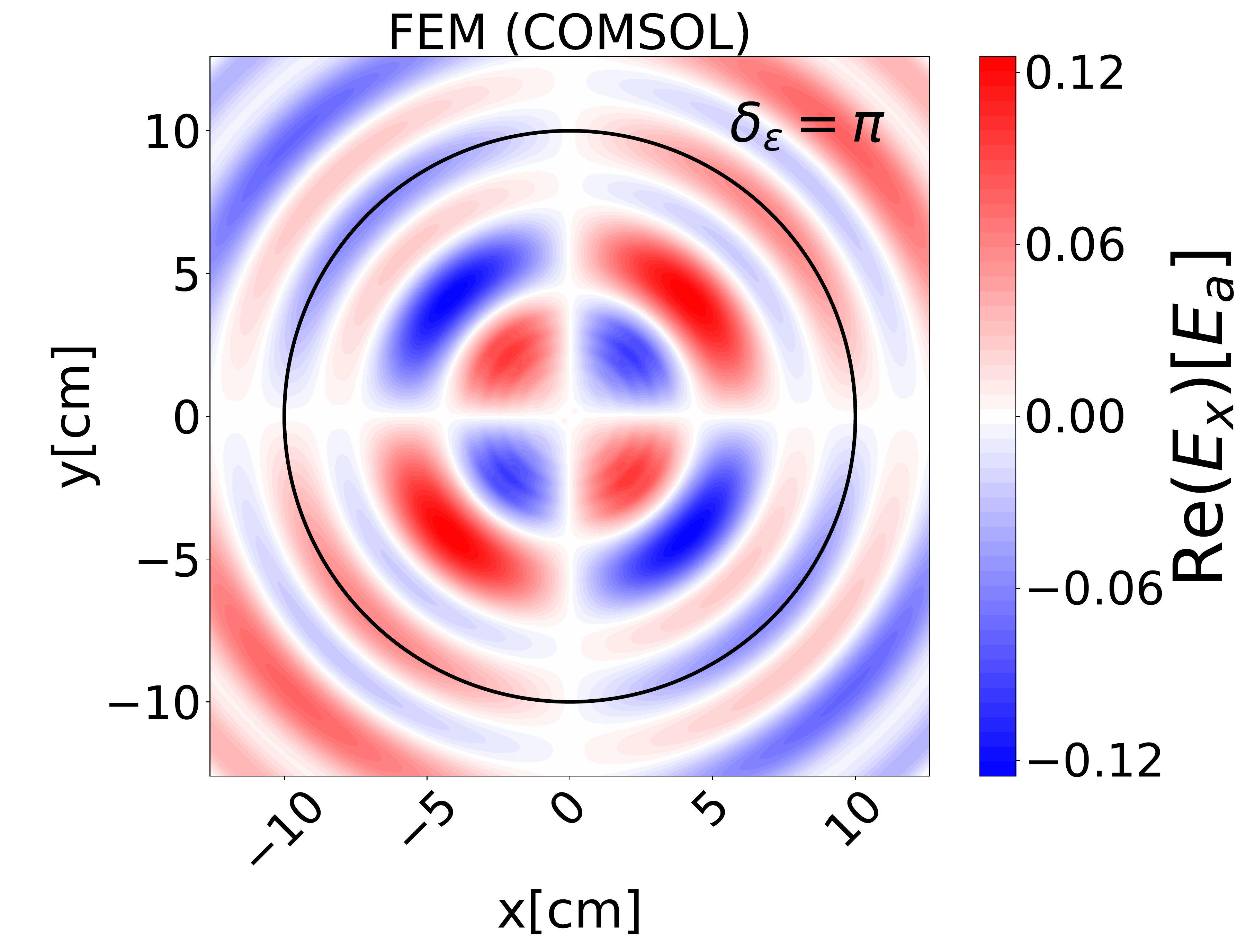}
   \includegraphics[width=0.49\textwidth]{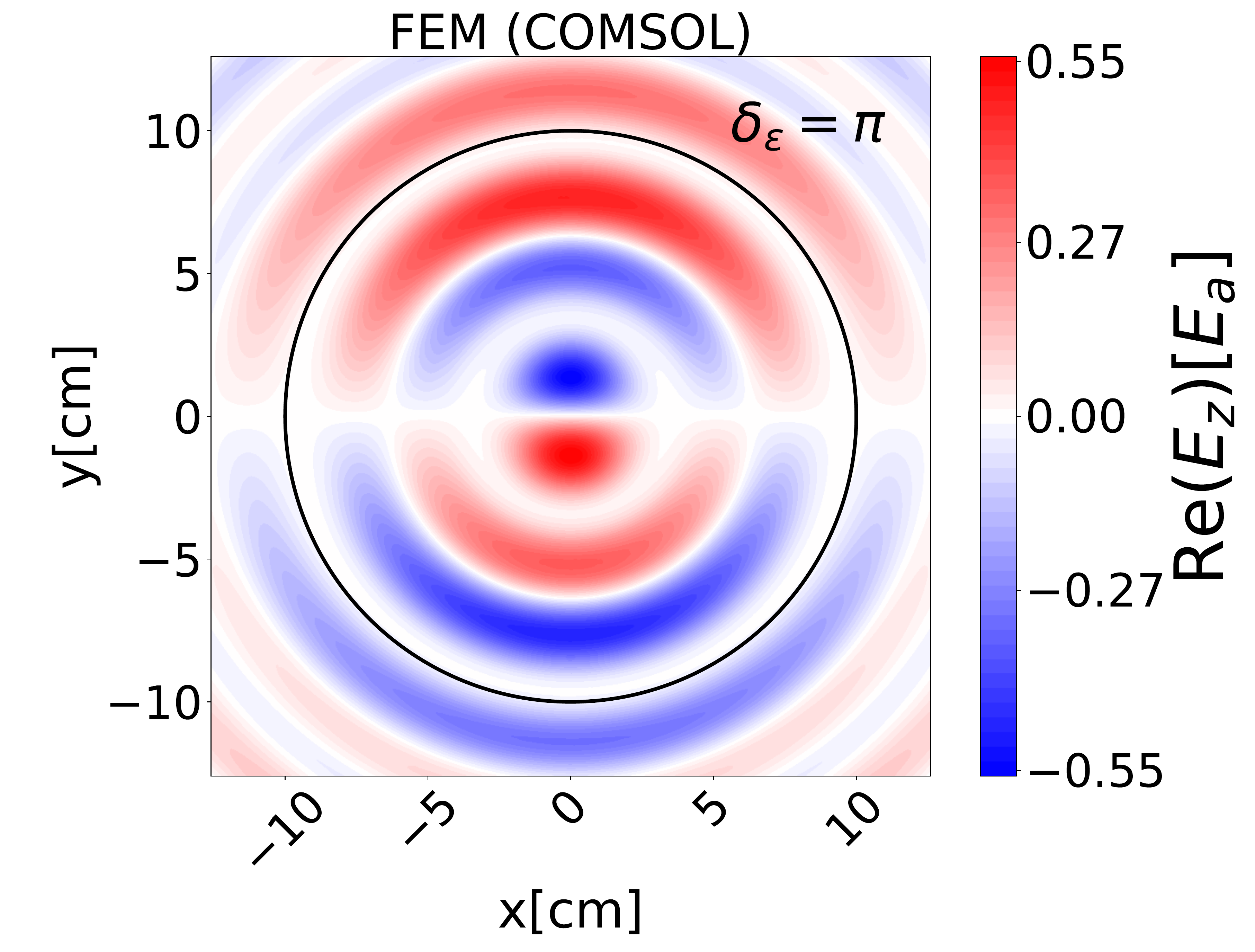}
  \caption{Radiation patterns for a minimal dielectric haloscope. We show the $x$-component (left) and the $z$-component (right) of the $E$-field in $xy$-slices. The $xy$-slices are evaluated $\SI{14}{\centi\metre}$ away from the minimal dielectric haloscope for \SI{10}{\giga\hertz}, i.e., $m_a \approx \SI{40}{\micro\electronvolt}$, and the external magnetic field points in $y$-direction. We show the slices for different phase depths: resonant case {$\delta_{\epsilon} = \pi/2$} (top), intermediate case {$\delta_{\epsilon} = 3\pi/4$} (middle), and transparent case {$\delta_{\epsilon} = \pi$} (bottom). The $x$-component has always a characteristic quadrupole structure, while the $z$-component has always a dipole structure. Note the similarity of the left column to figure~\ref{fig:PEC:NearFields}~(b). Compared to the $y$-component shown in figure~\ref{fig:Disk_Mirror:shapes_Ey}, the $x$ and $z$-components are small since the resonance condition for the considered minimal dielectric haloscope is tuned to enhance the $y$-component.}
  \label{fig:Disk_Mirror:shapes_xz} 
\end{figure}

Moreover, note that the radiation patterns obtained with the recursive Fourier propagation approach again are qualitatively the same than the ones obtained from COMSOL in figure~\ref{fig:Disk_Mirror:shapes_Ey}\,(left). They agree better than $85\%$ when calculating their correlation $|\int E_1^* E_2 dA|^2$ over the shown $xy$-slices with the normalized fields $E_1$ and $E_2$ from both approaches. The decomposition into plane waves in the Fourier propagation approach directly corresponds to the above argument on the transverse wavelength. Therefore, this good match confirms again that near field and boundary charge effects do not have crucial impact on the $y$-component of the emitted electric fields.
In addition, this is supported by looking at the details of the radiation pattern: We observe for example that the beam in the $zy$-slice is always dropping off to the boundaries of the disk, which indicates that boundary charges will be induced mainly by the axion induced field and only partly by the emitted fields which are boosted. In section~\ref{sec:pec} we saw that the $x$ and $z$-components of the $E$-field are due to near field effects and boundary charges. Therefore we expect the $x$ and $z$-components of the $E$-field to be small. This can be confirmed by looking at figure~\ref{fig:Disk_Mirror:shapes_xz} where we show the $E_x$ and $E_z$-fields. In all cases the $x$ and $z$-components of the $E$-fields are smaller than the $y$-components. The most extreme difference is observed in the resonant case, where the $x$ and $z$-components are roughly only $10\%$ of the $y$-component. This confirms, that boundary charges and near fields do affect the fields in the minimal dielectric haloscope, but they are not boosted because they do not fulfill the resonance condition and hence are smaller than the $E_y$-fields. Finally let us mention that we observe in the $x$($z$)-component a characteristic quadrupole (dipole) structure that we already observed in the case of a single PEC. The structure comes from the boundary charge fields and near fields which are now emitted from every disk and superimposed in the end.

%% file: conclusion.tex
In this paper we establish various methods to simulate the electromagnetic radiation generated by axion-induced fields in 3D settings of open axion haloscopes. Previous work for resonant cavities typically focuses on simulating cavity modes to predict their matching with the axion field, just assuming the driving of the resonance and its power loss implicitly~\cite{Hoang:2017bwp,Krawczyk:2018wma,Jeong:2017hqs}. In order to generalize this to open resonators, we explicitly include the axion source term in our simulations and calculate the out-propagation of power from the system in this work. We do so by directly including the axion as a current density term in the electromagnetic wave equation and by simulating for open situations the power propagating out of the simulation domain.
We evaluate different implementations of \emph{Finite Element Methods} (FEM) in COMSOL and Elmer. Furthermore, we reduce the full 3D FEM problem for a radial geometry to a 2D simulation even in the case where the external $B$-field is linearly polarized and breaks the rotational symmetry (\textit{2D3D approach}).
We validate the FEM results with analytically well understood cases in free space and for a dish antenna, i.e., a single perfectly electrically conducting (PEC) mirror. In addition, we gather analytical formalisms to describe diffraction and near fields in simple cases. This allows us to analytically investigate the far field behaviour of the radiation pattern for a dish antenna, also in the presence of velocity corrections.
Moreover, approximating diffraction effects by decomposing into plane waves recursively (\emph{Recursive Fourier Propagation approach}) and neglecting near field effects, we obtain a second fast method for the calculation of diffraction effects inside dielectric haloscopes.
The validation and the comparison of our methods allows for a fundamental understanding of the physics effects at play.
Furthermore, the 2D3D FEM and the Fourier propagation approaches are efficient numerical tools allowing for parameter sweeps like frequency scans in reasonable times. They pave the way to describe full size open axion haloscopes in 3D in the future.

Our results have direct implications on the design of dish antennas and dielectric haloscopes such as MADMAX.
First of all, our studies for dish antennas confirm that diffraction effects are getting important for small diameter-to-wavelength ratios. 
This shows explicitly that the dish antenna and dielectric haloscope approaches are limited towards lower axion masses due to diffraction, i.e., they need to be at least several photon wavelengths in diameter.
To prepare the 3D description of a dielectric haloscope, we show for a single dielectric disk with a diameter of 4 wavelengths that the results obtained with the 1D model essentially stay valid, i.e., stay within deviations of smaller than $\sim 10 \%$ in 3D.

In order to create large dielectric disks, it may be necessary to glue them together from smaller pieces. Therefore, we conduct first simulations of a tiled disk. Our results do not show deviations in the emitted power larger than naively expected from the reduced surface area. However, we see changes to the diffraction pattern of a tiled disk, which leaves the necessity of a more quantitative study of tiled disks in various configurations and with larger diameters for further work.

For a minimal dielectric haloscope setup consisting of a single dielectric disk and single PEC mirror both with a diameter of approximately 7~wavelengths, we show that diffraction losses are a main limitation to the power boost factor $\beta^2$. However, even in the most resonant configuration with a sapphire disk in front of the PEC, the diffraction loss is less than $25 \%$. Such a reduced power boost factor would still allow for an axion search with reasonable sensitivity. In addition, we again expect this effect to be smaller for larger disks and higher frequencies. Also note that the idea of a dielectric haloscope is explicitly not to operate in the most resonant configuration~\cite{millar2017dielectric,Brun:2019lyf}. Moreover, the comparison of the FEM solutions with the result from the Fourier propagation approach shows the important fact that near field and boundary charge effects are negligible. A pure scalar diffraction calculation can predict all relevant effects on the emitted power for the minimal dielectric haloscope.

It remains to apply the methods gathered here to more general cases with further geometrical inaccuracies such as more complicated disk tiling and tilts. Also extended settings with multiple dielectric disks are left for future work.

In summary, we obtain for the first time results on implications of 3D effects such as diffraction and near fields on dielectric haloscopes.
While the 1D calculations approximately stay valid for a single disk, diffraction losses can reduce the power boost factor of a dielectric haloscope. This effect is suppressed by going to large disk radii and high frequencies, as planned in dielectric haloscope searches such as MADMAX and LAMPOST.